\RequirePackage{lineno}
\documentclass[aps,twocolumn,superscriptaddress,showpacs,preprintnumbers,amsmath,amssymb]{revtex4}

\usepackage{enumerate}
\usepackage{graphicx} 
\usepackage{dcolumn}  
\usepackage[title,titletoc]{appendix}
\usepackage[english]{babel}
\usepackage{picinpar}
\usepackage{etoolbox}
\allowdisplaybreaks
\graphicspath{{ps}}

\makeatletter
\renewcommand{\p@subsection}{\Roman{section}.}
\makeatother

\patchcmd{\appendices}{\quad}{: }{}{}

\begin{document}
\preprint{\vbox{ \hbox{   }
					\hbox{Belle preprint \# 2015-8}
                     \hbox{KEK preprint \# 2015-5}
                     \hbox{Submitted to Physical Review D}}
         }
\title{ \quad\\[1.5cm] Study of $D^{**}$ production and light hadronic states in the $\bar{B}^0 \to D^{*+} \omega \pi^-$ decay}
\noaffiliation
\affiliation{University of the Basque Country UPV/EHU, 48080 Bilbao}
\affiliation{Beihang University, Beijing 100191}
\affiliation{University of Bonn, 53115 Bonn}
\affiliation{Budker Institute of Nuclear Physics SB RAS, Novosibirsk 630090}
\affiliation{Faculty of Mathematics and Physics, Charles University, 121 16 Prague}
\affiliation{Chonnam National University, Kwangju 660-701}
\affiliation{University of Cincinnati, Cincinnati, Ohio 45221}
\affiliation{Deutsches Elektronen--Synchrotron, 22607 Hamburg}
\affiliation{Justus-Liebig-Universit\"at Gie\ss{}en, 35392 Gie\ss{}en}
\affiliation{Gifu University, Gifu 501-1193}
\affiliation{SOKENDAI (The Graduate University for Advanced Studies), Hayama 240-0193}
\affiliation{Gyeongsang National University, Chinju 660-701}
\affiliation{Hanyang University, Seoul 133-791}
\affiliation{University of Hawaii, Honolulu, Hawaii 96822}
\affiliation{High Energy Accelerator Research Organization (KEK), Tsukuba 305-0801}
\affiliation{IKERBASQUE, Basque Foundation for Science, 48013 Bilbao}
\affiliation{Indian Institute of Technology Bhubaneswar, Satya Nagar 751007}
\affiliation{Indian Institute of Technology Guwahati, Assam 781039}
\affiliation{Indian Institute of Technology Madras, Chennai 600036}
\affiliation{Institute of High Energy Physics, Chinese Academy of Sciences, Beijing 100049}
\affiliation{Institute of High Energy Physics, Vienna 1050}
\affiliation{Institute for High Energy Physics, Protvino 142281}
\affiliation{INFN - Sezione di Torino, 10125 Torino}
\affiliation{Institute for Theoretical and Experimental Physics, Moscow 117218}
\affiliation{J. Stefan Institute, 1000 Ljubljana}
\affiliation{Kanagawa University, Yokohama 221-8686}
\affiliation{Institut f\"ur Experimentelle Kernphysik, Karlsruher Institut f\"ur Technologie, 76131 Karlsruhe}
\affiliation{Kennesaw State University, Kennesaw GA 30144}
\affiliation{King Abdulaziz City for Science and Technology, Riyadh 11442}
\affiliation{Department of Physics, Faculty of Science, King Abdulaziz University, Jeddah 21589}
\affiliation{Korea Institute of Science and Technology Information, Daejeon 305-806}
\affiliation{Korea University, Seoul 136-713}
\affiliation{Kyungpook National University, Daegu 702-701}
\affiliation{\'Ecole Polytechnique F\'ed\'erale de Lausanne (EPFL), Lausanne 1015}
\affiliation{Faculty of Mathematics and Physics, University of Ljubljana, 1000 Ljubljana}
\affiliation{Luther College, Decorah, Iowa 52101}
\affiliation{University of Maribor, 2000 Maribor}
\affiliation{Max-Planck-Institut f\"ur Physik, 80805 M\"unchen}
\affiliation{School of Physics, University of Melbourne, Victoria 3010}
\affiliation{Moscow Physical Engineering Institute, Moscow 115409}
\affiliation{Moscow Institute of Physics and Technology, Moscow Region 141700}
\affiliation{Graduate School of Science, Nagoya University, Nagoya 464-8602}
\affiliation{Kobayashi-Maskawa Institute, Nagoya University, Nagoya 464-8602}
\affiliation{Nara Women's University, Nara 630-8506}
\affiliation{National Central University, Chung-li 32054}
\affiliation{National United University, Miao Li 36003}
\affiliation{Department of Physics, National Taiwan University, Taipei 10617}
\affiliation{H. Niewodniczanski Institute of Nuclear Physics, Krakow 31-342}
\affiliation{Niigata University, Niigata 950-2181}
\affiliation{Novosibirsk State University, Novosibirsk 630090}
\affiliation{Osaka City University, Osaka 558-8585}
\affiliation{Pacific Northwest National Laboratory, Richland, Washington 99352}
\affiliation{Peking University, Beijing 100871}
\affiliation{Punjab Agricultural University, Ludhiana 141004}
\affiliation{University of Science and Technology of China, Hefei 230026}
\affiliation{Seoul National University, Seoul 151-742}
\affiliation{Soongsil University, Seoul 156-743}
\affiliation{University of South Carolina, Columbia, South Carolina 29208}
\affiliation{Sungkyunkwan University, Suwon 440-746}
\affiliation{School of Physics, University of Sydney, NSW 2006}
\affiliation{Department of Physics, Faculty of Science, University of Tabuk, Tabuk 71451}
\affiliation{Tata Institute of Fundamental Research, Mumbai 400005}
\affiliation{Excellence Cluster Universe, Technische Universit\"at M\"unchen, 85748 Garching}
\affiliation{Toho University, Funabashi 274-8510}
\affiliation{Tohoku University, Sendai 980-8578}
\affiliation{Department of Physics, University of Tokyo, Tokyo 113-0033}
\affiliation{Tokyo Institute of Technology, Tokyo 152-8550}
\affiliation{University of Torino, 10124 Torino}
\affiliation{CNP, Virginia Polytechnic Institute and State University, Blacksburg, Virginia 24061}
\affiliation{Wayne State University, Detroit, Michigan 48202}
\affiliation{Yamagata University, Yamagata 990-8560}
\affiliation{Yonsei University, Seoul 120-749}
  \author{D.~Matvienko}\affiliation{Budker Institute of Nuclear Physics SB RAS, Novosibirsk 630090}\affiliation{Novosibirsk State University, Novosibirsk 630090} 
  \author{A.~Kuzmin}\affiliation{Budker Institute of Nuclear Physics SB RAS, Novosibirsk 630090}\affiliation{Novosibirsk State University, Novosibirsk 630090} 
  \author{S.~Eidelman}\affiliation{Budker Institute of Nuclear Physics SB RAS, Novosibirsk 630090}\affiliation{Novosibirsk State University, Novosibirsk 630090} 
  \author{A.~Abdesselam}\affiliation{Department of Physics, Faculty of Science, University of Tabuk, Tabuk 71451} 
  \author{I.~Adachi}\affiliation{High Energy Accelerator Research Organization (KEK), Tsukuba 305-0801}\affiliation{SOKENDAI (The Graduate University for Advanced Studies), Hayama 240-0193} 
  \author{H.~Aihara}\affiliation{Department of Physics, University of Tokyo, Tokyo 113-0033} 
  \author{S.~Al~Said}\affiliation{Department of Physics, Faculty of Science, University of Tabuk, Tabuk 71451}\affiliation{Department of Physics, Faculty of Science, King Abdulaziz University, Jeddah 21589} 
  \author{K.~Arinstein}\affiliation{Budker Institute of Nuclear Physics SB RAS, Novosibirsk 630090}\affiliation{Novosibirsk State University, Novosibirsk 630090} 
  \author{D.~M.~Asner}\affiliation{Pacific Northwest National Laboratory, Richland, Washington 99352} 
  \author{V.~Aulchenko}\affiliation{Budker Institute of Nuclear Physics SB RAS, Novosibirsk 630090}\affiliation{Novosibirsk State University, Novosibirsk 630090} 
% ...
\collaboration{The Belle Collaboration}

\begin{abstract}
We report on the first observations of $\bar{B}^0 \to D_1(2430)^0 \omega$, $\bar{B}^0 \to D_1(2420)^0 \omega$ and $\bar{B}^0 \to D^*_2(2460)^0 \omega$ decays. The $\bar{B}^0 \to D^{*+} \rho(1450)^-$ decay is also observed. The branching fraction measurements are based on $(771.6 \pm 10.6)\times 10^6$ $B\bar{B}$ events collected at the $\Upsilon(4S)$ resonance with
the Belle detector at the KEKB asymmetric-energy $e^+e^-$ collider.
The fractions of longitudinal polarization of the $D^{**}$ states as well as partial wave fractions of the $D_1(2430)^0$ are obtained.
We also set a $90\%$ confidence level upper limit for the product of branching fractions of $\mathcal{B}(\bar{B}^0 \to D^{*+} b_1(1235)^-) \times \mathcal{B}(b_1(1235)^- \to \omega \pi^-)$.
The measurements show evidence of nontrivial final-state interaction phases for the $\rho$-meson-like amplitudes.
\end{abstract}

\maketitle

\tighten

{\renewcommand{\thefootnote}{\fnsymbol{footnote}}}
\setcounter{footnote}{0}

\section{ \bf INTRODUCTION}

Orbitally excited states of the $D$ meson ($D^{**}$ states) provide a good opportunity to test heavy
quark effective theory (HQET) \cite{neubert} and QCD sum rule predictions \cite{uraltsev}.
The simplest system consists of a charm quark and a light
antiquark in an orbital angular momentum $L=1$ ($P$-wave) state. Four such states are expected
with spin-parities $J^P=0^+$ ($j=1/2$), $1^+$ ($j=1/2$), $1^+$ ($j=3/2$) and $2^+$ ($j=3/2$), where $j$ is the
sum of the light quark spin and angular momentum $L$. All these states have been discovered \cite{pdg}.
They are $D^*_0(2400)$, $D_1(2430)$, $D_1(2420)$ and $D^*_2(2460)$.
The conservation of parity and angular momentum in strong interactions imposes constraints
on the decays of $D^{**}$ states to $D^{(*)} \pi$. The $j=1/2$ states are predicted to decay
mainly through an $S$-wave: $D^*_0(2400) \to D \pi$ and $D_1(2430) \to D^* \pi$.
The $j=3/2$ states are expected to decay mainly through a $D$-wave: $D_1(2420) \to D^* \pi$ and
$D^*_2(2460) \to D \pi$ and $D^* \pi$.
The $j=1/2$ states with $L=1$ are expected and proven to be broad
(hundreds of MeV$/c^2$), while the $j=3/2$ states are expected and proven to be narrow (tens of MeV$/c^2$).
The {\it BABAR} \cite{babar_spectr} and LHCb \cite{lhcb_spectr} collaborations have discovered other excited $D$ mesons interpreted as $n L=2S$ and $n L=1D$ states as well as a possible superposition of several $n L=1F$ states, where $n$ is the radial quantum number.  

Since HQET is violated, the physical $D^{**}$ state with $J^P=1^+$ can contain admixtures of the states with $j=1/2$ and $j=3/2$ \cite{mixing}.

A similar spectroscopy exists for the $D_{sJ}$ states \cite{pdg}. However, the observed masses for the $D^*_{s0}(2317)$ and $D_{s1}(2460)$ resonances with $j=1/2$ are significantly smaller than predicted \cite{dsmasses}. The $D_{s1}(2536)$ and $D_{s1}(2460)$ states with $J^P=1^+$ can mix with each other. This effect is observed in an angular analysis of the $D_{s1}(2536)^+ \to D^{*+} K^0_S$ decay \cite{balagura}.

Precise knowledge of the properties of the
$D^{**}$ states is important to reduce uncertainties
in the measurements of the semileptonic decays and thus in
the determination of the Cabibbo-Kobayashi-Maskawa matrix elements
$|V_{cb}|$ and $|V_{ub}|$ \cite{ckm}.

The $D^{**}$ mesons have been observed in both semileptonic \cite{bsemil} and hadronic $B$ decays \cite{b0todstpi,b-todstpi,lhcbtod3pi,lhcb0todpipi}. 
The recent LHCb study \cite{lhcb0todpipi} shows the first observation of the $\bar{B}^0 \to D^*_3(2760)^+ \pi^-$ decay as well.
The dynamic properties of $D^{**}$ production
are determined by the Wilson operator product expansion \cite{wilson}. In color-favored $\bar{B}^0 \to D^{**+} \pi^-$ decays \cite{b0todstpi,lhcb0todpipi}, dominance of the narrow $D^{**}$ states over the broad ones is observed. A study of $B^- \to D^{**0} \pi^-$ decays \cite{b-todstpi} with the color-favored and color-suppressed possibilities shows approximately equal production of the broad and narrow $D^{**}$ mesons.
It can be explained by a significant suppression  of the narrow states in the color-suppressed channel.
Calculations based on HQET and quark models \cite{fddst} predict such suppression.

\begin{figure*}[ht]
\begin{tabular}{c c}
\includegraphics[width=0.5\textwidth]{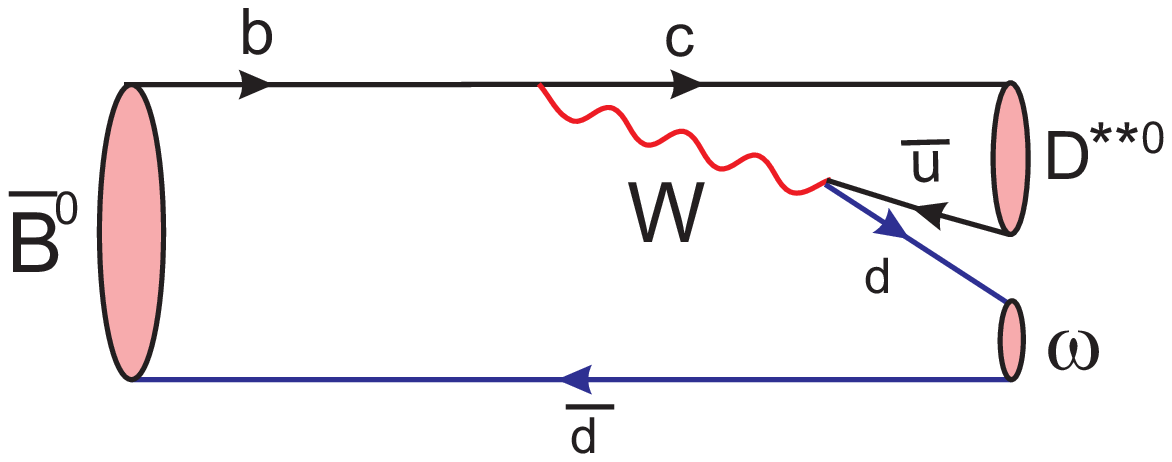} &
\includegraphics[width=0.5\textwidth]{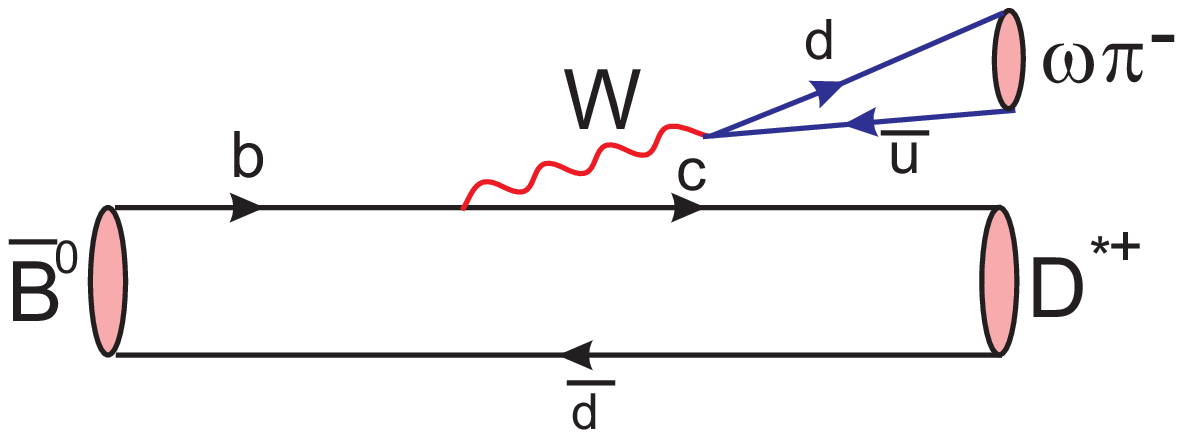}  \\
{\bf (a)} & {\bf (b)} \\
\end{tabular}
\caption{(color online). (a) Color-suppressed and (b) color-favored tree diagrams for the  production of $D^{**}$ and $\omega\pi$ states
in $\bar{B}^0 \to D^{*+} \omega \pi^-$ decays, respectively.}
\label{fig:diag}
\end{figure*}

In this paper, we perform an amplitude analysis of the $\bar{B}^0 \to D^{*+} \omega \pi^-$ decay to measure the decay fractions to $D^{**}$ states produced via the color-suppressed channel [Fig.~\ref{fig:diag}(a)] and to study the $D^{**}$ properties.

This decay is sensitive not only to the vector $D_1(2430)^0$ and $D_1(2420)^0$ states but also to the tensor $D^*_2(2460)^0$ state.
Although the $\bar{B}^0$ decay to $D^*_2(2460)^0 \omega$ is prohibited under the naive factorization hypothesis, it can
nevertheless be produced via final-state interactions (FSI) and/or nonfactorizable contributions.
In soft-collinear effective theory (SCET), color-suppressed decays to the $D^*_2(2460)^0$ state can receive a factorizable contribution at the leading order in $\Lambda/m_B$ \cite{scet}.
This mechanism leads to the equality of branching fractions and strong phases in the decays $\bar{B}^0 \to D^*_2(2460)^0 M$ and $\bar{B}^0 \to D_1(2420)^0 M$, where $M$ is a light meson. Possible deviations from this equality can be attributed to subleading effects \cite{scet}. A discussion of $D^{**}$ production in hadronic $B$ decays can be found in Ref.~ \cite{revddst}.

The color-favored mode of the studied decay [Fig.~\ref{fig:diag}(b)] is saturated by light $\omega\pi$ resonances.
Hadronic weak currents can be classified as either first- or second-class, depending on the combination of spin $J$ and the $P$- and $G$-parities of the $\omega\pi$ system \cite{fscc}. In the standard model, first-class currents (FCC) have $J^{PG}=0^{++},0^{--},1^{+-}$ or $1^{-+}$ and are expected to dominate.
Second-class currents (SCC) have $J^{PG}=0^{+-},0^{-+},1^{++}$ or $1^{--}$ and are associated with a decay constant proportional to the mass difference between the up and down quarks.
Thus, they are expected to vanish in the limit of perfect isospin symmetry.
The decay $\bar{B}^0 \to D^{*+} \omega \pi^-$ is expected to proceed predominantly through the FCC, mediated by $\rho$-meson-like resonances, such as
off-shell $\rho(770)^-$ and $\rho(1450)^-$. In contrast, the SCC may be mediated by $b_1(1235)^-$. SCC searches have
been performed extensively in nuclear $\beta$ decays \cite{betadecay} and $\tau$ decays \cite{taudecay}, with no evidence found.

The structure of the $\rho$-meson-like states is not yet completely clear. The $\rho(1450)$ has a mass consistent with that of a radial $2S$ excitation \cite{rhoradial} but its
decays show characteristics of hybrids \cite{rhohybrid} and suggest that this state may be a $2S$-hybrid mixture \cite{rhomixture}. The observation of the $\rho(1450)$ in $B$-meson decays and the study of its interference with the $\rho(770)$ would lead to a better understanding of the properties of the $\rho$-meson-like states.

Another aim of this study is a test of the factorization hypothesis in the $D^{**}$ production region.
The factorization hypothesis, widely used in heavy-quark physics for hadronic two-body decays, assumes that two hadronic currents may be
treated independently of each other, neglecting FSI. The factorization can be tested by examining the polarization in $B$-meson decays
into two vector mesons. The idea is that, under the factorization, certain hadronic decays are analogous to similar semileptonic decays evaluated at a fixed
value of the momentum transfer, $q^2=M^2_{l \bar{\nu}}$ \cite{factortest}.  Based on the polarization measurements of the decays $\bar{B}^0 \to D^{*0} \omega$ \cite{b0todstomega} and
$B \to \phi K^*$ \cite{btophiks}, we can conclude that nonfactorizable QCD effects are essential in color-suppressed decays. The significant transverse polarizations measured in these decays may arise from the existence of effects from nontrivial long distance contributions, as predicted by SCET studies \cite{scetddst}.
The longitudinal polarizations of similar decays $\bar{B}^0 \to D_{1}(2430)^0 \omega$, $\bar{B}^0 \to D_1(2420)^0\omega$ and $\bar{B}^0 \to D^*_2(2460)^0 \omega$
are measured in our study.

The studied decay has been first observed by the CLEO~\cite{cleo} and
{\it BABAR}~\cite{babar} collaborations, the latter finding an enhancement
in the $D^*\pi$ mass broadly distributed around $2.5$ GeV$/c^2$.

\section{\bf EXPERIMENT AND DETECTOR}

This study uses a data sample containing $771.6 \pm 10.6$ million $B\bar{B}$ events collected at the $\Upsilon(4S)$ resonance with the Belle detector at the KEKB asymmetric-energy $e^+e^-$ collider \cite{kekb}.
The Belle detector, which is a large-solid-angle magnetic
spectrometer based on a 1.5 T superconducting solenoid magnet,
consists of several subdetectors.

Charged particle tracking is provided by a 4-layer silicon vertex detector (SVD)
and a 50-layer central drift chamber (CDC).
The charged particle acceptance covers laboratory polar angles between $\theta\,=\,{\rm 17}^{\circ}$ and ${\rm 150}^{\circ}$, corresponding to about ${\rm 92}\%$ of the total solid angle in the $e^+e^-$ center-of-mass (c.m.) frame.

Charged hadron identification is provided by the ionization energy-loss $dE/dx$ measurements in the CDC,
an array of aerogel threshold Cherenkov counters (ACC), and a barrel-like arrangement of time-of-flight
scintillation counters (TOF). The information from these three subdetectors is combined to form likelihood ratios (PID), which are then used
for pion, kaon and proton discrimination. An electromagnetic calorimeter (ECL),
comprised of 8736 CsI(Tl) crystals and covering the same solid angle as the charged particle tracking system, serves for the detection of electrons and photons.
Electron identification is based on a combination of $dE/dx$ measurements in the CDC, the response of the ACC and energy-to-momentum ratio of an ECL shower with a track as well as a transverse shape of this shower.
An iron flux-return located outside of
the coil (KLM) is instrumented to detect $K_L^0$ mesons and to identify
muons.  The detector
is described in detail elsewhere~\cite{belledet}.

The EvtGen event generator, \cite{evtgen}  with PHOTOS \cite{photos} for radiative corrections and a GEANT-based Monte Carlo (MC) simulation \cite{geant} to model the response of the detector and determine the acceptance, are used in this analysis. The MC simulation includes run-dependent detector performance and background conditions.

\section{\bf EVENT SELECTION}

Candidate $\bar{B}^0 \to D^{*+} \omega \pi^-$ events as well as charge-conjugate combinations are selected. The $D^{*+}$ candidates are reconstructed in the $D^{*+} \to D^0 \pi^+$ mode. The $D^0$ candidates are selected using the $D^0 \to K^- \pi^+$ mode. Other $D^0$ decay modes, which lead to significantly smaller signal-to-noise ratios, are not used in this analysis. The $\omega$ candidates are reconstructed in the $\omega \to \pi^+ \pi^- \pi^0$ mode.

Charged tracks are selected with a set of track quality requirements based on the average hit residuals and impact parameters to the interaction point. To reduce the low momentum combinatorial background, we also require that the track momentum transverse to the beam direction be greater than $100$ MeV$/c$ for all tracks except for the slow pion candidate in the $D^{*+} \to D^0 \pi^+$ decay, for which we apply a looser cut of  $50$ MeV$/c$.

A PID requirement is applied for kaon candidates but not for pion candidates. The kaon identification efficiency is about $90\%$ and the pion misidentification rate is less than $10\%$.
All tracks that are positively identified as electrons are rejected.

Photons are identified as ECL clusters that are not associated with charged tracks and have a minimum energy of $70$ MeV in both the barrel and endcap regions.

$D^{0}$ candidates are reconstructed from $K^- \pi^+$ combinations with an invariant mass within $15$ MeV$/c^2$ of the nominal $D^0$ mass \cite{pdg}.
This window corresponds to approximately $\pm 3$ times the mass resolution.
$D^{*+}$ candidates are selected by combining $D^0$ candidates with an additional track, assumed to be a $\pi^+$. The mass difference $m_{D\pi} -m_{D^0}$ is required to be within $2$ MeV$/c^2$ of its nominal value; the resolution of this quantity is about $0.5$ MeV$/c^2$.

Neutral pion candidates are formed from photon pairs that have an invariant mass within $11.25$ MeV$/c^2$ of the nominal $\pi^0$ mass, which corresponds to about $\pm 2.5$ times the reconstructed mass resolution. To reduce the combinatorial background, the total energy of the photons is required to be greater than $250$ MeV. 

The $\omega$ candidates are formed from a pair of oppositely-charged tracks, assumed to be a $\pi^+\pi^-$ pair, and a $\pi^0$. The invariant mass of the $\pi^+\pi^-\pi^0$ combinations is required to be within $73.5$ MeV$/c^2$ of the nominal $\omega$ mass. This very loose cut retains sideband candidates for background estimation.
The instrumental resolution on the $\omega$ candidates is about $7.3$ MeV$/c^2$.

\begin{figure}[ht]
\includegraphics[scale=0.41]{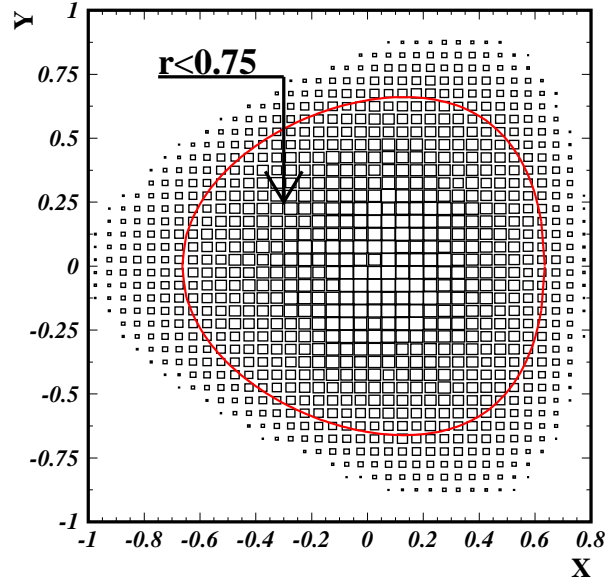}
\caption{(color online). Simulated $(X,Y)$ Dalitz distribution of the $\omega$ signal events. The curve bounds the area selected for further study.}
\label{fig:omdalitz}
\end{figure}

To reduce the number of false $\omega$ candidates formed from random combinations of pions, we impose an
additional requirement in the $\omega$ Dalitz plane, motivated by the $\omega$ decay dynamics and spin-parity in the $\bar{B}^0 \to D^{*+} \omega \pi^-$ decay \cite{jhep}. We define two orthogonal coordinates $X\,=\,3 T_0/Q-1$ and $Y\,=\,\sqrt{3}(T_+-T_-)/Q$, where $T_{\pm,0}$ are the kinetic energies of the pions in the $\omega$ rest frame and $Q=T_0+T_-+T_+$ is the energy release in the $\omega$ decay. Further we define a variable $r$ properly scaled to the kinematic limit as
\begin{linenomath*}
\begin{equation}
r\,=\,\frac{\sqrt{X^2+Y^2}}{r_b}{,}
\end{equation}
\end{linenomath*}
where $r_b$ is the distance from $(0,0)$ to the boundary in the direction of $(X,Y)$.
Since the Dalitz plot density peaks at $r=0$ for the $\omega$ signal, we impose the requirement $r<0.75$. This requirement eliminates about $41\%$ of the background while retaining about $84\%$ of the signal.
In Fig.~\ref{fig:omdalitz} we show the simulated $(X,Y)$ Dalitz plane
of the $\omega$ signal events and the restriction on $r$ variable. 

$B$ candidates are reconstructed by combining a $D^{*+}$ candidate, an $\omega$ candidate, and an additional negatively charged track.
All $B$ candidates are identified using two kinematic variables: the energy difference $\Delta E\,=\,\sum_i\sqrt{|\mathbf{p}^*_i|^2 c^2 + m^2_i c^4}-E^*_{\rm beam}$ and the beam-constrained mass $M_{{\rm bc}}\,=\,\sqrt{E^{*2}_{\rm beam}/c^4-|\sum_i \mathbf{p}^*_i|^2/c^2}$, where the summation is over all particles forming the $B$ candidate, $\mathbf{p}^*_i$ and $m_i$ are their three-momenta and masses, respectively, and $E^*_{\rm beam}$ is the beam energy. All quantities are defined in the $e^+e^-$ c.m.\ frame. We select events with a tight cut on $M_{\rm bc}$ of $5.2725\,{\rm GeV}/c^2\,<\,M_{{\rm bc}}\,<\,5.2845\,{\rm GeV}/c^2$, corresponding to about $\pm 2$ times the mass resolution, and a loose cut on $\Delta E$ of $|\Delta E|< 0.22\,{\rm GeV}$.
To suppress possible continuum events ($e^+e^- \to q\bar{q}$, where $q=u\,, d\,, s\,, c$),
we limit the angle between the thrust
of the $B$ candidate and that of the rest of the event by requiring $|\cos\Theta_{{\rm thrust}}| < {\rm 0.8}$ \cite{brandt}.

In this study, we perform an amplitude analysis that accounts for the kinematic properties of the decay matrix element.
The matrix element should be symmetrized relative to the  exchange of two identical particles in the final state (two $\pi^-$ mesons in our decay mode \cite{pi+}) according to the identity principle. Such symmetrization leads to an interference term in the squared matrix element.
This term consists of two $\omega$ decay amplitudes with different $\pi^+\pi^-\pi^0$ combinations in the $D^{*+} \pi^+\pi^-\pi^0\pi^-$ final state.
Since the $\omega$ is a relatively narrow resonance, the interference term is essential only in the overlapping region of the $\pi^+\pi^-\pi^0$ invariant masses, which is of the order of the $\omega$ width. To correctly describe the angular distributions in this interference region, the internal degrees of freedom of the $\omega$ decay should be taken into account \cite{jhep}.
However, due to lack of statistics, which prevents a full analysis in such a case, we exclude this interference region without significant loss of statistical power.

In order to reduce smearing from detector resolution, a simultaneous fit constraining the $\gamma\gamma$, $K^-\pi^+$, $D^0\pi^+$ and $D^{*+} \omega \pi^-$ invariant masses to match the known $\pi^0$, $D^0$, $D^{*+}$ and $\bar{B}^0$ masses, respectively, is performed. The $\pi^+\pi^-\pi^0$ invariant mass is not constrained to the $\omega$ mass in the fit because of the non-negligible width of the $\omega$ meson.

There are events for which two or more candidates pass all the selection criteria. According to MC simulation, this occurs primarily because of the misreconstruction of one of the pions from the $\omega \to \pi^+\pi^-\pi^0$ decay. To ensure that no $B$ decay is counted more than once, a best-candidate selection is performed based on a $\chi^2$
defined as the sum of three terms. The first determines the deviation of the $\pi^0$ invariant mass from its nominal value, the second represents the deviation of $M_{\rm bc}$ from the nominal $\bar{B}^0$ mass and the third uses the distribution of the difference between the $z$ coordinate at the interaction point of the track corresponding to the primary pion ($\pi^-$) from the $\bar{B}^0$ signal decay and the average $z$ coordinate for the tracks corresponding to the decay products ($K^-$ and $\pi^+$) from the $D^0$ meson decay.
We retain only the $z$ coordinate information  because $B$ mesons are boosted along $z$ and the vertex resolution is worse in that direction. 
We omit the $\omega$ candidate mass in this procedure in order to avoid any bias in the $\omega$ mass distribution since this distribution is used extensively for the background description.

The signal sample is composed of two components---correctly reconstructed (CR) and self cross-feed (SCF)---that are distinguished by whether or not the kinematic variables of the $D^{*+} \omega \pi^-$ decay are well reconstructed. MC simulation shows that the SCF component predominantly occurs due to the combinatorial background for the $\omega$.
To define the CR and SCF components, we
use the following $\chi^2$ describing the deviation of the reconstructed momenta of the final particle system ({\rm rec}) from the generated momenta ({\rm gen}):
\begin{linenomath*}
\begin{equation}
\chi^2\, = \,\sum_{i}\sum_{k=1}^{3}\frac{(x^{(i)}_{k\, {\rm gen}}-x^{(i)}_{k\, {\rm rec}})^2}{\sigma^{2}(x^{(i)}_{k\,{\rm gen}})}{,}
\label{chi2genrec}
\end{equation}
\end{linenomath*}
where $x^{(i)}_{1,2,3}=(p^{(i)},\,\theta^{(i)},\,\varphi^{(i)})$
are the spherical momentum coordinates of the $i$th particle in the final state,
$\sigma(x^{(i)}_{k})$ is the corresponding detector resolution, and the summation is over all tracks and $\pi^0$ forming the $B$ candidate.
We choose to define the CR (SCF) component by the condition $\chi^2<C$ ($\chi^2>C$).
The value of $C=300$ is determined by examining the shapes of the distributions of the difference between the reconstructed and generated kinematic variables.
Variations of the value of $C$ are considered as a source of systematic uncertainty.

Figure~\ref{fig:devsom} shows the distribution of the selected events in the $(\Delta E, M(\pi^+\pi^-\pi^0))$ plane, where we define the following four regions to distinguish between signal and background:
\begin{figure}[ht!]
\includegraphics[scale=0.41]{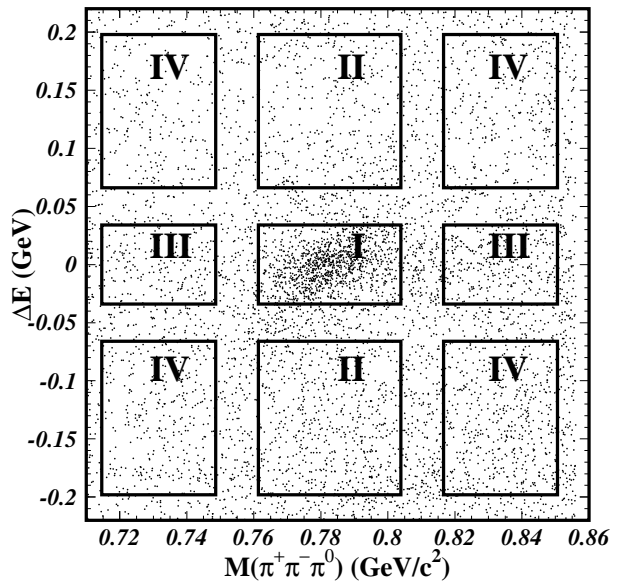}
\caption{Distribution of $\Delta E$ versus $M(\pi^+\pi^-\pi^0)$ for the selected $\bar{B}^0 \to D^{*+} \omega \pi^-$ candidates. The signal Region (${\rm I}$) and sideband Regions (${\rm II}$, ${\rm III}$ and ${\rm IV}$) are shown.
A clear correlation between the variables is seen in Region ${\rm I}$.}
\label{fig:devsom}
\end{figure}
\begin{linenomath*}
\begin{eqnarray}
{\rm I} & \to & |\Delta E|<34\,{\rm MeV}; \nonumber \\ && |M(\pi^+\pi^-\pi^0)-m_{\omega}|<21.25\,{\rm MeV}/c^2{,}\nonumber\\
{\rm II} & \to & 66\,{\rm MeV}<|\Delta E|<198\,{\rm MeV}; \nonumber \\ && |M(\pi^+\pi^-\pi^0)-m_{\omega}|<21.25\,{\rm MeV}/c^2{,}\nonumber\\
{\rm III} & \to & |\Delta E|<34\,{\rm MeV}; \nonumber  \\ && |M(\pi^+\pi^-\pi^0)-m_{\omega}|\,\in\, [34;68]\,{\rm MeV}/c^2{,}\nonumber\\
{\rm IV} & \to & 66\,{\rm MeV}<|\Delta E|<198\,{\rm MeV}; \nonumber \\
&&  |M(\pi^+\pi^-\pi^0)-m_{\omega}|\,\in\,[34;68]\,{\rm MeV}/c^2{.}\nonumber
\end{eqnarray}
\end{linenomath*}
Here $m_{\omega}$ is the nominal $\omega$ mass. Region ${\rm I}$ is the signal region while the others are sideband regions. 
A clear correlation between the $\Delta E$ and $M(\pi^+\pi^-\pi^0)$ variables is seen in Region ${\rm I}$ due to the experimental resolution.
The signal window for the $\omega$ invariant mass corresponds to $\pm 2.5$ times the world-average $\omega$ width of $8.5$ MeV$/c^2$.

\begin{figure}[ht!]
\includegraphics[scale=0.41]{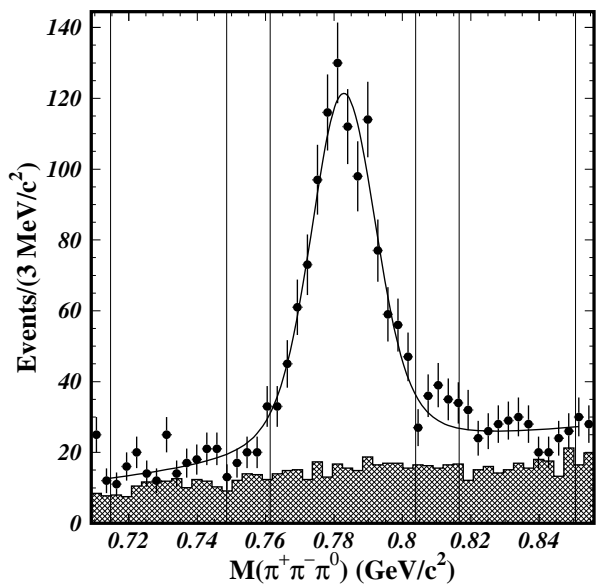}
\caption{$M(\pi^-\pi^+\pi^0$) distribution of the $\bar{B}^0 \to D^{*+} \omega \pi^-$ candidates in the $\Delta E$ signal region (points with error bars) and sideband (hatched histogram). The sideband distribution is normalized to the size of the $\Delta E$ signal region. The $M(\pi^+\pi^-\pi^0)$ signal region and sideband are indicated by the vertical lines. The curve is the result of the fit described in the text.}
\label{fig:mom}
\end{figure}

Figure~\ref{fig:mom} shows the $M(\pi^+\pi^-\pi^0)$ distributions in the $\Delta E$ signal and sideband regions defined above. The curve corresponds to the sum of a Voigtian function (the convolution of a Breit-Wigner function with a Gaussian function) and a linear background function. The $\omega$ mass, the Gaussian resolution $\sigma$, and the parameters of the linear function are free in the fit but the Breit-Wigner width is fixed to the world-average decay width of the $\omega$ \cite{pdg}.
The difference between the number of observed events away from the $M(\pi^+\pi^-\pi^0)$ peak and the number of events predicted from the $\Delta E$ sideband is explained by the $\Delta E$ peaking background component, corresponding to $\bar{B}^0 \to D^{*+} \pi^+ \pi^- \pi^0 \pi^-$ decays.

\section{\bf TOTAL BRANCHING FRACTION}

\begin{figure*}[ht!]
\begin{tabular}{c c}
\includegraphics[scale=0.41]{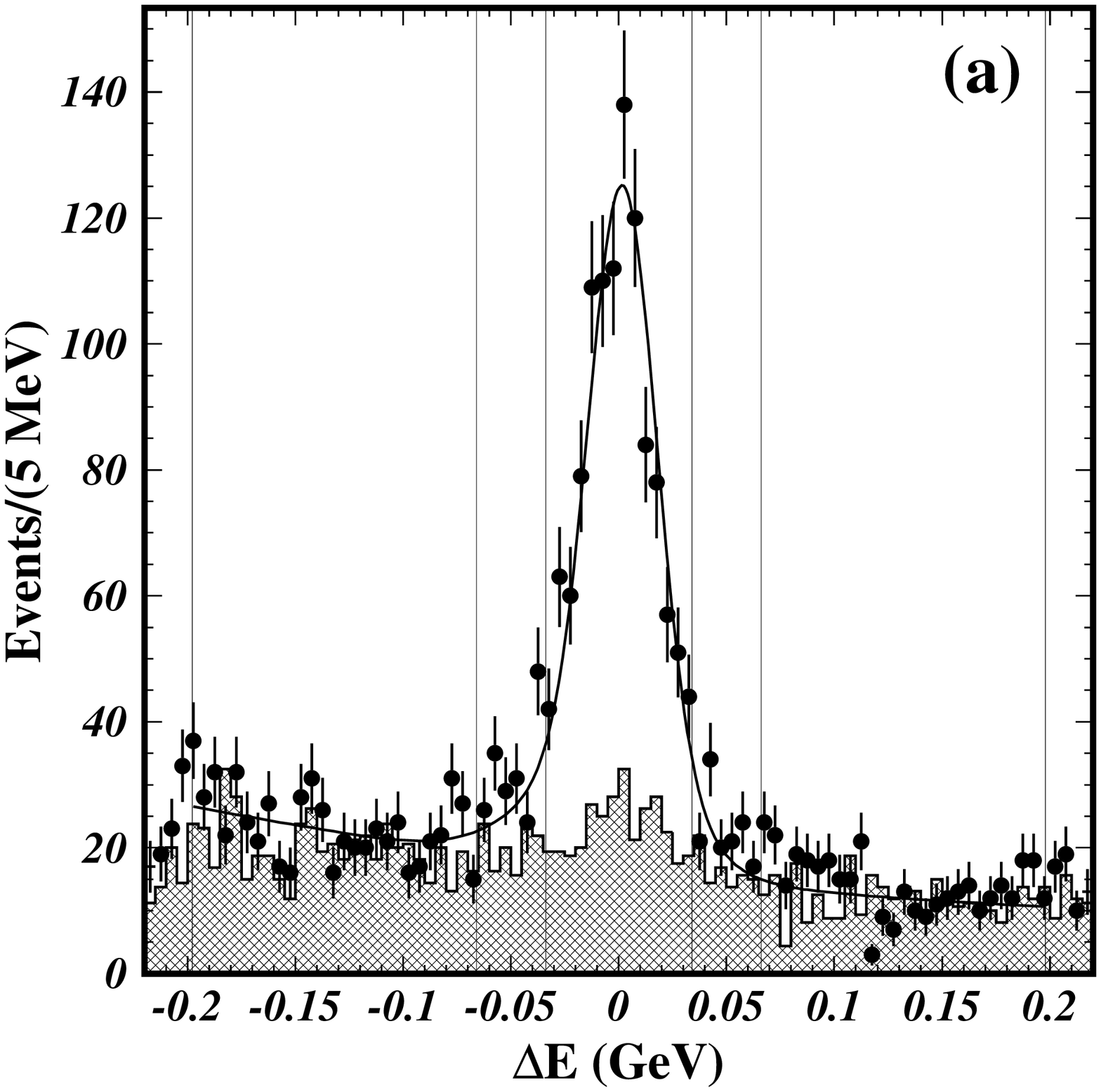} &
\includegraphics[scale=0.41]{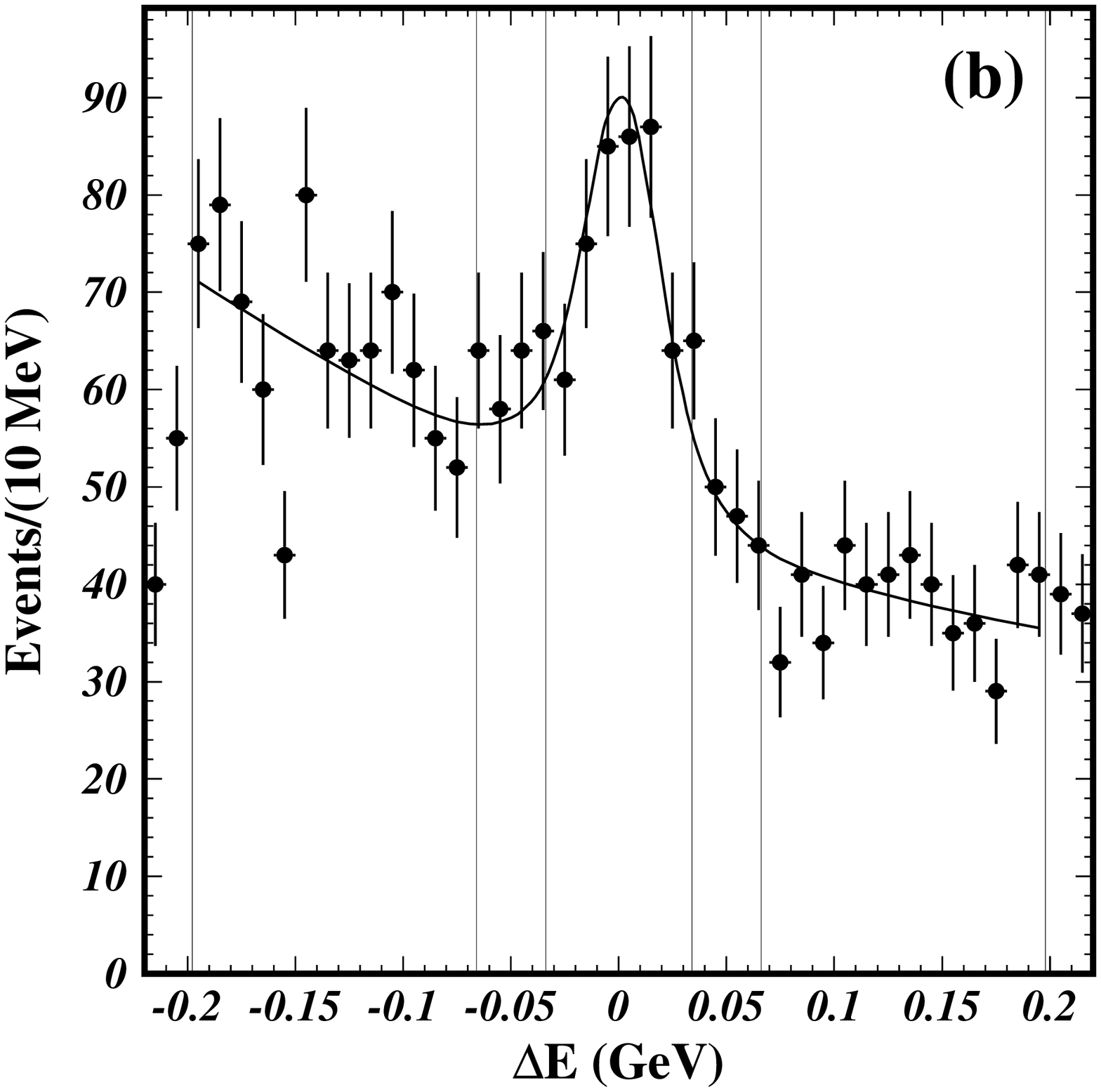} \\
\end{tabular}
\caption{$\Delta E$ distributions of the $\bar{B}^0 \to D^{*+} \omega \pi^-$ candidates in the (a) signal and (b) sideband regions of $M(\pi^+\pi^-\pi^0)$. The hatched histogram in (a) represents the $M(\pi^+\pi^-\pi^0)$ sideband normalized to the size of the signal region. The $\Delta E$ signal region and sideband are indicated by the vertical lines. The curves are the results of the fit described in the text.}
\label{fig:de}
\end{figure*}

The signal yield is obtained from a binned $\chi^2$ fit to the $\Delta E$ distribution using a function describing the CR and SCF components together with a smooth combinatorial background. Since
the $\bar{B}^0 \to D^{*+} \pi^+ \pi^- \pi^0 \pi^-$ events observed in Fig.~3 produce a peak in $\Delta E$, the fit is performed separately in the  $M(\pi^+\pi^-\pi^0)$ signal and sideband regions defined above on the $(\Delta E, M(\pi^+\pi^-\pi^0))$ plane. MC simulation shows that these events have the same shape as the CR component.
In the fit, the CR component is described by a double-Gaussian function with distinct means and widths, the SCF component is described by the sum of a Gaussian function and a second-order polynomial, and the combinatorial background is described by another second-order polynomial.
The means, widths and relative normalizations of the CR and SCF functions are fixed to the values obtained from the signal MC simulation, while the signal normalization and the parameters of the polynomial background function are treated as free parameters.
The differences between MC and data values for the fixed parameters in the fit are found to be within MC statistical errors.
The fit results are shown in Fig.~\ref{fig:de} in both the $M(\pi^+\pi^-\pi^0)$ signal and sideband regions. The fitted signal yield is found to be $919 \pm 37$ for the $M(\pi^+\pi^-\pi^0)$ signal region and $157 \pm 21$ for the sideband region.
The final yield $N_S=821 \pm 39$ is computed as the difference between these two yields, taking into account the ratio of $5/8$ between the widths of the $M(\pi^+\pi^-\pi^0)$ signal and sideband regions.

The fraction of neutral $B$ mesons decaying to the studied final state is expressed as
\begin{linenomath*}
\begin{equation}
\mathcal{B}\,=\,\frac{N_{\rm S}}{\epsilon_{\rm S} \eta N_{\rm B} \mathcal{B}_{\rm sec}}{,}
\label{bf}
\end{equation}
\end{linenomath*}
where
$\epsilon_{\rm S}=(2.11 \pm 0.02)\%$ is the detection efficiency determined from a MC simulation that uses a Dalitz plot distribution generated according to the signal model described below,
$\eta=0.941 \pm 0.029$ is the efficiency correction factor that accounts for the difference between data and MC and obtained from the momentum-dependent corrections for the $\pi^0$ and slow pion from the $D^*$ decay and the PID corrections for the kaon,
$N_{B}=(771.6 \pm 10.6) \times 10^6$ is the total number of neutral $B$ mesons in the data \cite{NBB} and
$\mathcal{B}_{\rm sec}=(2.32 \pm 0.04)\%$ is the product of the secondary branching fractions.
Using Eq.~(\ref{bf}), we obtain
\begin{linenomath*}
\begin{equation}
\mathcal{B}\,=\,(2.31 \pm 0.11\, ({\rm stat.}) \pm 0.14\, ({\rm syst.})) \times 10^{-3}{,} \nonumber
\end{equation}
\end{linenomath*}
which is consistent with the CLEO value \cite{cleo} within $1.2\sigma$ and the {\it BABAR} value \cite{babar} within $1.5\sigma$.
\begin{table}
\caption{Sources of relative systematic error in the branching fraction measurement.}
\begin{ruledtabular}
\begin{tabular}{l l}
Source & Error (\%) \\
\hline
Signal yield, $N_{\rm S}$ \\
---$M(\pi^+\pi^-\pi^0)$ signal region & $1.3$ \\
---Definition of SCF and CR components & $0.9$ \\
---$\Delta E$ signal shape & $2.2$ \\
---$\Delta E$ background shape & $1.3$ \\
Signal efficiency, $\epsilon_{\rm S}$ \\
---Track reconstruction efficiency & $3.9$ \\
---$\pi^0$ reconstruction efficiency & $2.3$ \\
---Kaon identification efficiency & $0.9$ \\
---$\bar{B^0}$ signal decay model & $1.1$ \\
---MC statistics & $0.8$ \\
Number of neutral $B$ mesons, $N_B$ & $1.4$ \\
Secondary branching fractions, ${\cal B}_{\rm sec}$ & 1.7 \\
\hline
Quadratic sum & $6.1$ \\
\end{tabular}
\label{t:sysbf}
\end{ruledtabular}
\end{table}
The total systematic error of $6.1\%$ summarized in Table~\ref{t:sysbf} arises from the following sources:
\begin{enumerate}[(i)]
\item An uncertainty of $1.3\%$ due to the choice of the signal window for the $M(\pi^+\pi^-\pi^0)$ invariant mass is estimated by reducing the size of the window from $21.25$ to $12.75$ MeV$/c^2$.
The reduced window corresponds to $1.5$ times the world average $\omega$ width.
\item An uncertainty of $0.9\%$ related to the definition of the SCF and CR components is estimated by changing the requirement on the $\chi^2$ defined in Eq.~(\ref{chi2genrec}) to $C=200$ or $C=400$.
\item An uncertainty of $2.2\%$ related to the $\Delta E$ shape description is estimated by varying the shape parameters fixed from MC simulation in accordance with their MC statistical errors.
\item An uncertainty of $1.3\%$ due to the background description in the $\Delta E$ shape is estimated by adding higher-order polynomial terms or keeping a linear term only.
\item A dominant uncertainty of $3.9\%$ is assigned to the total reconstruction efficiency of all charged tracks in the decay. For a single track, this uncertainty depends on the transverse momentum $p_T$ of the track \cite{tracking}. For low momentum tracks (with $p_T<200$ MeV$/c$), it is estimated using the decays $B^0 \to D^{*-}\pi^+$ and $B^+ \to \bar{D}^{*0} \pi^+$; for high momentum tracks, a study of the tracking efficiency is based on partially reconstructed $D^{*+} \to D^0 (K^0_S\pi^+\pi^-)\pi^+$ decays. The total tracking error is the linear sum of the errors corresponding to the individual tracks.
\item An uncertainty of $2.3\%$ in the reconstruction efficiency of neutral pions is estimated using the $\tau^- \to \pi^- \pi^0 \nu_{\tau}$ branching fraction and events where the other $\tau$ decay is tagged \cite{pi0}.
\begin{figure*}[ht!]
\begin{tabular}{c c}
\includegraphics[width=0.5\textwidth]{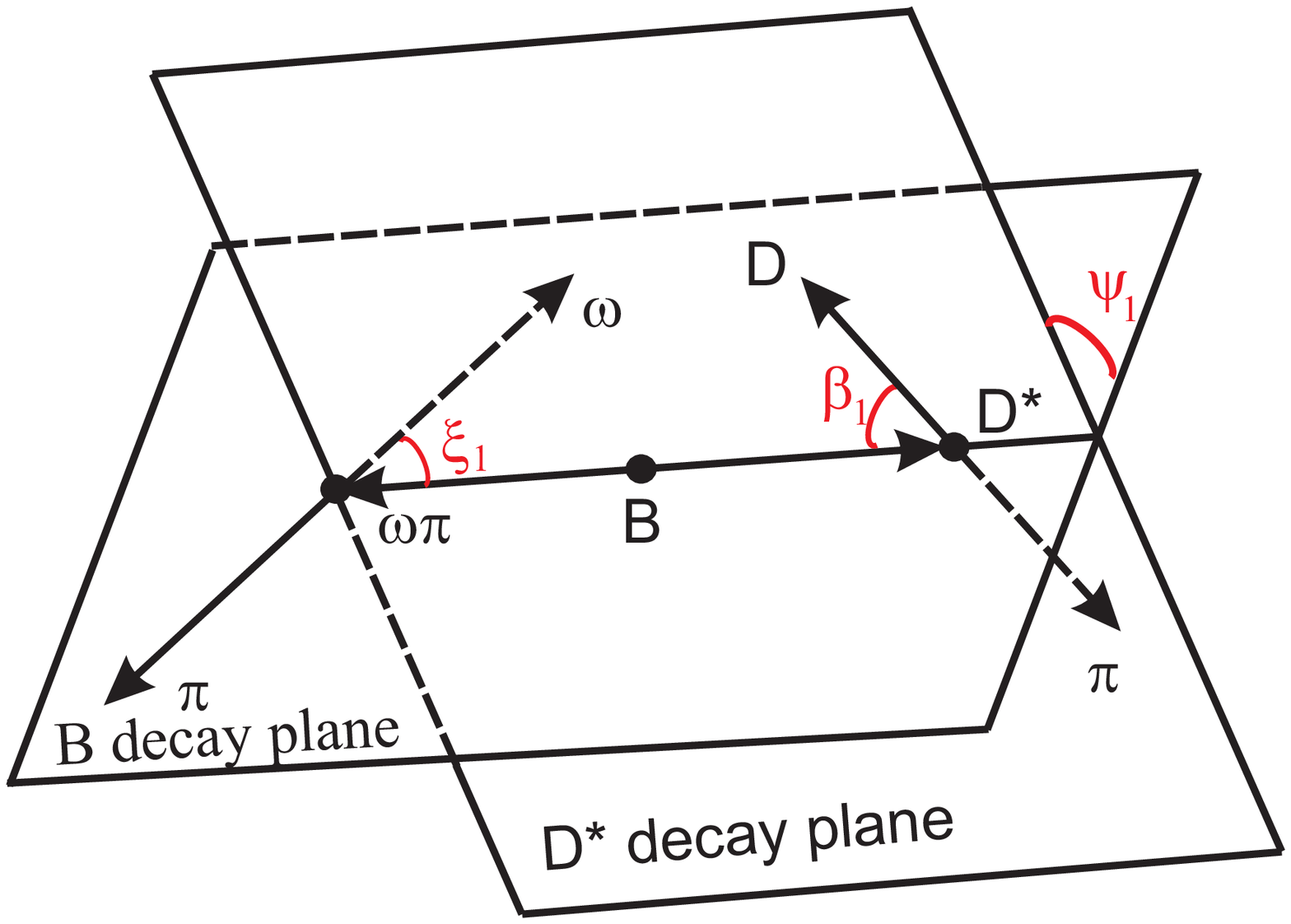} &
\includegraphics[width=0.5\textwidth]{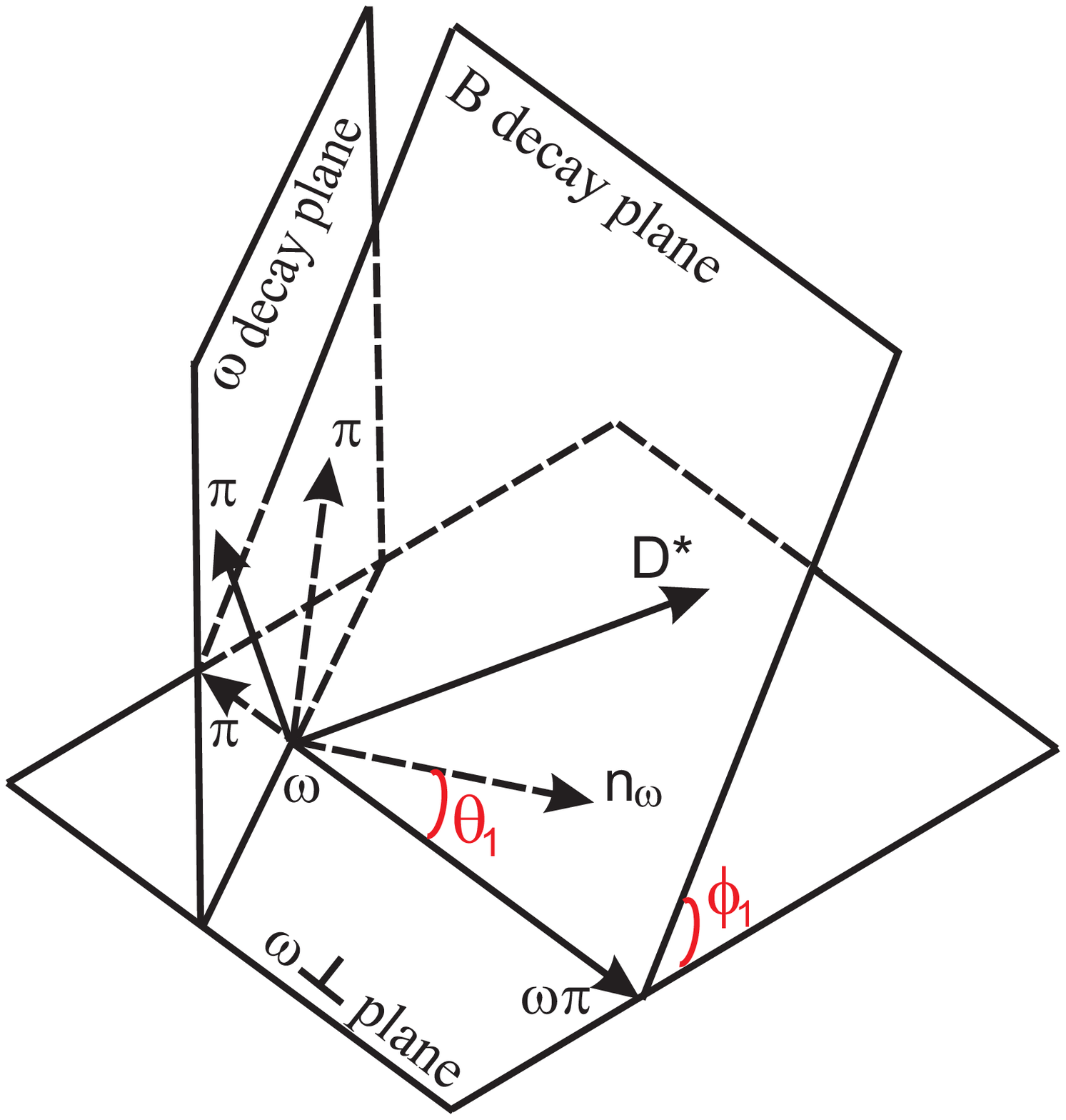} \\
{\bf (a)} & {\bf (b)}
\end{tabular}
\caption{(color online). Kinematics of a $\bar{B}^0 \to D^{*+} \omega \pi^-$ decay mediated by an $\omega\pi^-$ intermediate resonance. The diagram in (a) defines two polar angles $\xi_1$ and $\beta_1$ and one azimuthal angle $\psi_1$. The diagram in (b) defines one polar angle $\theta_1$ and one azimuthal angle $\phi_1$. The direction $n_{\omega}$ in (b) corresponds to the vector normal to the $\omega$ decay plane.}
\label{fig:angles}
\end{figure*}
\item An uncertainty of $0.9\%$ in the efficiency of the kaon particle identification requirement is obtained using a control sample of $D^{*+}\to D^0 (K^-\pi^+)\pi^+$ decays \cite{pid}.
\item An uncertainty of $1.1\%$ is assigned due to the model dependence of the signal reconstruction efficiency. The signal model with the best description of the data is constructed in Secs.~\ref{sec:VB} and \ref{sec:VC} in the frame of the amplitude analysis.
The model parameters obtained from the fit in Sec. \ref{sec:VC} 
have statistical uncertainties. These are propagated as a systematic uncertainty on the signal efficiency, taking into account the full covariance matrix.
\item A binomial uncertainty of $0.8\%$ due to the limited Monte Carlo sample size arises in the efficiency calculation.
\item An uncertainty of $1.4\%$ in the number of $B$ mesons is estimated from Ref.~\cite{NBBsys}.
\item An uncertainty of $1.7\%$ is associated with the measured branching fractions of the $D^*$, $D$ and $\omega$ \cite{pdg}.
\end{enumerate}

\section{\bf AMPLITUDE ANALYSIS}

To study the resonant structure of the $\bar{B}^0 \to D^{*+} \omega \pi^-$ decay, we perform an amplitude analysis. Using an unbinned likelihood method, we simultaneously fit the data in the six-dimensional phase space according to Ref.~\cite{jhep}.

We define two sets of kinematic variables: [$M^2(\omega\pi)$, $\cos\theta_1$, $\phi_1$, $\cos\beta_1$, $\psi_1$ and $\cos\xi_1$] and [$M^2(D^*\pi)$, $\cos\theta_2$, $\phi_2$, $\cos\beta_2$, $\psi_2$ and $\cos\xi_2$], corresponding to the $\omega\pi$ and $D^{**}$ productions (color-favored and -suppressed diagrams of Fig.~\ref{fig:diag}), respectively.

The masses $M(\omega\pi)$ and $M(D^*\pi)$ are the invariant masses of the $\omega\pi$ and $D^*\pi$ combinations. 
The angular variables, [$\cos\theta_1$, $\phi_1$, $\cos\beta_1$, $\psi_1$ and $\cos\xi_1$], describing $\omega\pi$ production, are defined in Fig.~\ref{fig:angles}.
The polar and azimuthal angles, $\theta_1$ and $\phi_1$, defined in the $\omega$ rest frame, are the angle between the normal $n_{\omega}$ to the $\omega$ decay plane and the $\omega \pi$ direction, and the angle between the $B$-decay plane and the plane formed by the $n_{\omega}$ and $\omega\pi$ directions, respectively.
The polar and azimuthal angles, $\beta_1$ and $\psi_1$, defined in the $D^{*}$ rest frame, are the angle between the $D$ and the $\omega\pi$ flight directions, and the angle between the $B$- and $D^*$-decay planes, respectively.  
The polar angle $\xi_1$ is the angle between the $D^{*}$ and $\omega$ flight directions in the $\omega \pi$ rest frame.

The angular variables, $\theta_2$ and $\phi_2$ as well as $\beta_2$ and $\psi_2$, describing the  $D^{**}$ production, are defined in the same manner as angles for the $\omega\pi$ production but with the $D^*\pi$ flight direction instead of the $\omega \pi$.
The polar angle $\xi_2$ corresponds to the angle $\xi_1$ but in the $D^*\pi$ rest frame.    
The $\cos{\xi_1}$ variable is related to $M^2(D^*\pi)$ whereas the $\cos{\xi_2}$ is related to $M^2(\omega\pi)$.

Each set of variables (denoted below with the six-dimensional vector $\vec{x}$) fully defines the kinematics of the decay chain, either in the color-favored or the color-suppressed channel \cite{jhep}.
The probability density function (PDF) in the signal region, which is the sum of signal and background components, is constructed in such a way that the kinematic dependence of the efficiency can be omitted in the minimization \cite{b-todstpi}:
\begin{linenomath*}
\begin{align}
\label{pdf}
{\rm PDF}(\vec{x},\vec{a})\,=&\,\frac{\epsilon(\vec{x})}{n_{\rm s}+\sum_j n_{{\rm bkg}\,j}}\times{} \nonumber \\
&\left\{
n_{\rm s} \frac{|M(\vec{x},\vec{a})|^2}{\epsilon_{\rm s}(\vec{a})} + \sum_j n_{{\rm bkg}\,j} \frac{B_j(\vec{x})}{\epsilon_{{\rm bkg}\,j}}\right\}{,}
\end{align}
\end{linenomath*}
where the sum is over the background components estimated in the sideband Regions ${\rm II}$, ${\rm III}$ and ${\rm IV}$ (see Fig.~\ref{fig:devsom}),
and the efficiencies $\epsilon_{{\rm s}}$ and $\epsilon_{{\rm bkg}\,j}$ correspond to average signal and background efficiencies, respectively, in the signal Region ${\rm I}$ integrated over the phase space.
In Eq.~(\ref{pdf}), $\vec{a}$ is the vector of parameters determined from the unbinned likelihood fit; $n_{\rm s}$ is the expected number of the signal events in the signal Region ${\rm I}$ distributed according to
the matrix element squared $|M(\vec{x},\vec{a})|^2$;
$\epsilon(\vec{x})$ is the reconstruction efficiency for the $\bar{B}^0 \to D^{*+} \omega \pi^-$ CR events in Region ${\rm I}$ depending on the decay kinematics and slowly varying within the scale of resolution of the observables; and
$n_{{\rm bkg}\,j}$ is the expected number of background events in the signal Region I distributed according to
the function $B_j(\vec{x})$.
We neglect the convolution with the resolution function in Eq.~(\ref{pdf}) due to the small invariant mass resolutions ($4$ MeV$/c^2$ for $\omega\pi$ and $3$ MeV$/c^2$ for $D^{**}$) in comparison with the resonance widths (more than $150$ MeV$/c^2$ for the $\rho$-meson-like resonance and more than $25$ MeV$/c^2$ for the $D^{**}$ states).

An unbinned likelihood fit to the $\bar{B}^0 \to D^{*+} \omega \pi^-$ phase space is performed to minimize the negative log-likelihood function $\mathcal{L}(\vec{a})$:
\begin{linenomath*}
\begin{align}
\label{likelihood}
\mathcal{L}(\vec{a})\,=&\,-\sum_{{\rm events}}{\rm \ln}\,{{\rm PDF}} +{} \nonumber \\
& \frac{(n_{\rm s}+\sum_j n_{{\rm bkg}\,j}-n_{{\rm tot}})^2}{2 (n_{{\rm tot}}+\sigma_{{\rm bkg}}^2)}{,}
\end{align}
\end{linenomath*}
where $n_{\rm tot}$ is the total number of events in the signal Region ${\rm I}$ and
$\sigma_{\rm bkg}$ is the uncertainty of the total number of background events $\sum_j n_{{\rm  bkg}\,j}$.
The second term in Eq.~(\ref{likelihood}) takes into account our knowledge of the background contribution in the signal region.

The function $\mathcal{L}(\vec{a})$ does not incorporate the interference between the $D^* 4\pi$ peaking background and the $D^* \omega \pi$ signal.
This effect is expected to be small (see Section~\ref{sec:e}).

\subsection{\bf Background description}

The background components of Eq.~(\ref{pdf}) can be addressed using the $(\Delta E, M(\pi^+\pi^-\pi^0))$ scatter plot (Fig.~\ref{fig:devsom}).
The combinatorial background with misreconstructed $\omega$ candidates saturates Region ${\rm IV}$.
The $\bar{B}^0 \to D^{*+} \pi^+ \pi^- \pi^0 \pi^-$ events without $\omega$ in the intermediate state can be found in Region ${\rm III}$.
The combinatorial background with a correctly reconstructed $\omega$ falls into Region ${\rm II}$. In addition, the SCF events lie in all regions.

We determine the six-dimensional shapes of the background PDFs $B_j(\vec{x})$ by performing an unbinned-likelihood fit in the sideband regions. For details, see Appendix \ref{sec:appA}.

\begin{figure*}[!htbp]
\begin{center}
\begin{tabular}{c c}
\includegraphics[scale=0.35]{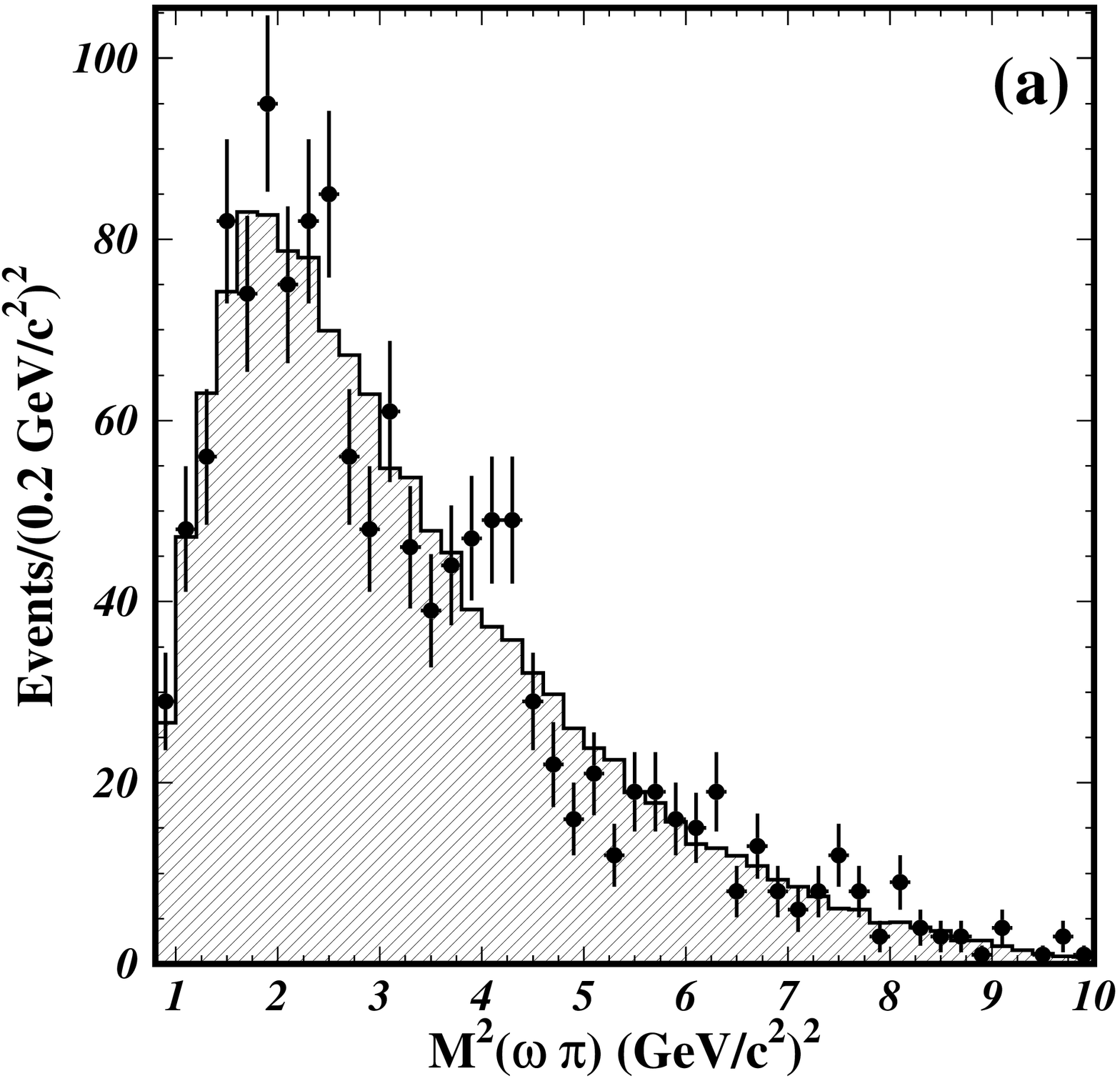} &
\includegraphics[scale=0.35]{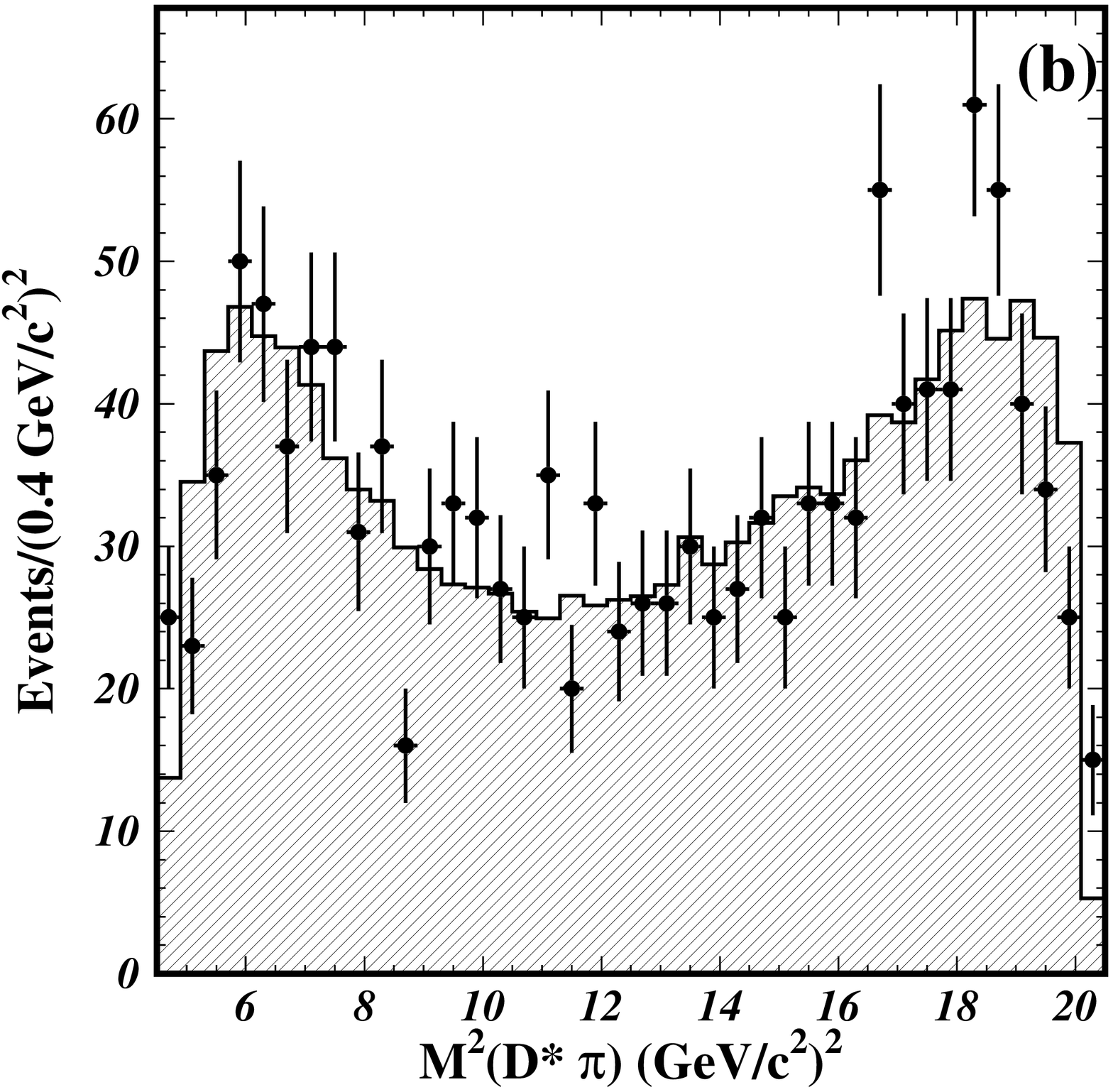} \\
\includegraphics[scale=0.35]{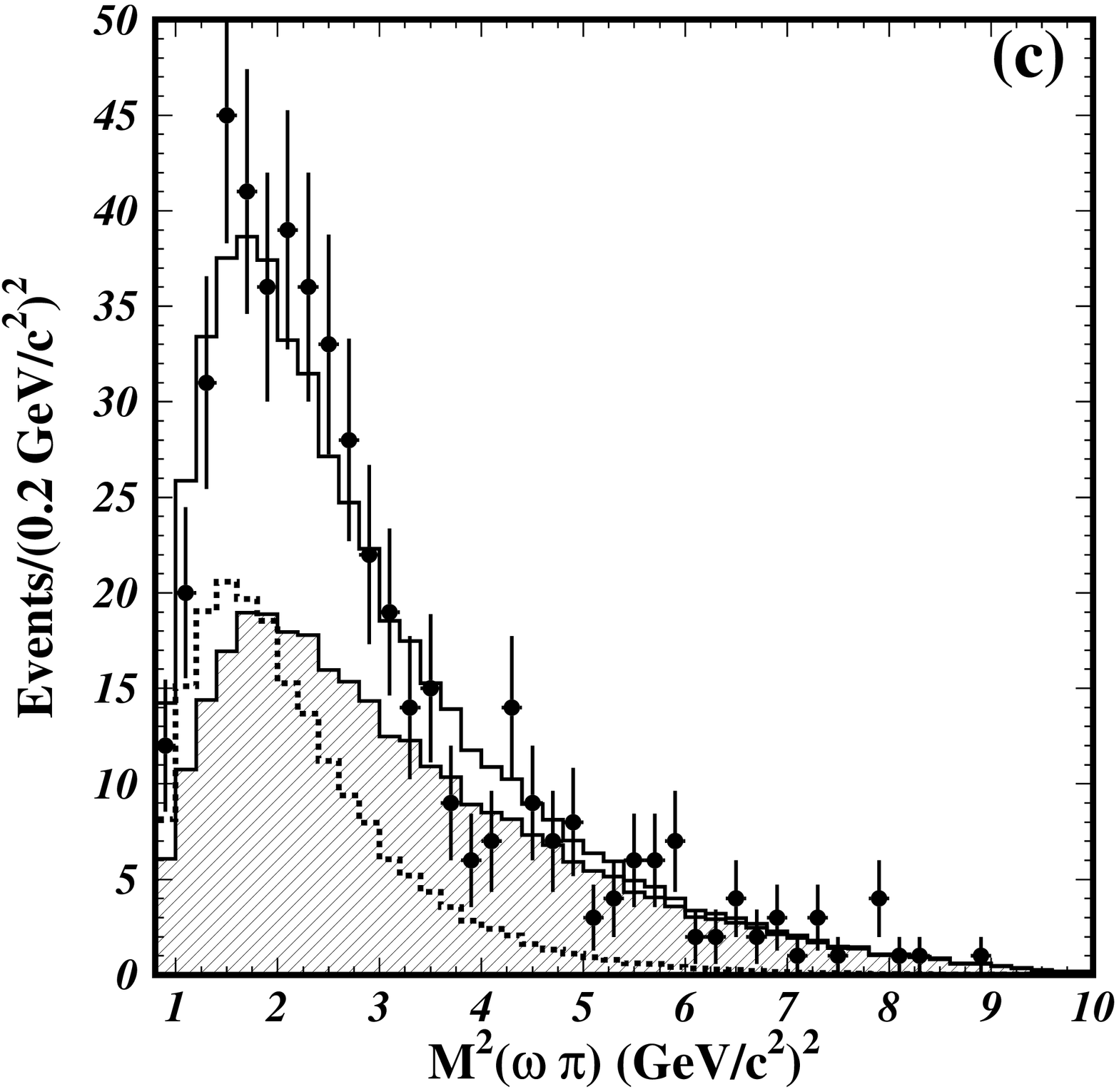} &
\includegraphics[scale=0.35]{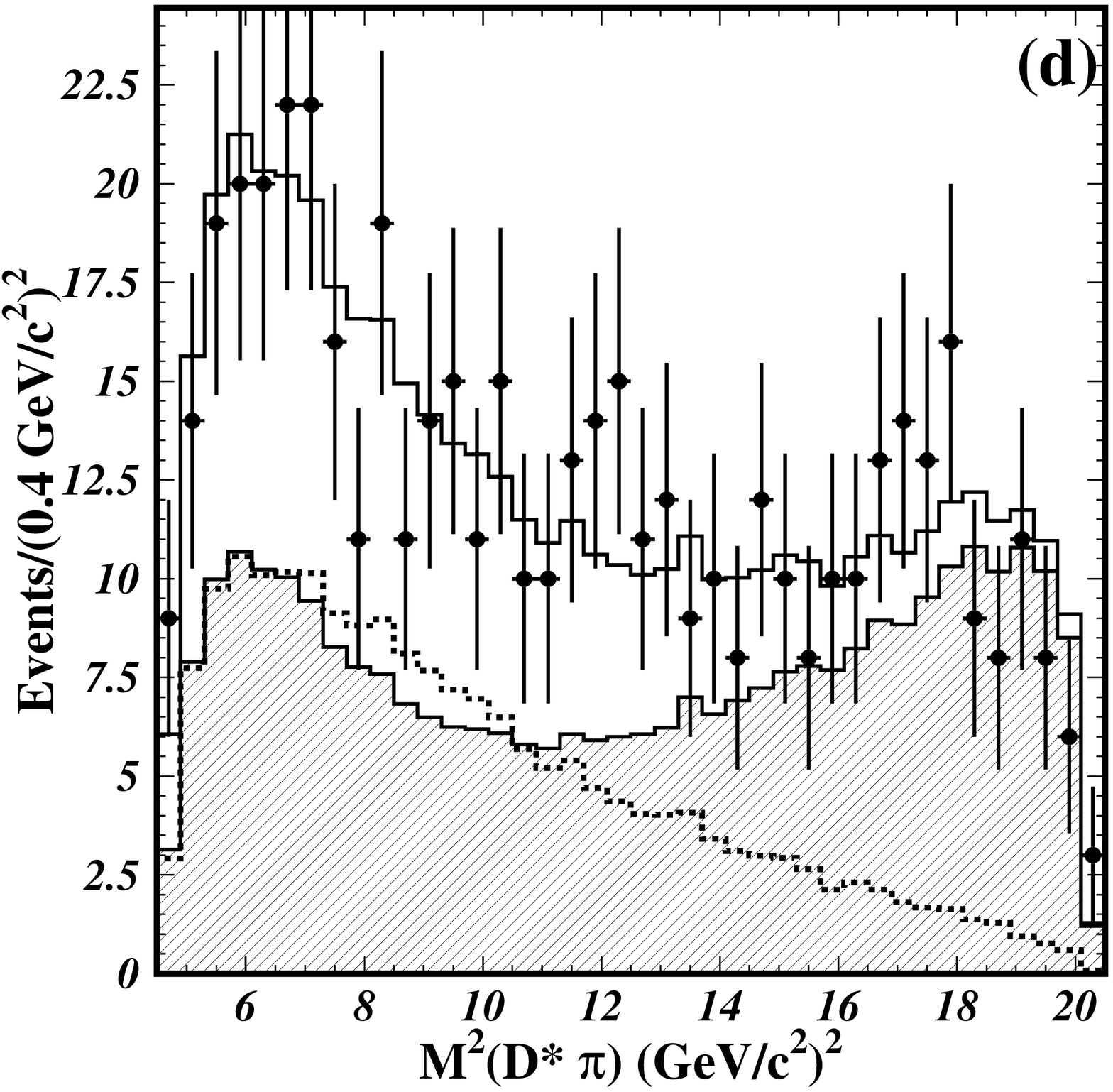} \\
\includegraphics[scale=0.35]{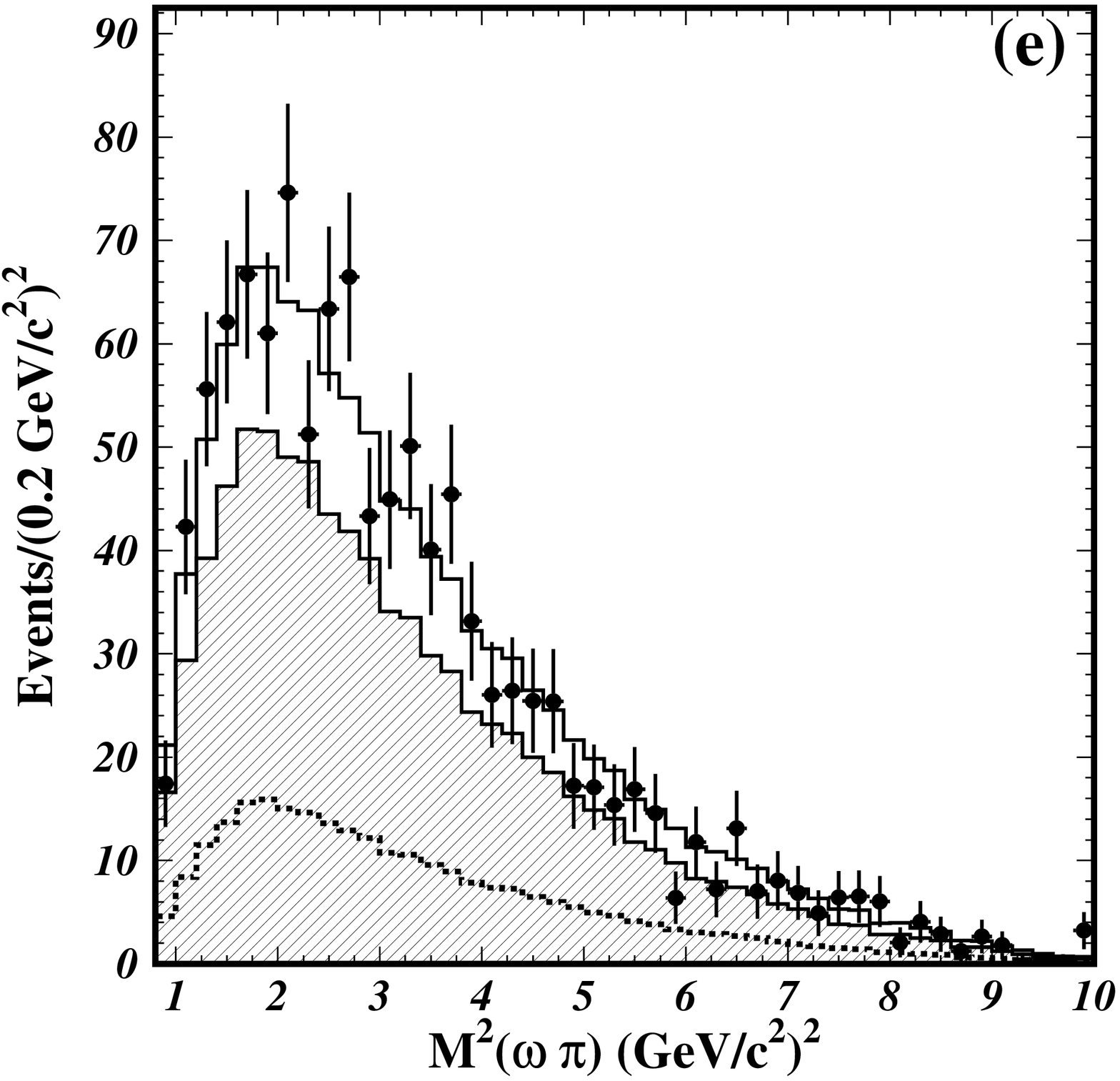} &
\includegraphics[scale=0.35]{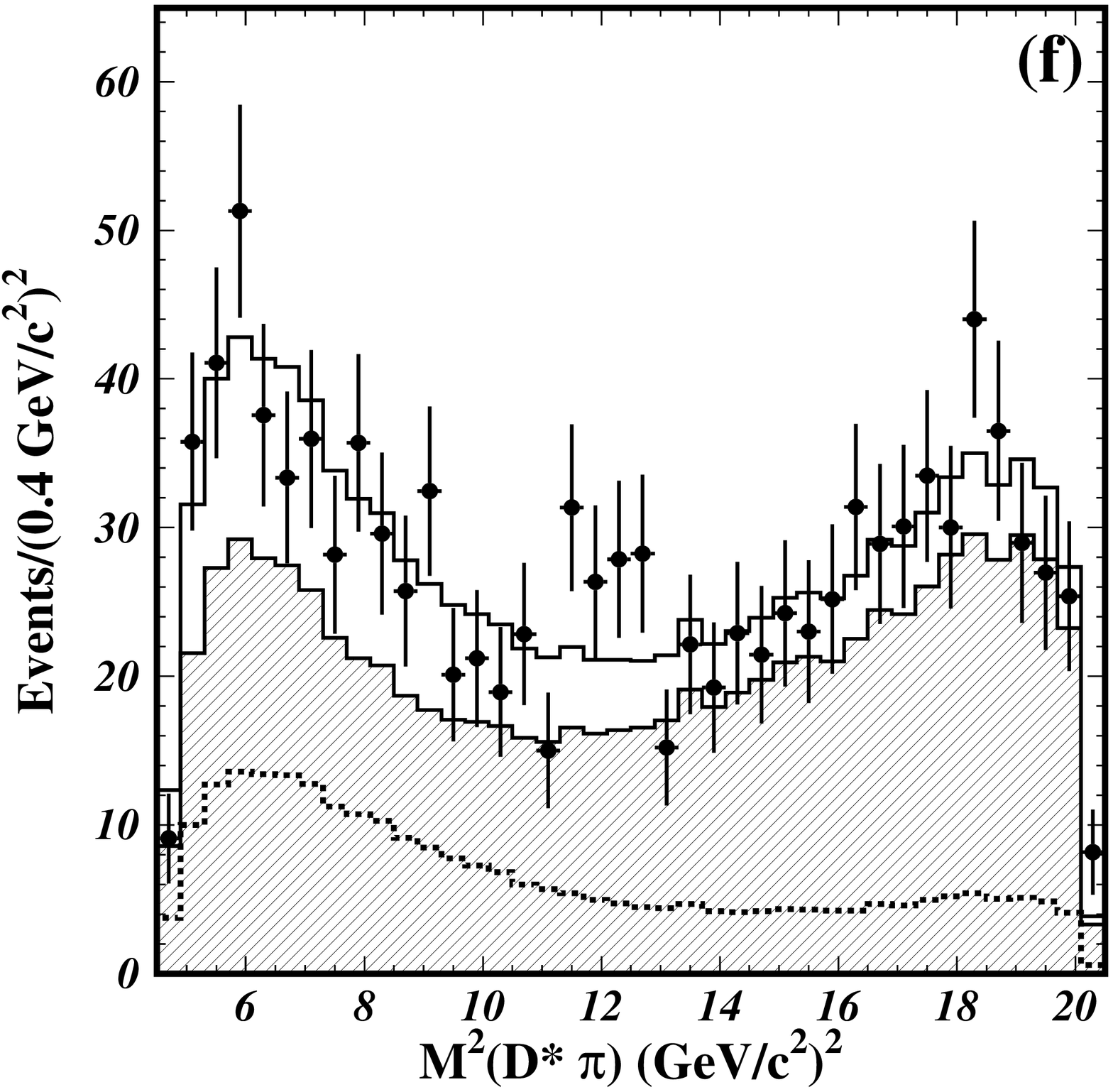} \\
\end{tabular}
\caption{$M^2(\omega\pi)$ and $M^2(D^*\pi)$ distributions of the $\bar{B}^0\to D^{*+} \omega \pi^-$ candidates in the $(\Delta E, M(\pi^+\pi^-\pi^0))$ sideband regions (a,b) ${\rm IV}$, (c,d) ${\rm III}$ and (e,f) ${\rm II}$.
Points with error bars are data;
hatched histograms correspond to the contribution from $B_{\rm IV}(\vec{x})$;
dotted histograms represent the component described by the function $B_{\rm III}(\vec{x})$ in (c) and (d) and $B_{\rm II}(\vec{x})$ in (e) and (f);
open histograms correspond to the total fit results in Regions ${\rm III}$ and ${\rm II}$.}
\label{fig:bkg}
\end{center}
\end{figure*}

The projections on the $M^2(\omega\pi)$ and $M^2(D^*\pi)$ variables and the corresponding background fits are shown in Fig.~\ref{fig:bkg}.
The result of the unbinned-like\-li\-hood fit in Region ${\rm IV}$ determining the function $B_{\rm IV}(\vec{x})$ is shown in Figs.~\ref{fig:bkg} (a) and (b).
Figures~\ref{fig:bkg} (c) and (d) correspond to Region ${\rm III}$.
The backgrounds in this region are described by the function $B_{\rm III}(\vec{x})$ plus a contribution components described by $B_{\rm IV}(\vec{x})$.
In a similar way, Region ${\rm II}$ includes the  background components described by $B_{\rm IV}(\vec{x})$ and $B_{\rm II}(\vec{x})$; these components are shown in Figs.~\ref{fig:bkg} (e) and (f).

\begin{figure*}[!htbp]
\begin{tabular}{c c}
\includegraphics[scale=0.41]{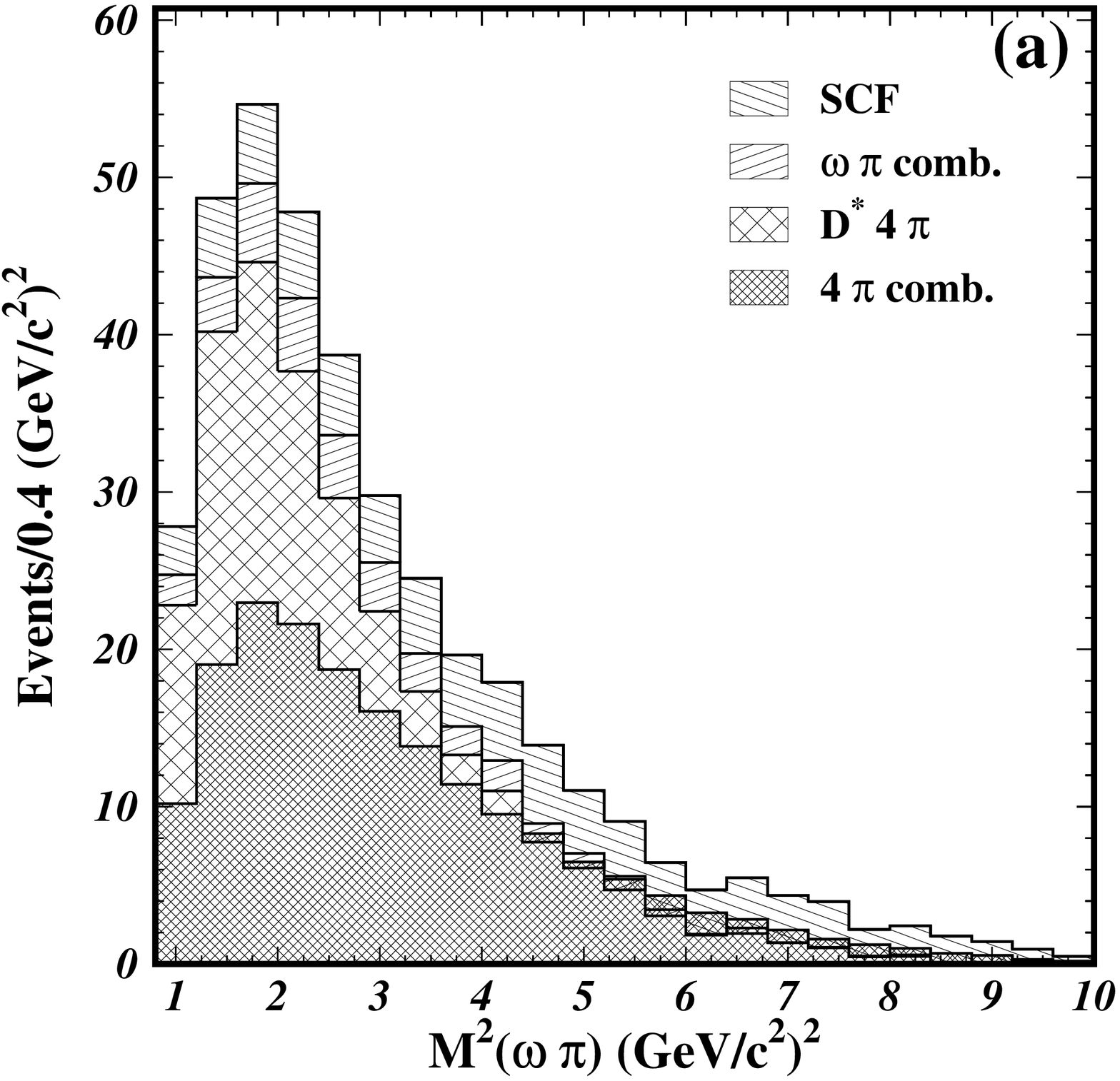} &
\includegraphics[scale=0.41]{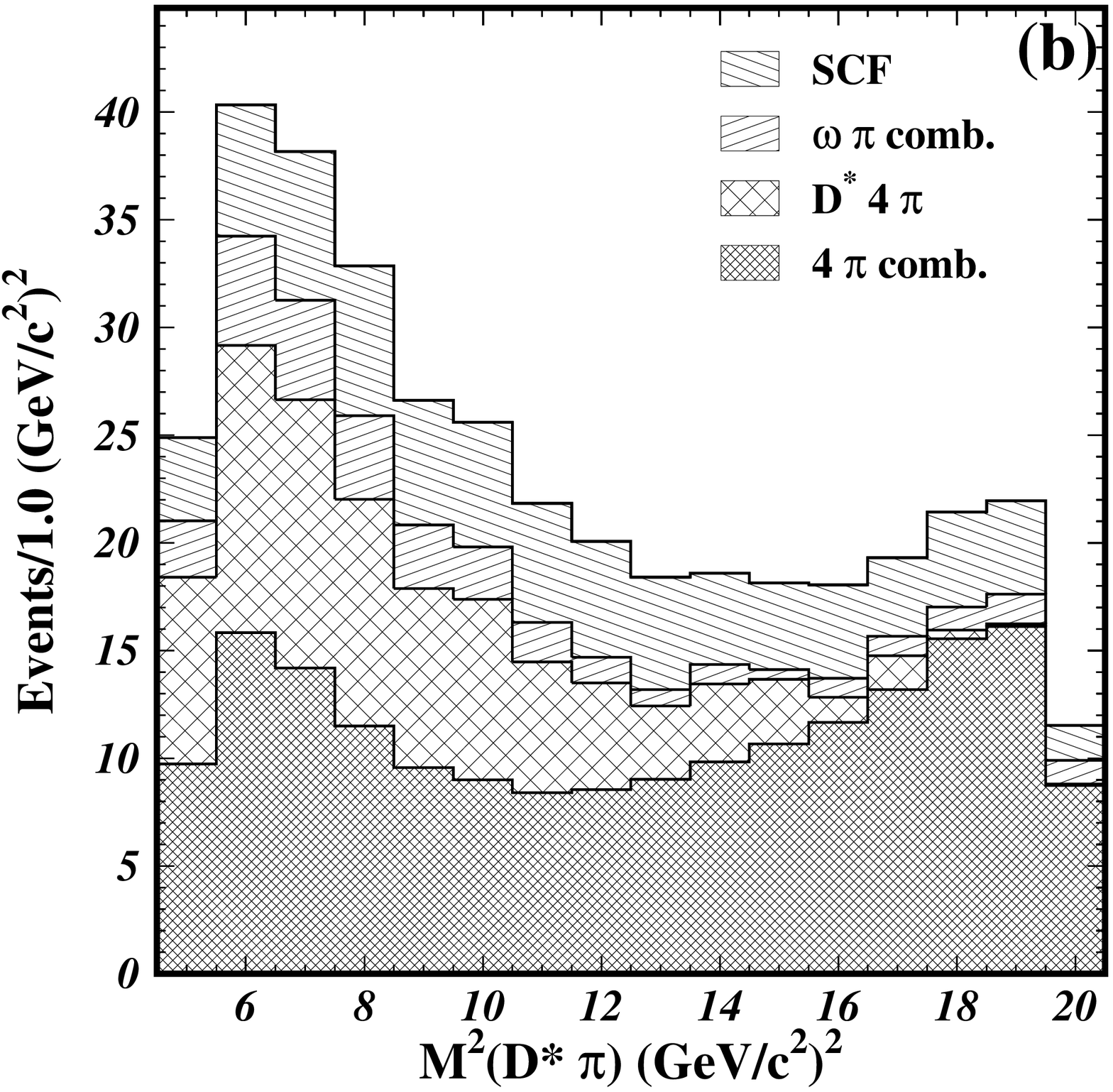}
\end{tabular}
\caption{(a) $M^2(\omega\pi)$ and (b) $M^2(D^*\pi)$ distributions of the background components in the signal region. The histograms are stacked on top of each other.}
\label{fig:bkgstruct}
\end{figure*}

The $M^2(\omega\pi)$ and $M^2(D^*\pi)$ distributions of the background in the signal Region ${\rm I}$ are shown in Fig.~\ref{fig:bkgstruct}.
These distributions are the sum of the SCF distribution in Region ${\rm I}$ obtained from the MC study and the distributions describing the backgrounds defined above.
The latter distributions are the differences between the $B_{\rm II}(\vec{x})$, $B_{\rm III}(\vec{x})$ and $B_{\rm IV}(\vec{x})$ distributions and the SCF distributions in Regions ${\rm II}$, ${\rm III}$ and ${\rm IV}$, respectively.
Figure~\ref{fig:bkgstruct} illustrates the dominant contribution due to the combinatorial background with a misreconstructed $\omega$. The $D^{*+} \pi^+ \pi^- \pi^0 \pi^-$ component estimated from Region ${\rm III}$ is also significant in Region ${\rm I}$. The combinatorial background with a correctly reconstructed $\omega$ and SCF component obtained from the MC study have lower fractions but are also included in the description.

\subsection{\bf Signal description}
\label{sec:VB}

The description of the $D^{*+} \omega \pi^-$ signal events in phase space is based on the study of Ref.~\cite{jhep}. Since charge conjugation is taken into account, the total matrix element is calculated as:
\begin{linenomath*}
\begin{equation}
M\,=\,\frac{1+Q}{2}M_++\frac{1-Q}{2}M_-{,}
\end{equation}
\end{linenomath*}
where $Q=+1$ ($-1$) for $\bar{B}^0$ ($B^0$) decays and $M_-$ differs from $M_+$ by the sign of the P-violating terms.
Following the isobar model formulation \cite{isobar} with quasi-two-body resonant amplitudes, the matrix element $M_{\pm}$ is given by:
\begin{linenomath*}
\begin{equation}
M_{\pm}\,=\,\sum_{R} a_R e^{i \phi_R} M_{R\, \pm}{,}
\end{equation}
\end{linenomath*}
where $a_R$ and $\phi_R$ are relative amplitudes and phases of the intermediate resonances and $R$ is an index numbering all the $\omega\pi$ and $D^{**}$ resonances. The full description of the resonant matrix elements $M_{R\,\pm}$ can be found in Appendix \ref{sec:appB}. The parametrization of the form factors used in the matrix elements $M_{R\,\pm}$ is presented in Appendix \ref{sec:appC}.
The fraction $f_R$ of the total three-body signal attributed to a particular quasi-two-body intermediate state is defined as
\begin{linenomath*}
\begin{equation}
f_R\,=\,\frac{\int a^2_{R} |M_{R\, \pm}(\vec{x})|^2 \rho(\vec{x}) d\vec{x}}{\int |M_{\pm}(\vec{x})|^2 \rho(\vec{x}) d\vec{x}}{,}
\end{equation}
\end{linenomath*}
where $\rho(\vec{x})$ is the phase space density of events determined from the kinematic conditions of the decay \cite{jhep}.
The sum of the fit fractions for all components is not necessarily unity because of interference effects.

The fraction $f^L_R$ of resonance $R$ produced in partial wave $L$ is determined as
\begin{linenomath*}
\begin{equation}
f^L_{R}\,=\,\frac{\int |M^L_{R\,\pm}(\vec{x})|^2\rho(\vec{x})d\vec{x}}{\int |M_{R\,\pm}(\vec{x})|^2\rho(\vec{x}) d\vec{x}}{,}
\end{equation}
\end{linenomath*}
where $M^L_{R\, \pm}$ is the matrix element describing the production of resonance $R$ in partial wave $L$ and the sum $\sum_L f^L_R$ is unity by definition.

The observable determined from the amplitude analysis is the longitudinal polarization $\mathcal{P}_R$ of resonance $R$.
This variable is calculated as
\begin{linenomath*}
\begin{equation}
\mathcal{P}_R\,=\,\frac{|H_0|^2}{|H_0|^2+|H_+|^2+|H_-|^2}{,}
\end{equation}
\end{linenomath*}
where $H_0$, $H_+$ and $H_-$ represent three complex helicity amplitudes which can be expressed via invariant and partial wave form factors (see Appendix \ref{sec:appC}).

\subsection{\bf Fitting the $\mathbf{\bar{B}^0 \to D^{*+} \omega \pi^-}$ signal}
\label{sec:VC}

\begin{figure*}[!htbp]
\begin{tabular}{c c}
\includegraphics[scale=0.415]
{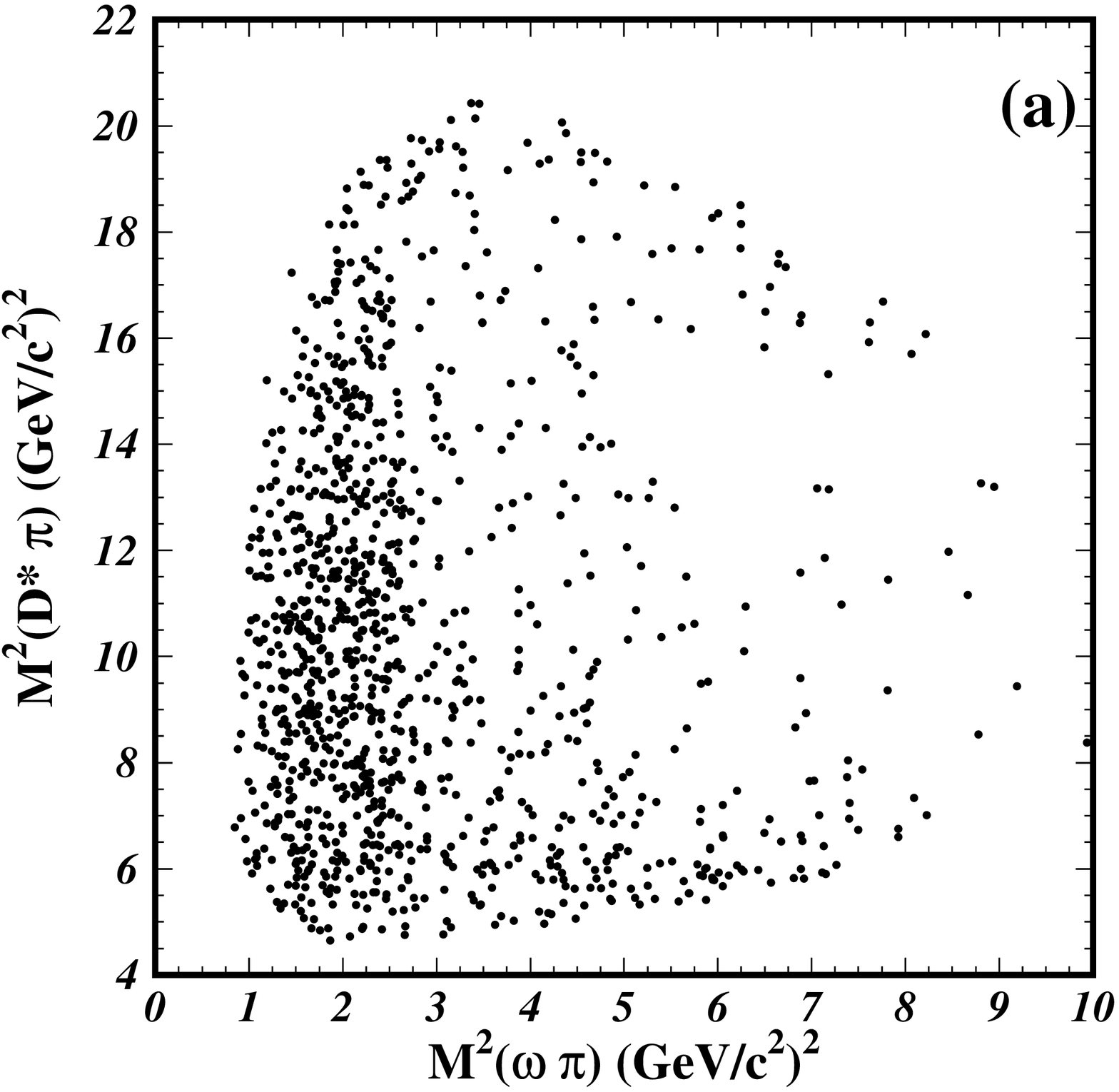} &
\includegraphics[scale=0.415]{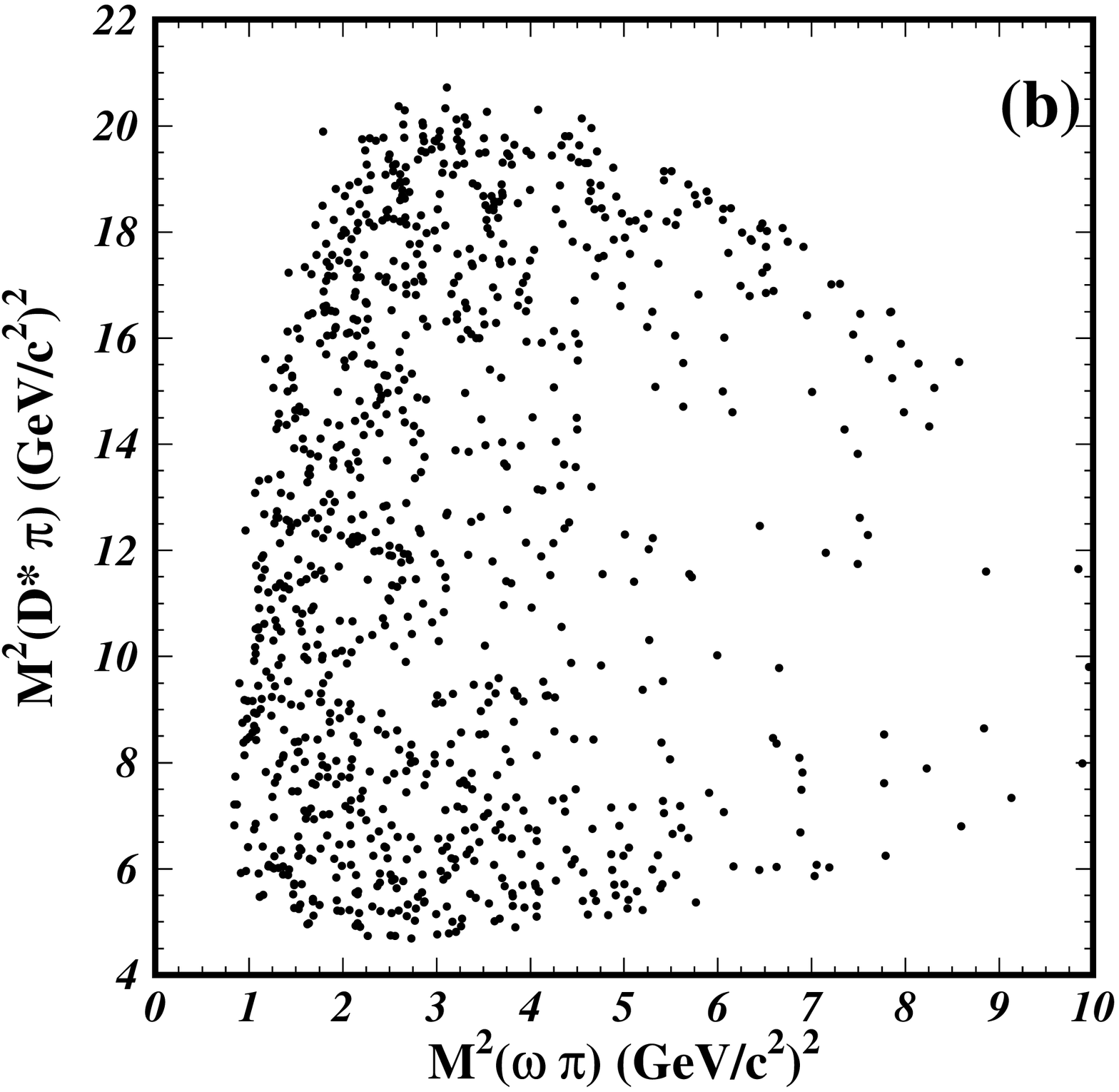} \\
\includegraphics[scale=0.415]{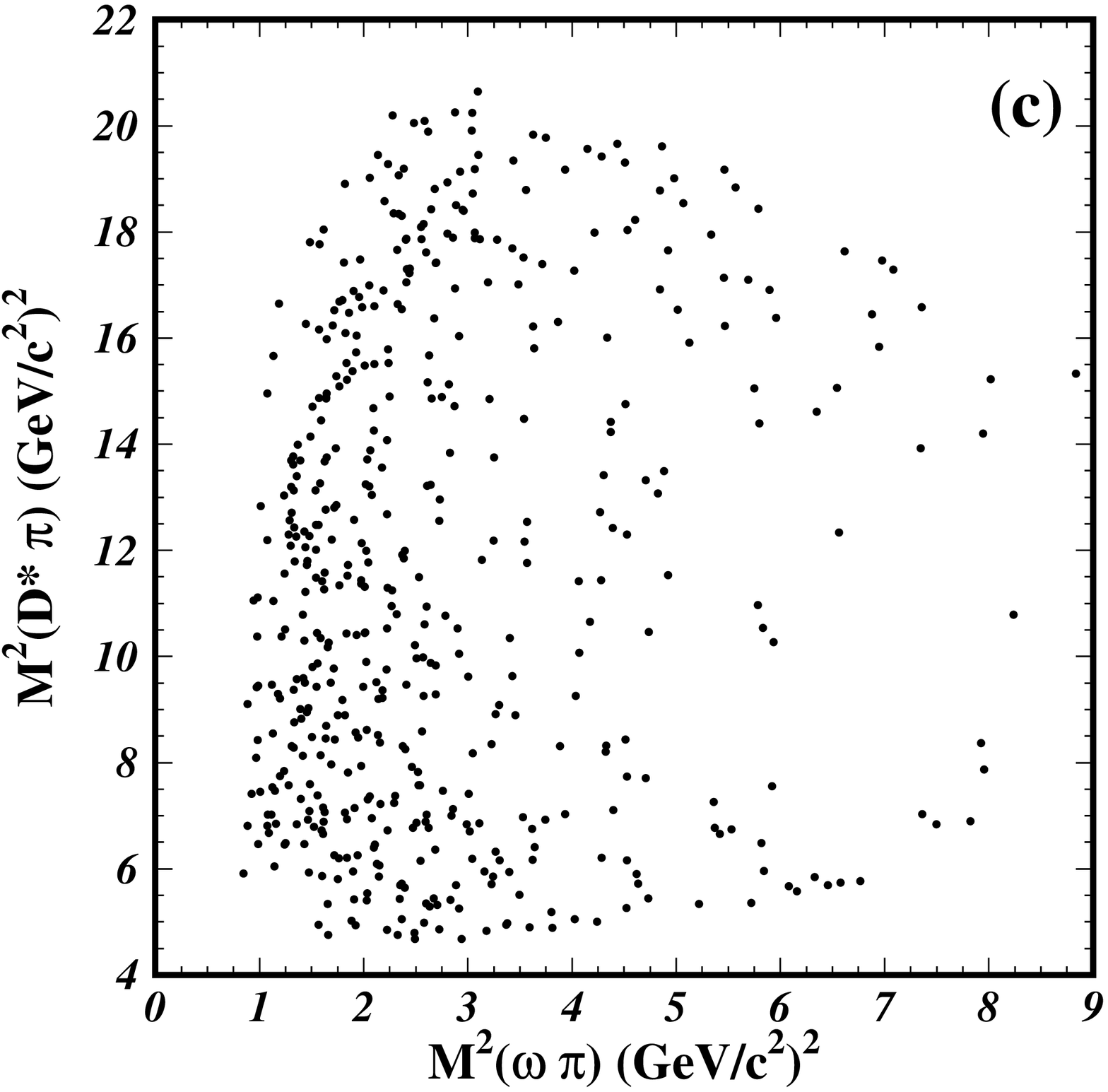} &
\includegraphics[scale=0.415]{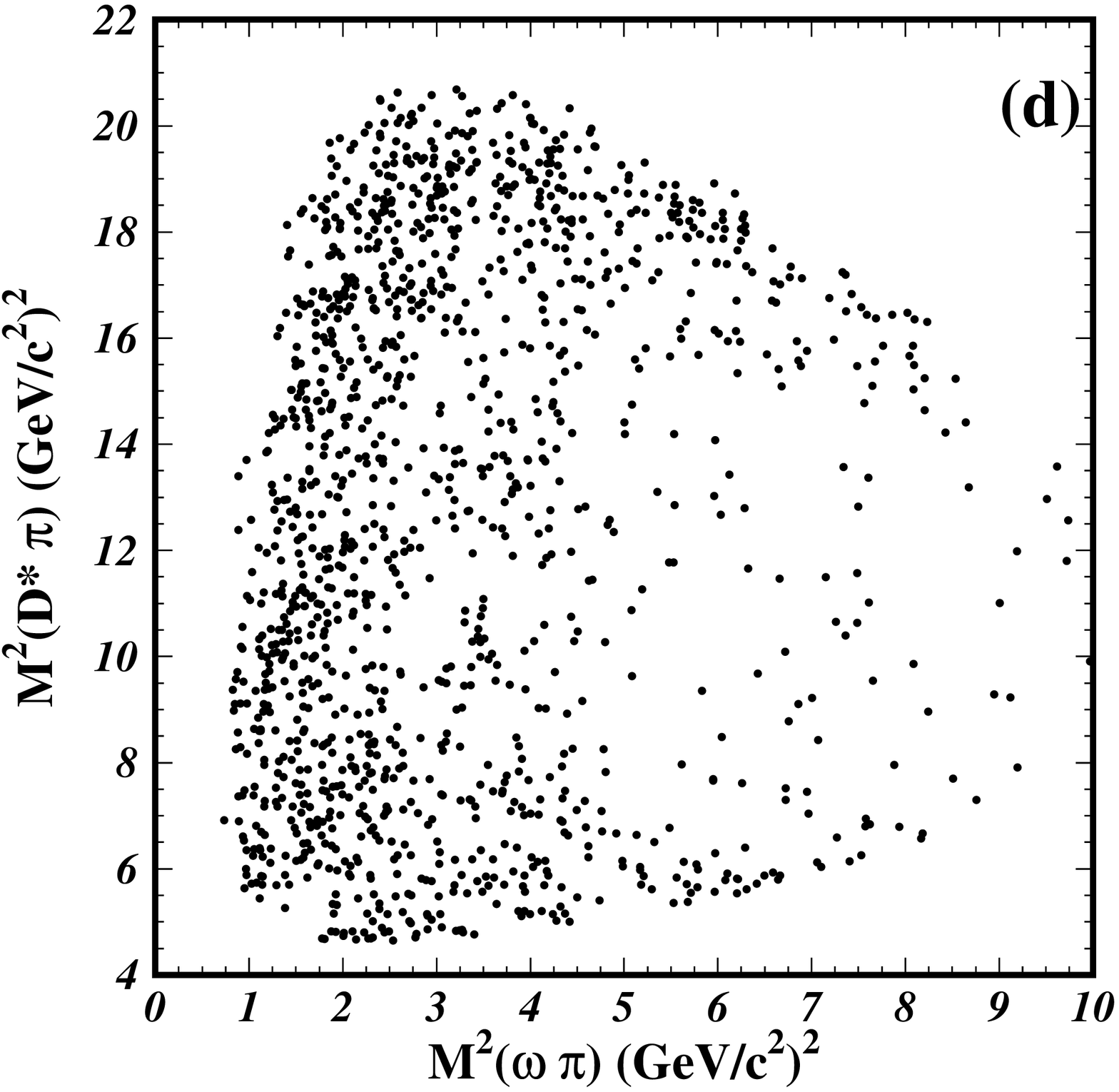} \\
\end{tabular}
\caption{Dalitz distributions of the $D^{*+} \omega \pi^-$ candidates in (a) signal Region ${\rm I}$, (b) sideband Region ${\rm II}$, (c) sideband Region ${\rm III}$ and (d) sideband Region ${\rm IV}$.}
\label{fig:dp}
\end{figure*}

Figure~\ref{fig:dp} shows the two-dimensional Dalitz distributions in signal Region ${\rm I}$ and sideband Regions ${\rm II}$, ${\rm III}$ and ${\rm IV}$.
There are $1129$ events in the signal region that satisfy all the selection criteria.

To describe all the features of the Dalitz plot, we use the following set of resonances: off-shell $\rho(770)^-$, $\rho(1450)^-$, $D_1(2430)^0$, $D_1(2420)^0$ and $D^*_2(2460)^0$. A CLEO analysis \cite{cleo} showed the dominance of the $\rho(1450)^-$ resonance in this final state. In a {\it BABAR} study \cite{babar}, a $D^* \pi$ enhancement was observed that was interpreted as a $D_1(2430)^0$ signal. Our data require additional resonances.
We take into account an off-shell $\rho(770)^-$ contribution, as suggested by the $e^+e^- \to \omega \pi^0$ data \cite{karda}.
To improve the description, we also include the amplitudes of the narrow resonances $D_1(2420)^0$ and $D^*_2(2460)^0$.
When both resonances are simultaneously included in the matrix element rather than just one of them, the statistical significance of the signal, given by $\sqrt{2(\mathcal{L}_R-\mathcal{L}_0)}$, where $\mathcal{L}_R$ ($\mathcal{L}_0$) is the negative log-likelihood value with the signal from the resonance $R$ fixed at zero (with the nominal signal yield), increases very significantly ($>5\,\sigma$ effect).
We also include in the fit a SCC contribution with the $b_1(1235)^-$ resonance. This contribution has a significance below $3.0 \sigma$ and we obtain an upper limit for the fraction of the SCC in $\bar{B}^0 \to D^{*+} \omega \pi^-$ decays.
To determine the upper limit, we generate pseudoexperiments (see Sec.~\ref{sec:D}).

\begin{table*}
\caption{Summary of the fit results to the $D^{*+} \omega \pi^-$ candidates in the signal region. Each column of results corresponds to a different signal model. The notations $\rho'=\rho(1450)$, $D'_1=D_1(2430)$, $D_1=D_1(2420)$ and $D^*_2=D^*_2(2460)$ are used. Quoted uncertainty is statistical only.
$\Delta \mathcal{L}=\mathcal{L}-\mathcal{L}_0$, where $\mathcal{L}$ defined in Eq.~(\ref{likelihood}) corresponds to the signal model for which this variable is calculated and $\mathcal{L}_0$ is the negative log-likelihood function calculated for the signal model with $\rho$, $\rho'$, $D'_1$, $D_1$ and $D^*_2$ resonances.}
\label{tab:models}
\begin{ruledtabular}
\begin{tabular}{c c c c c c c}
Contribution & Parameter & $\rho,\rho'$ &  $\rho,\rho'$ &  $\rho,\rho'$ &  $\rho,\rho'$ &
$\rho,\rho',b_1$ \\
&  &  $D'_1$  & $D'_1,D_1$ & $D'_1,D^*_2$ & $D'_1,D_1,D^*_2$ & $D'_1,D_1,D^*_2$ \\
\hline
$\rho(770)^-D^{*+}$ & Resonance phase & $0$ (fixed) & $0$ (fixed) & $0$ (fixed) & $0$ (fixed)  & $0$ (fixed)  \\
   & Resonance fraction, \%  & $65.7 \pm 10.2$ & $65.2 \pm 11.7$  & $61.6 \pm 11.9$  & $64.2 \pm 10.7$   &  $60.6 \pm 12.1$  \\
   \hline
$\rho(1450)^-D^{*+}$ & Resonance phase & $2.63 \pm 0.11$ & $2.55 \pm 0.11$ & $2.62 \pm 0.11$ & $2.56 \pm 0.12$ & $2.54 \pm 0.11$  \\
   & Resonance coupling &  $0.18^{+0.02}_{-0.05}$ & $0.18^{+0.02}_{-0.05}$ & $0.20^{+0.03}_{-0.06}$ & $0.18^{+0.02}_{-0.06}$ & $0.19^{+0.03}_{-0.06}$ \\
   & Mass, MeV$/c^2$ & $1549 \pm 22$ & $1546 \pm 23$ & $1543 \pm 23$ & $1544 \pm 22$  & $1540 \pm 22$ \\
   & Width, MeV$/c^2$  & $303^{+30}_{-50}$ & $305^{+31}_{-51}$ &  $316^{+30}_{-54}$  & $303^{+31}_{-52}$ & $302^{+30}_{-52}$ \\
   & $R_1$ & $1.40$ (fixed) & $1.40$ (fixed) & $1.40$ (fixed) & $1.40$ (fixed) & $1.40$ (fixed) \\
   & $R_2$ & $0.87$ (fixed) & $0.87$ (fixed) & $0.87$ (fixed) & $0.87$ (fixed) & $0.87$ (fixed) \\
   & $\rho^2$ & $0.79$ (fixed) & $0.79$ (fixed) & $0.79$ (fixed) & $0.79$ (fixed) & $0.79$ (fixed) \\
   & Resonance fraction, \% & $46.7^{+7.9}_{-11.9}$ & $44.5^{+6.9}_{-12.0}$ & $50.4^{+10.6}_{-13.1}$ &  $46.3^{+6.0}_{-13.4}$   & $47.5^{+9.3}_{-12.1}$  \\
   & $S$-wave fraction, \% & $76.9^{+4.2}_{-1.4}$
& $75.3^{+4.7}_{-1.9}$ & $76.7^{+4.3}_{-1.4}$ &
$75.1^{+4.4}_{-2.1}$ & $75.3^{+5.0}_{-1.8}$ \\
   & $P$-wave fraction, \% & $12.0 \pm 0.7$
& $12.8 \pm 1.1$ & $12.2 \pm 0.9$ &
$12.9 \pm 0.9$ & $12.7 \pm 0.8$ \\
   & $D$-wave fraction, \% & $11.0 \pm 0.4$
& $11.8 \pm 0.6$ & $11.0 \pm 0.6$ &
$11.9 \pm 0.5$ & $11.8 \pm 0.6$ \\
& $\phi_+$ phase &$0.66 \pm 0.33$ & $0.86 \pm 0.30$ & $0.67 \pm 0.37$  & $0.87 \pm 0.29$ & $0.85 \pm 0.31$  \\
   & $\phi_-$ phase & $-0.14 \pm 0.17$ & $-0.02 \pm 0.15$ & $-0.15 \pm 0.19$ &  $-0.02 \pm 0.13$  & $-0.02 \pm 0.15$ \\
   & Long. polarization, \% & $66.4 \pm 0.6$ &
$66.5 \pm 0.6$ & $66.5 \pm 0.6$ & $66.5 \pm 0.6$ & $66.6 \pm 0.6$ \\
&  FCC fraction, \% &$79.1 \pm 2.5$  &$82.6 \pm 2.4$ & $79.0 \pm 2.4$ & $82.2 \pm 2.2$  & $81.6 \pm 2.3$ \\
\hline
$D_1(2430)^0 \omega$ & Resonance phase & $0.91 \pm 0.26$ & $1.03 \pm 0.28$ & $1.11 \pm 0.29$ & $1.24 \pm 0.28$ & $1.27 \pm 0.35$ \\
   & $S$-wave phase & $0.26 \pm 0.20$ & $0.19 \pm 0.23$ & $0.14 \pm 0.23$  & $-0.05 \pm 0.25$  & $-0.09 \pm 0.26$   \\
   & $P$-wave phase & $2.71 \pm 0.21$ & $2.41 \pm 0.27$ & $2.56 \pm 0.24$ & $2.24 \pm 0.29$ &  $2.23 \pm 0.32$  \\
   & Resonance fraction, \%  & $13.6 \pm 2.1$ & $11.2 \pm 1.8$ & $12.6 \pm 1.8$ & $10.8 \pm 1.8$   & $11.6 \pm 2.0$    \\
   & $S$-wave fraction, \%  & $29.7 \pm 8.6$ & $33.6 \pm 9.5$ & $35.8 \pm 10.1$ &  $38.9 \pm 10.8$ &  $38.9 \pm 10.5$   \\
   & $P$-wave fraction, \% & $37.0 \pm 8.6$ & $34.1 \pm 9.2$ & $34.0 \pm 8.9$  & $33.1 \pm 9.5$  &   $29.1 \pm 9.1$  \\
   & $D$-wave fraction, \% & $33.5 \pm 8.8$ & $32.6 \pm 9.2$ & $30.5 \pm 9.2$ &  $28.3 \pm 8.9$  &   $32.2 \pm 9.2$  \\
   & Long. polarization, \% & $60.9 \pm 8.2$ & $63.4 \pm 8.9$ & $63.0 \pm 8.2$ & $63.0 \pm 9.1$  &  $67.6 \pm 9.2$ \\
   \hline
$D_1(2420)^0 \omega$ & Resonance phase & & $1.92 \pm 0.34$ & & $2.12 \pm 0.34$ & $2.16 \pm 0.42$ \\
   & $S$-wave phase &  & $-0.06 \pm 0.34$ & & $-0.07 \pm 0.43$ & $-0.10 \pm 0.43$  \\
   & $P$-wave phase & & $0.04 \pm 0.41$ &  & $-0.25 \pm 0.46$  & $-0.24 \pm 0.49$  \\
   & Resonance fraction, \%  & & $3.7 \pm 1.1$ & & $2.9 \pm 0.8$   &  $2.8 \pm 0.8$  \\
    & $S$-wave fraction, \%  &  & $35.6 \pm 13.2$ &  &  $34.0 \pm 13.4$ &  $35.8 \pm 13.0$   \\
   & $P$-wave fraction, \% &  & $36.6 \pm 11.8$ & & $31.2 \pm 11.4$  & $30.3 \pm 11.0$  \\
   & $D$-wave fraction, \% & & $27.9 \pm 11.0$ & &  $34.9 \pm 13.4$  & $34.0 \pm 13.1$  \\
   & Long. polarization, \%  & & $60.2 \pm 12.0$ & & $67.1 \pm 11.7$ &  $67.4 \pm 16.1$  \\
   \hline
$D^*_2(2460)^0 \omega$ & Resonance phase & & & $1.69 \pm 0.57$ & $2.31 \pm 0.50$  & $2.39 \pm 0.42$ \\
   & $P$-wave phase & & & $-0.67 \pm 0.54$ & $-0.77 \pm 0.62$ & $-0.84 \pm 0.52$ \\
   & $D$-wave phase & & & $-1.10 \pm 0.71$ & $-1.85 \pm 0.59$ & $-1.96 \pm 0.58$  \\
   & Resonance fraction, \% & & & $2.1 \pm 0.7$ & $1.8 \pm 0.6$   &  $1.8 \pm 0.6$   \\
    & $P$-wave fraction, \%  &  &  & $34.3 \pm 16.6$ &  $29.5 \pm 16.9$ &  $30.0 \pm 16.7$   \\
   & $D$-wave fraction, \% & & & $45.7 \pm 17.4$  & $40.2 \pm 17.7$  & $38.2 \pm 17.3$  \\
   & $F$-wave fraction, \% & & & $19.4 \pm 15.8$ &  $29.4 \pm 19.3$  &   $31.1 \pm 19.2$  \\
   & Long. polarization, \%  & & & $74.1 \pm 16.5$ & $76.0^{+18.3}_{-8.5}$  & $74.7 \pm 16.1$  \\
   \hline
$b_1(1235)^-D^{*+}$ & Resonance phase &  & & & & $0.52 \pm 0.42$  \\
   & Resonance fraction, \%  & & & &  &  $< 3.1\, (90\%\, {\rm C.L.})$  \\
   \hline
&  $\Delta \mathcal{L}$  & $+33.3$ & $+12.9$ &  $+16.4$ & $0$ & $-2.4$ \\
& Variation, $\sigma$ & $8.2$ & $5.1$ & $5.7$ & $0$ & $2.2$
\end{tabular}
\end{ruledtabular}
\end{table*}
The results of the fit are summarized in Table~\ref{tab:models}.
Together with the individual decay fractions $f_R$, we show the FCC fraction $f_{\rho+\rho'}$, which represents the decay fraction of the coherent sum of the $\rho(770)^-$ and $\rho(1450)^-$ states.

We also show the partial wave fractions describing the $\rho(1450)^-$ and $D^{**}$ production in the specific partial waves and the longitudinal polarizations of these resonances.
We see that the $\rho(1450)^-$ state is produced dominantly via $S$ wave, but that $D_1(2430)^0$ production requires approximately equal fractions of all partial waves.
The partial wave fractions of the $D_1(2420)^0$ and $D^*_2(2460)^0$ are not statistically significant.
In our analysis, the longitudinal polarization and partial wave fractions of the $\rho(1450)^-$ are fixed in part from the requirement on the relative normalizations of the helicity amplitudes, $R_1$, $R_2$ and $\rho^2$ (see Appendix~\ref{sec:appC}), and can be relaxed in the fit due to the free mass and width of the $\rho(1450)^-$. Large longitudinal polarizations of the $D^{**}$ states indicate violation of the factorization hypothesis but the statistical
uncertainties are large.

The final-state interaction phases $\phi_+$ and $\phi_-$ defined in Appendix~\ref{sec:appC} are taken into account in the description of the $\rho$-meson-like states. The fit gives a nontrivial value for the $\phi_+$ phase.

One must also consider the statistical errors on the fit fractions, partial wave fractions and longitudinal polarizations. These errors are determined with a pseudoexperiment technique (see Sec.~\ref{sec:D}).

The masses and widths of all resonances except for the $\rho(1450)$ are fixed at their PDG values \cite{pdg}.
Our measurements for the $\rho(1450)$ shape parameters do not contradict previous experimental observations \cite{pdg}, although they differ slightly from the CLEO results \cite{cleo}. This situation is expected because the broad $\rho$-meson-like states overlap strongly with each other and the Breit-Wigner description is not accurate.

Mixing between the $D_1(2430)^0$ and $D_1(2420)^0$ states is expected to be small and is therefore neglected. If we take into account the mixing effect, the mixing angles defined in Appendix~\ref{sec:appC} are found to be $\omega = -0.03 \pm 0.02\, ({\rm stat.})$ and $\varphi =  -0.27 \pm 0.75\, ({\rm stat.})$. Within errors, these angles are consistent with the previous Belle measurement \cite{b-todstpi}.

\begin{figure*}[!htbp]
\begin{center}
\begin{tabular}{c c}
\includegraphics[scale=.33]{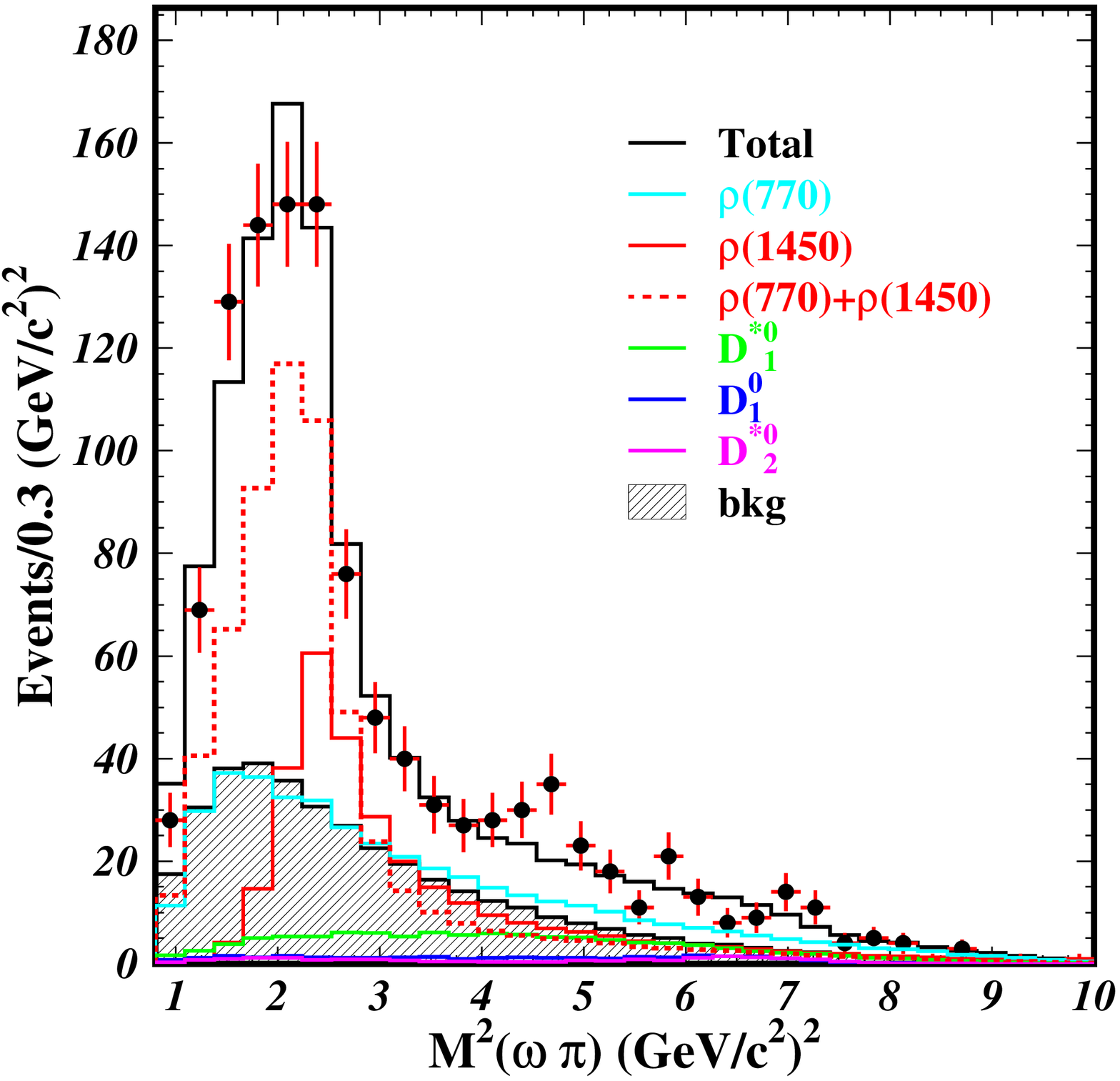} &
\includegraphics[scale=.33]{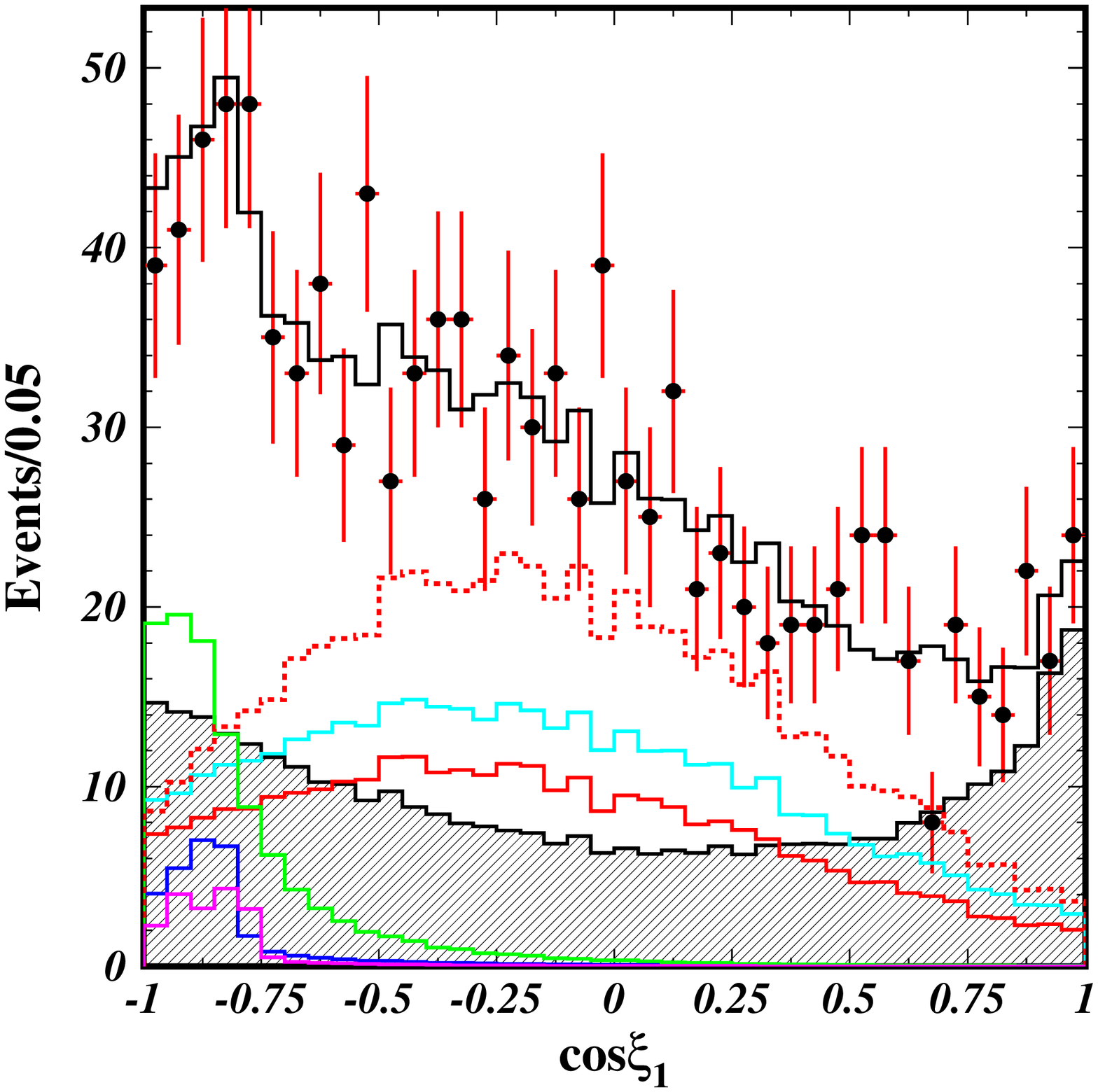} \\
\includegraphics[scale=.33]{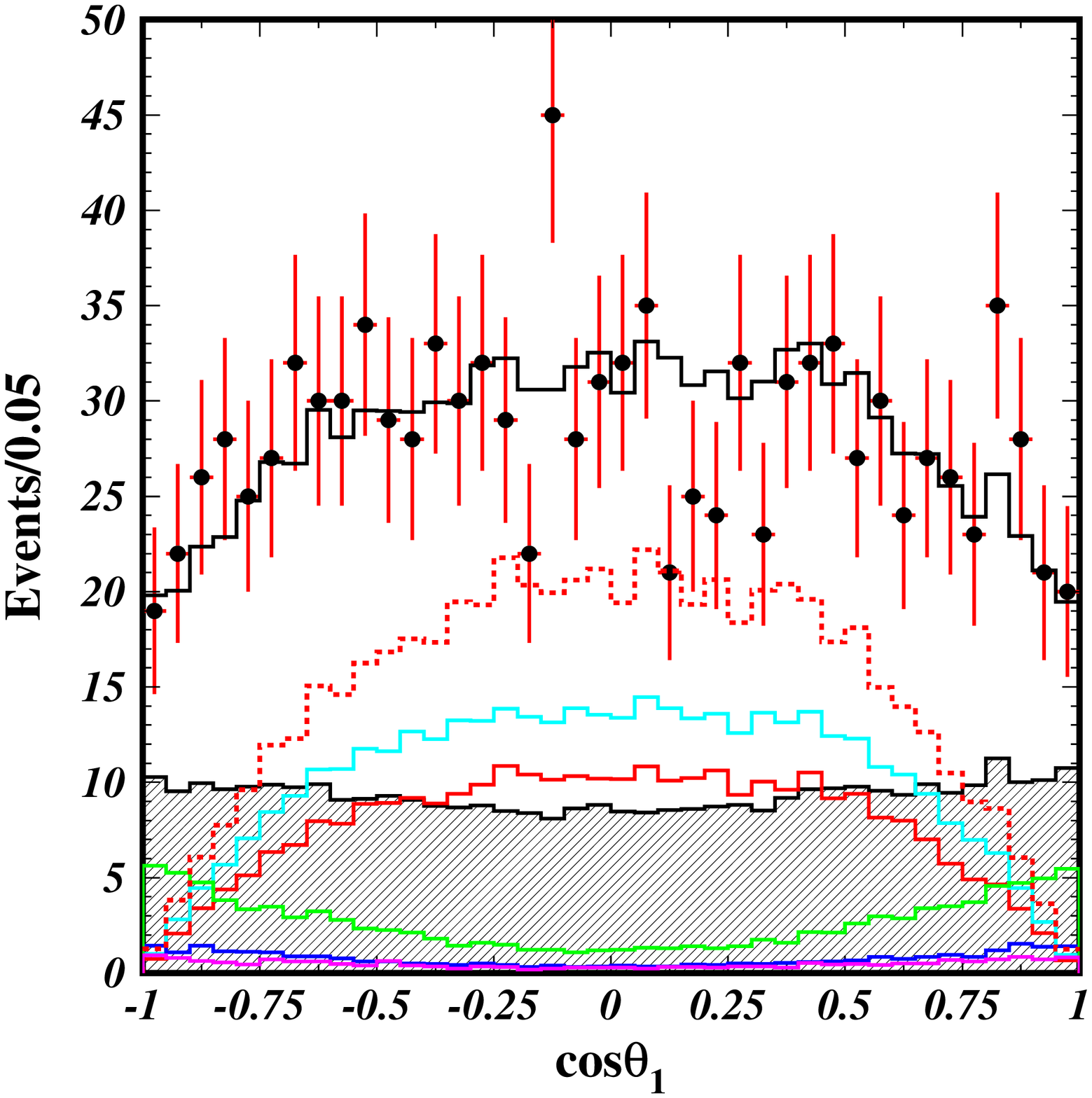} &
\includegraphics[scale=.33]{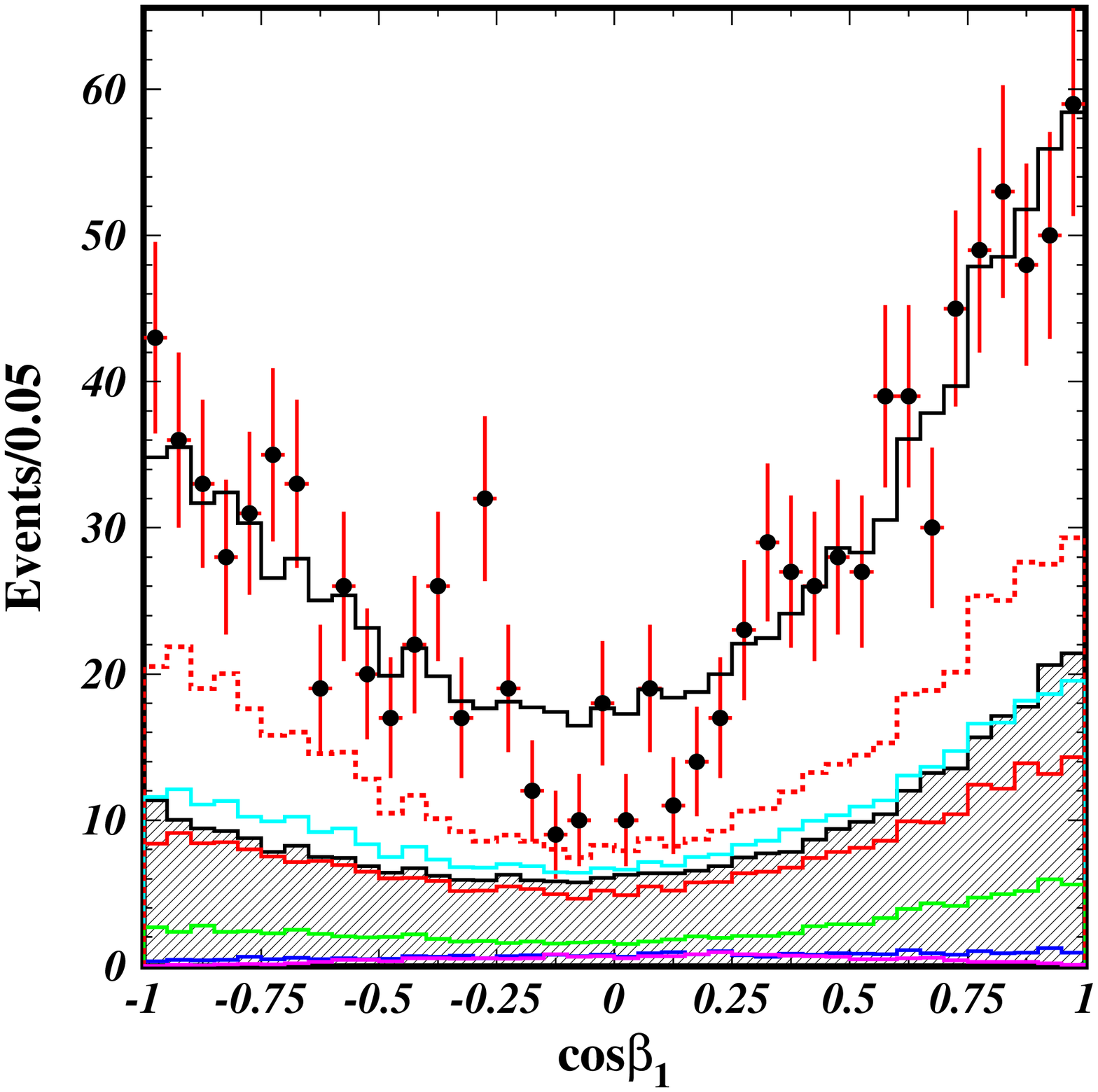} \\
\includegraphics[scale=.33]{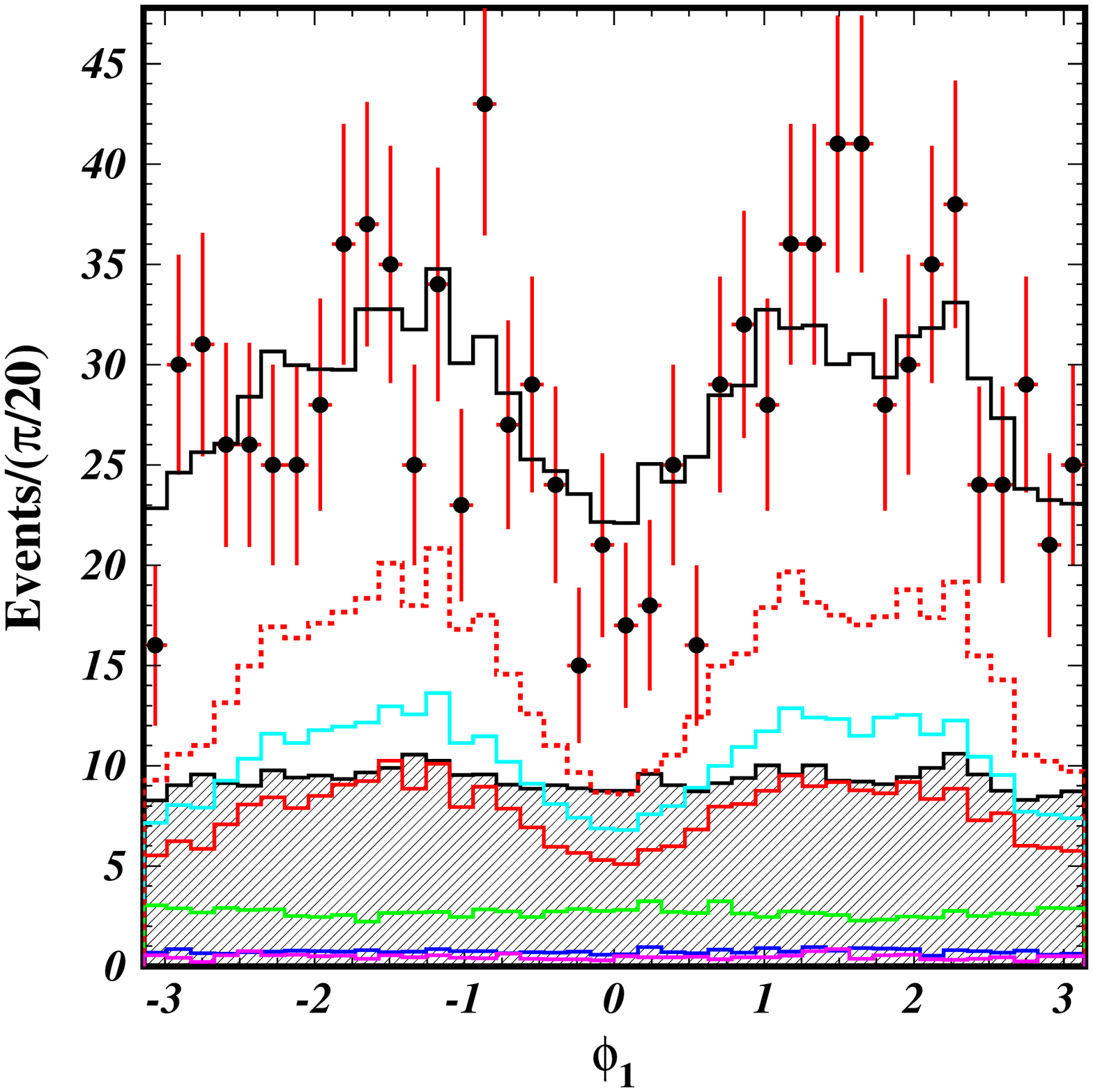} &
\includegraphics[scale=.33]{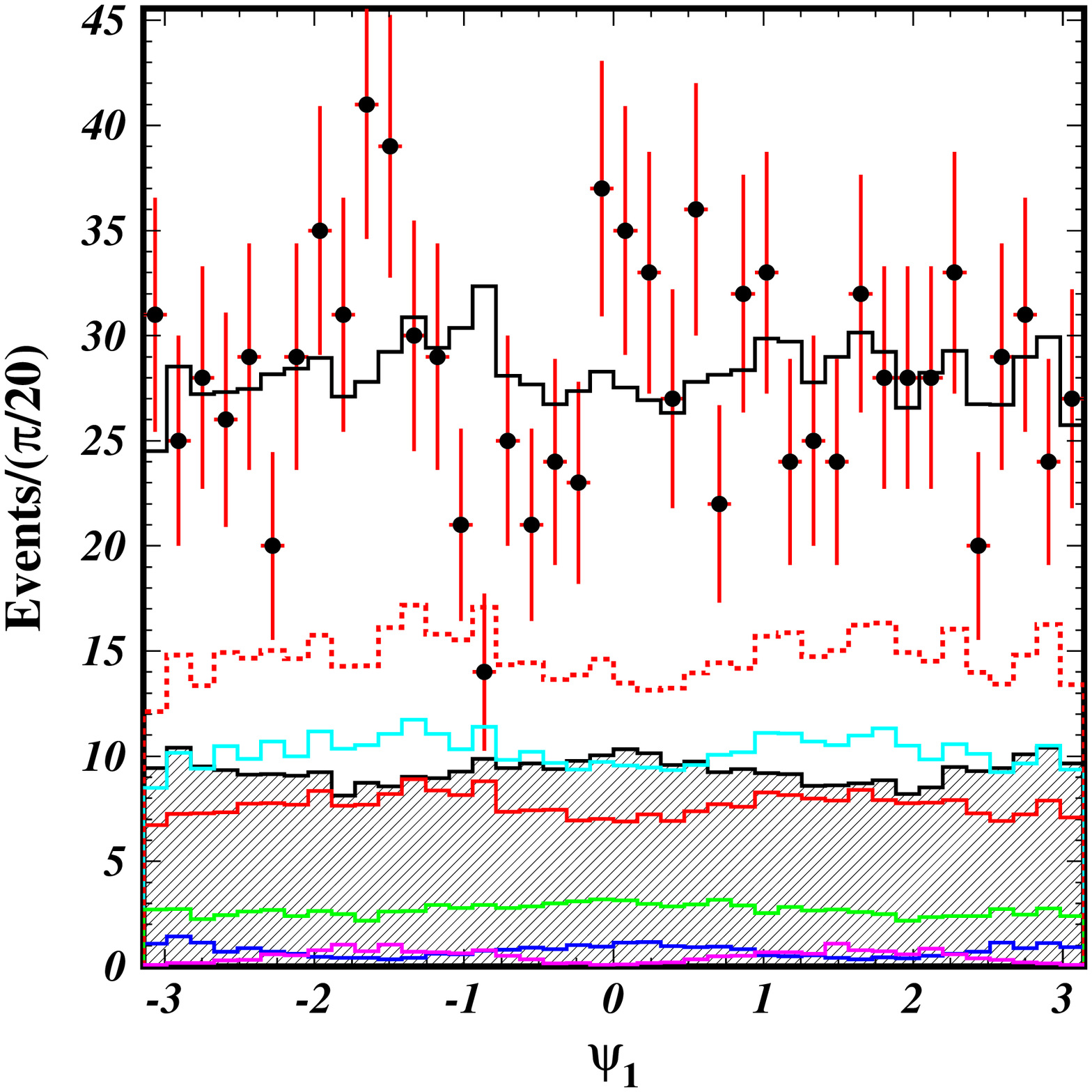}
\end{tabular}
\caption{(color online). Distribution of the six $\boldsymbol{\omega\pi}$ variables for $D^* \omega \pi$ candidates in the signal region (points with error bars). The histograms represent the results of the fit (black), including the following components: $\rho(770)$ (cyan), $\rho(1450)$ (red), $\rho(770)$ and $\rho(1450)$ together (red dashed), $D_1(2430)^0$ (green), $D_1(2420)^0$ (blue), $D^*_2(2460)^0$ (magenta) and background (hatched).}
\label{fig:fitresompi}
\end{center}
\end{figure*}

\begin{figure*}[!htbp]
\begin{center}
\begin{tabular}{c c}
\includegraphics[scale=.33]{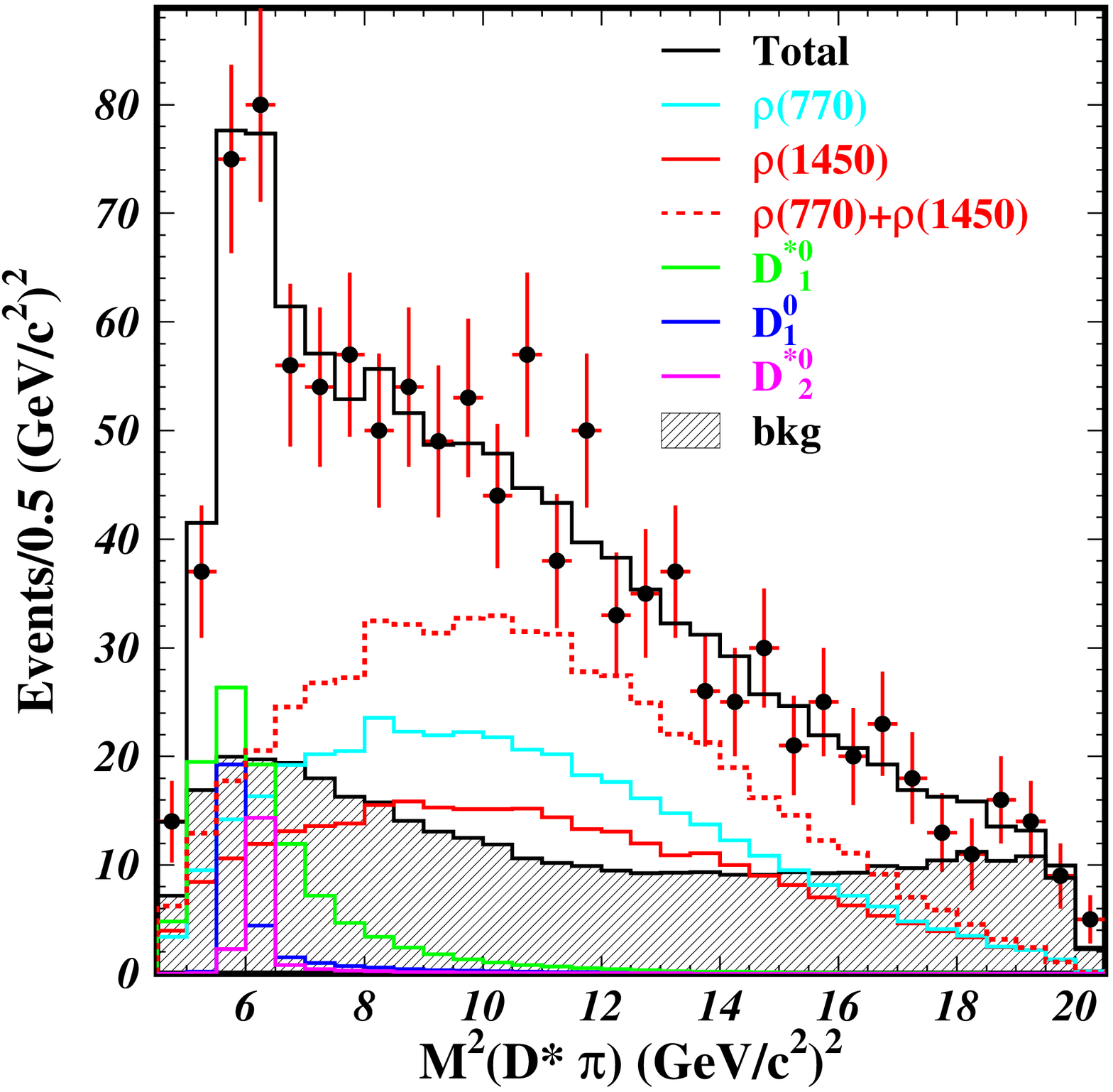} &
\includegraphics[scale=.33]{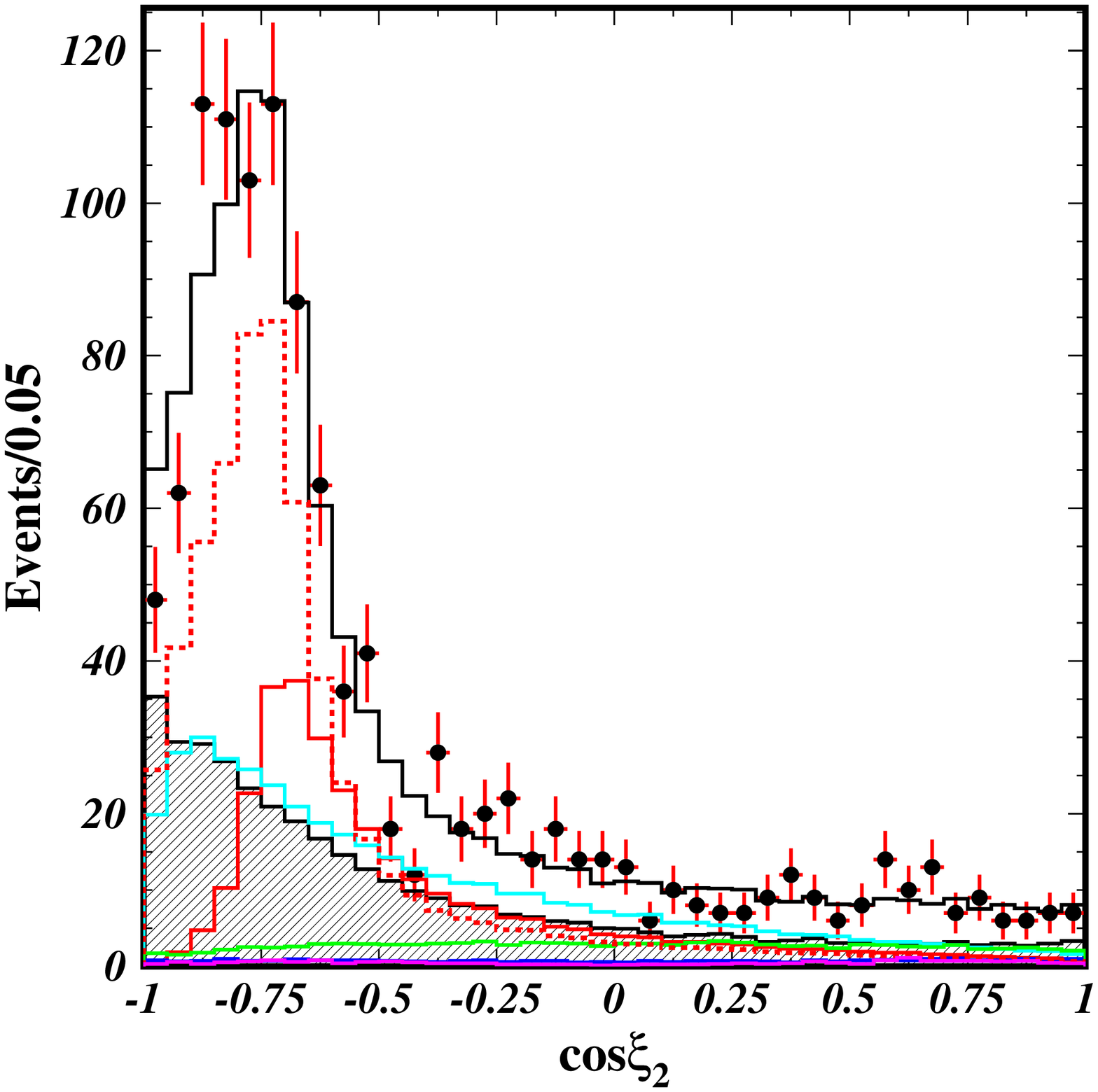} \\
\includegraphics[scale=.33]{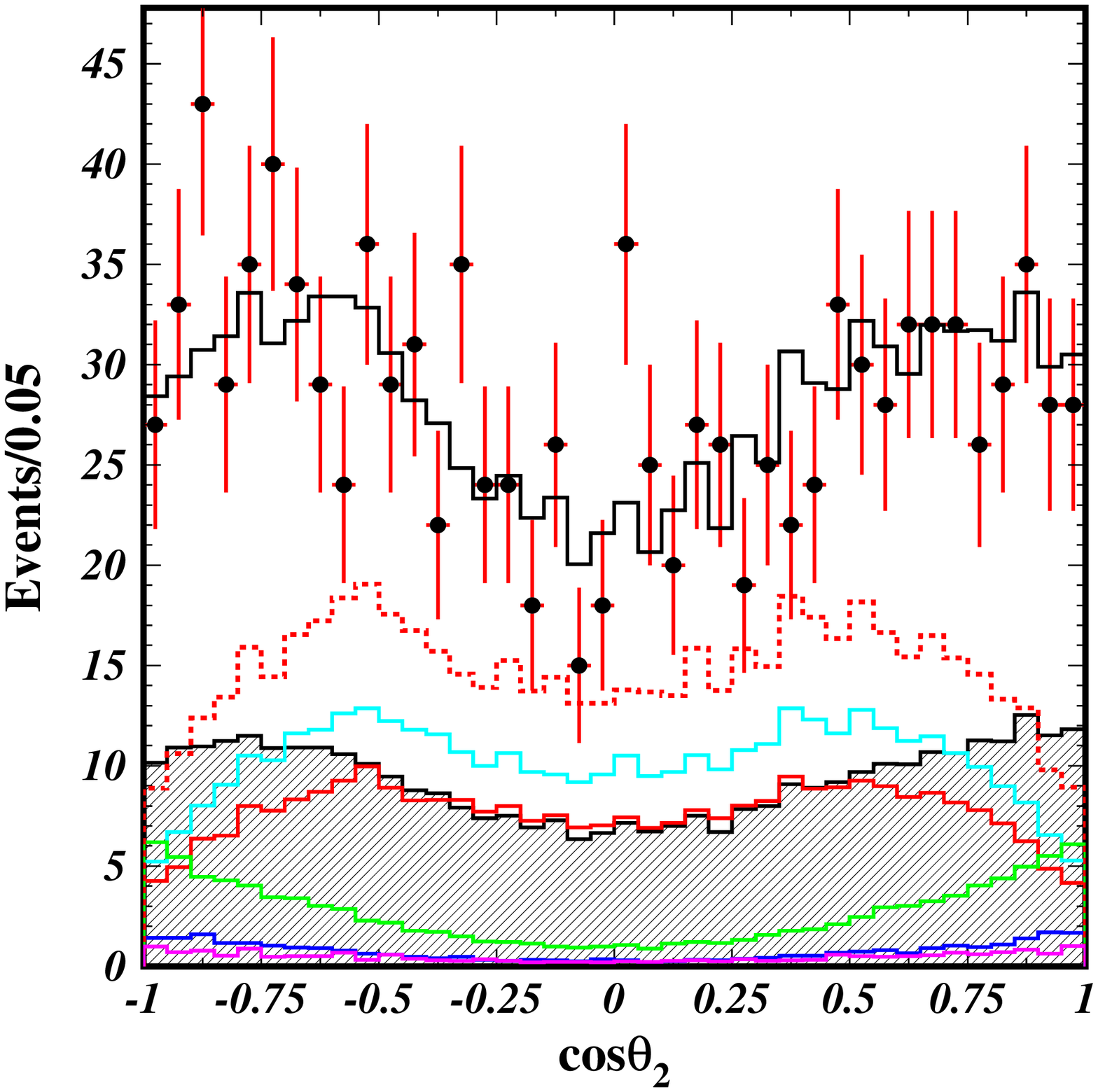} &
\includegraphics[scale=.33]{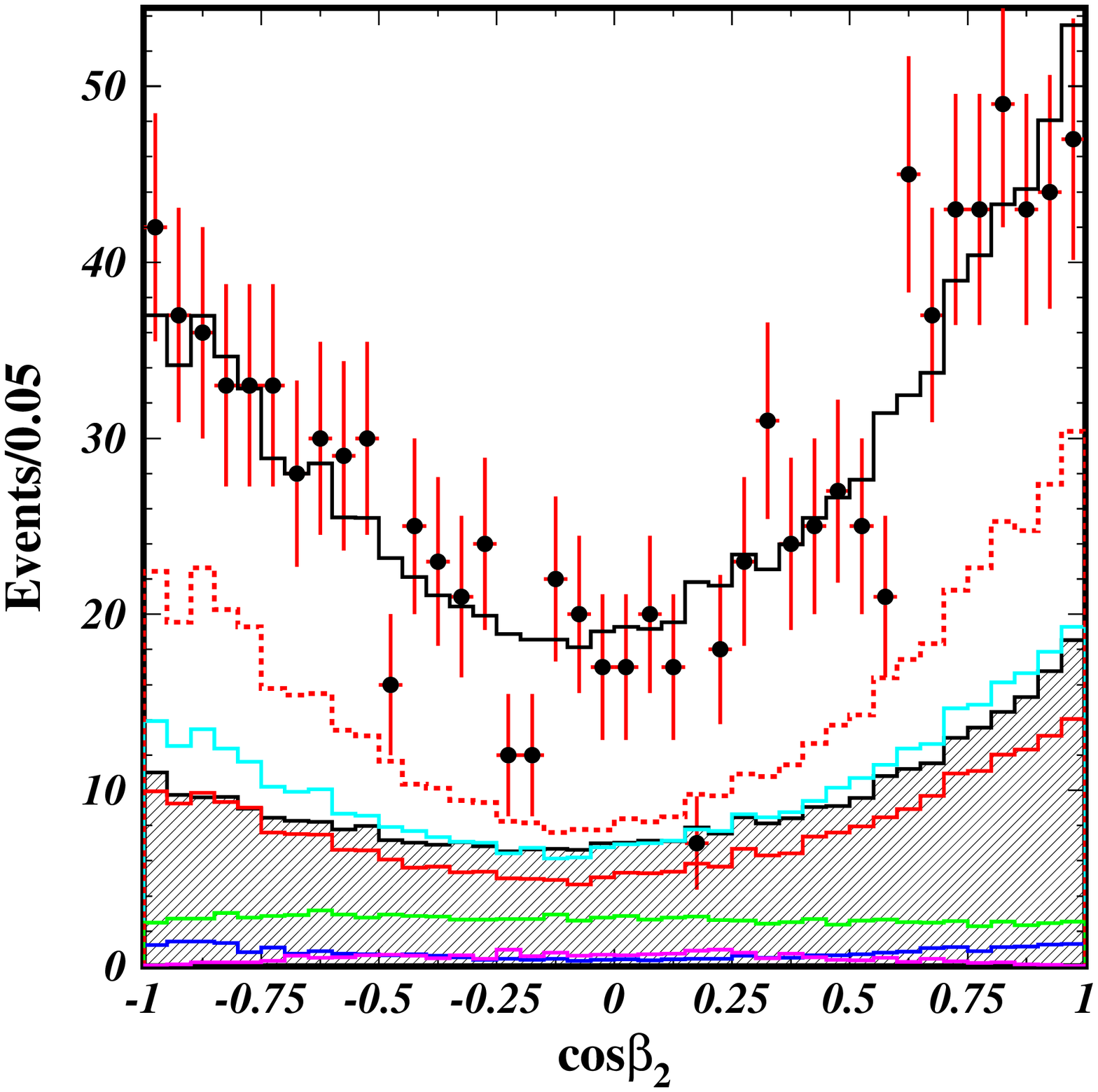} \\
\includegraphics[scale=.33]{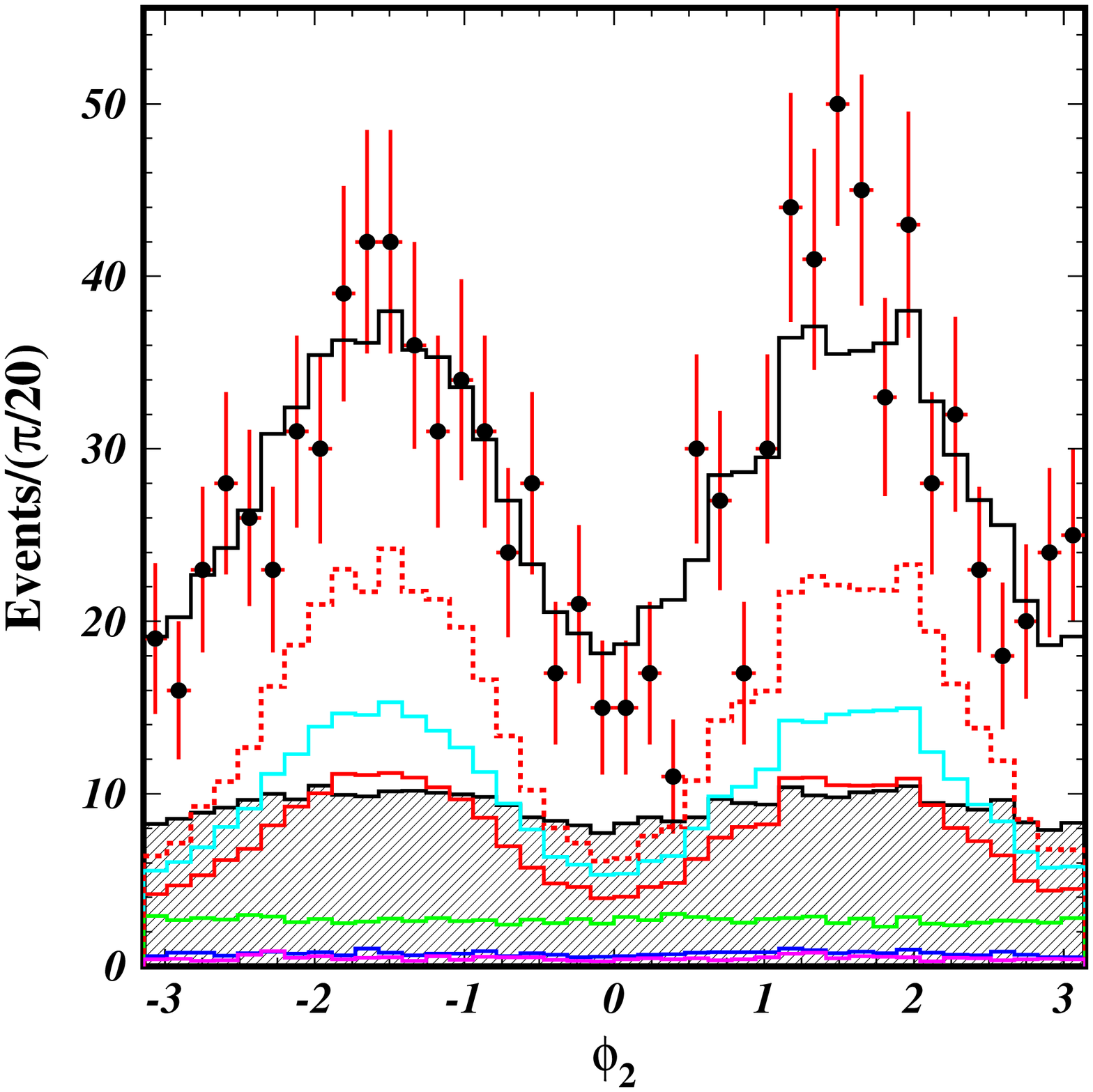} &
\includegraphics[scale=.33]{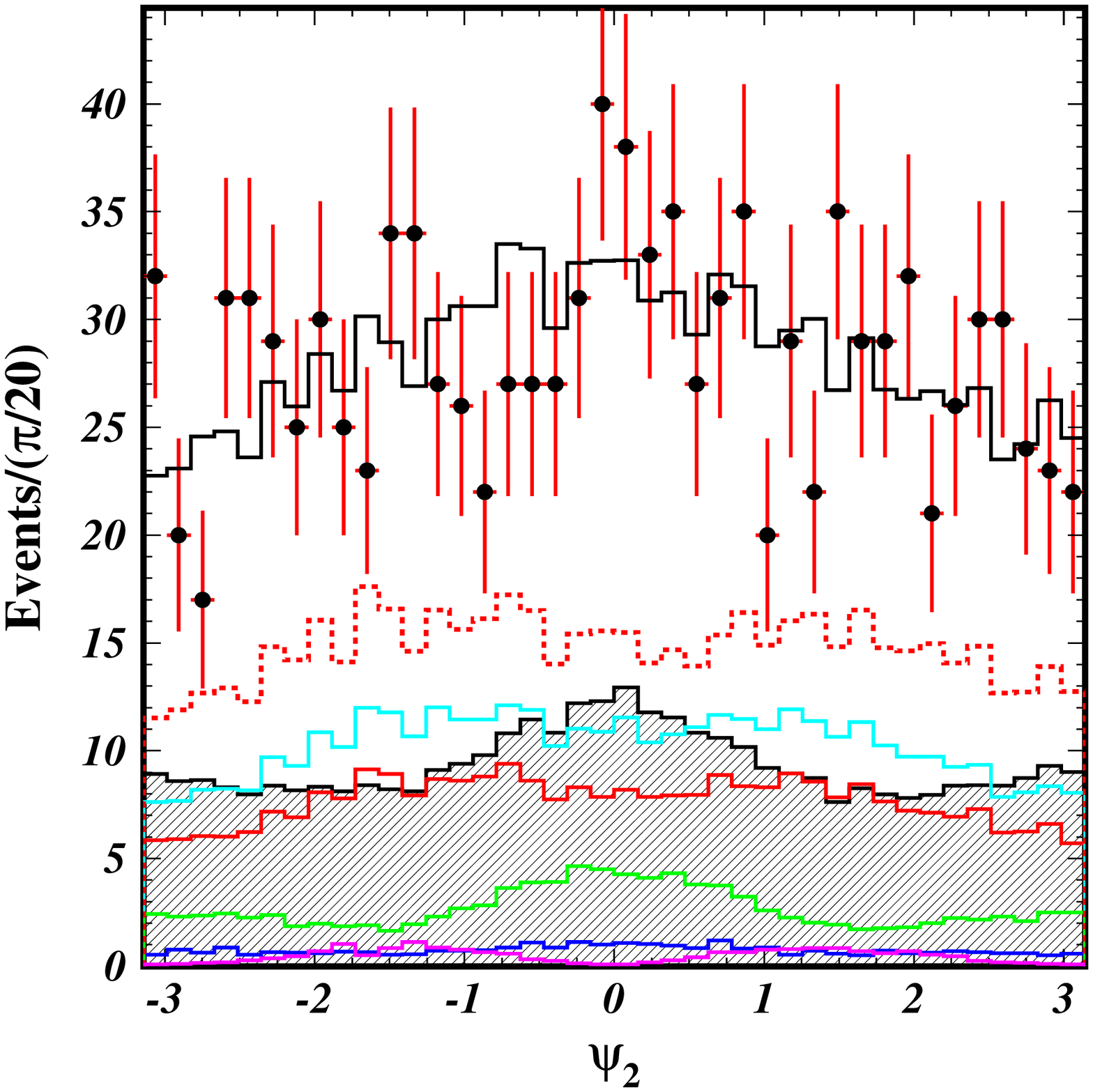}
\end{tabular}
\caption{(color online). Distribution of the six $\boldsymbol{D^*\pi}$ variables for $D^* \omega \pi$ candidates in the signal region (points with error bars). The histograms represent the results of the fit (black), including the following components: $\rho(770)$ (cyan), $\rho(1450)$ (red), $\rho(770)$ and $\rho(1450)$ together (red dashed), $D_1(2430)^0$ (green), $D_1(2420)^0$ (blue), $D^*_2(2460)^0$ (magenta) and background (hatched).}
\label{fig:fitresdstpi}
\end{center}
\end{figure*}

\begin{figure*}[!htbp]
\begin{center}
\begin{tabular}{c c}
\includegraphics[scale=.33]{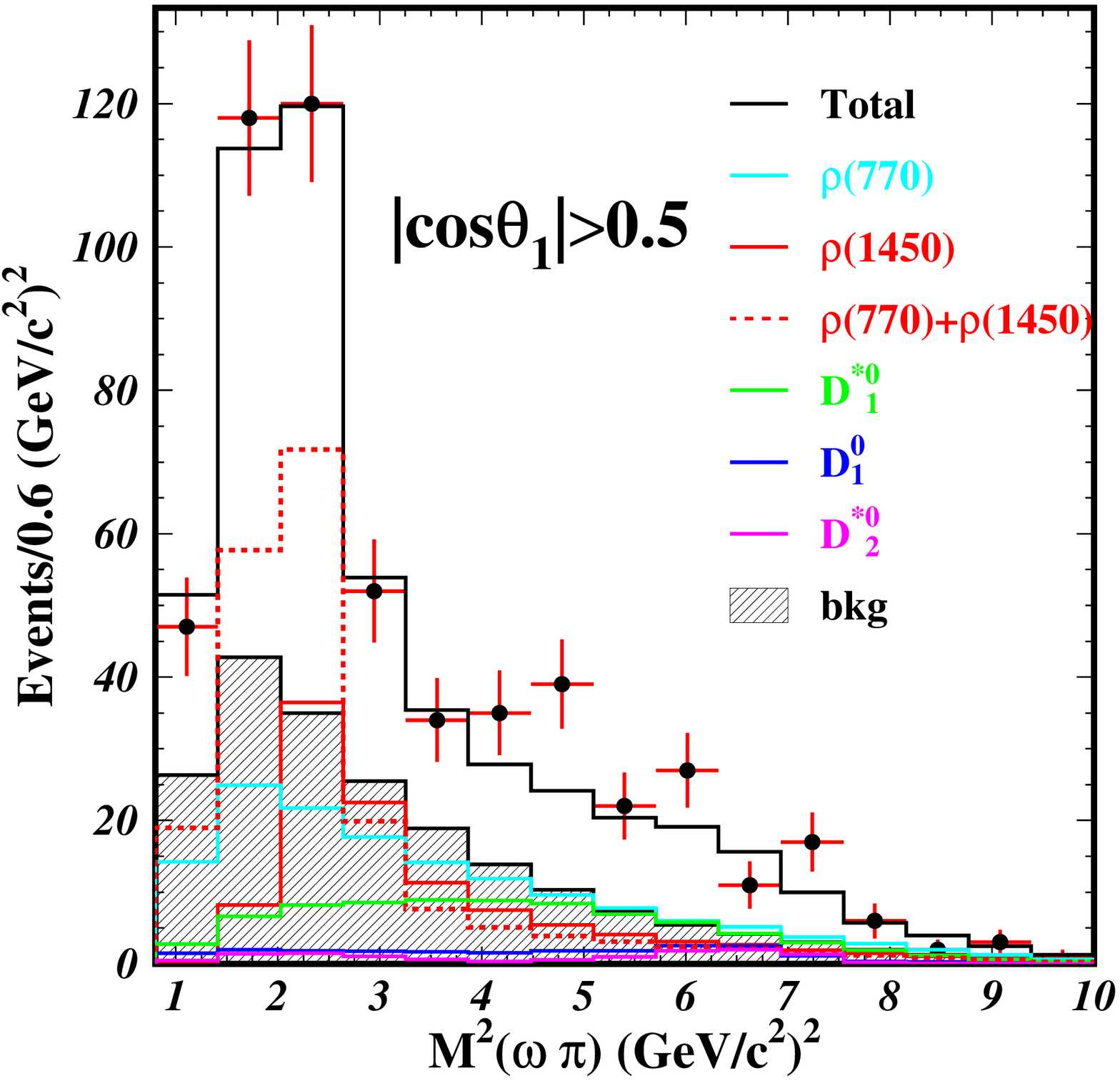} &
\includegraphics[scale=.33]{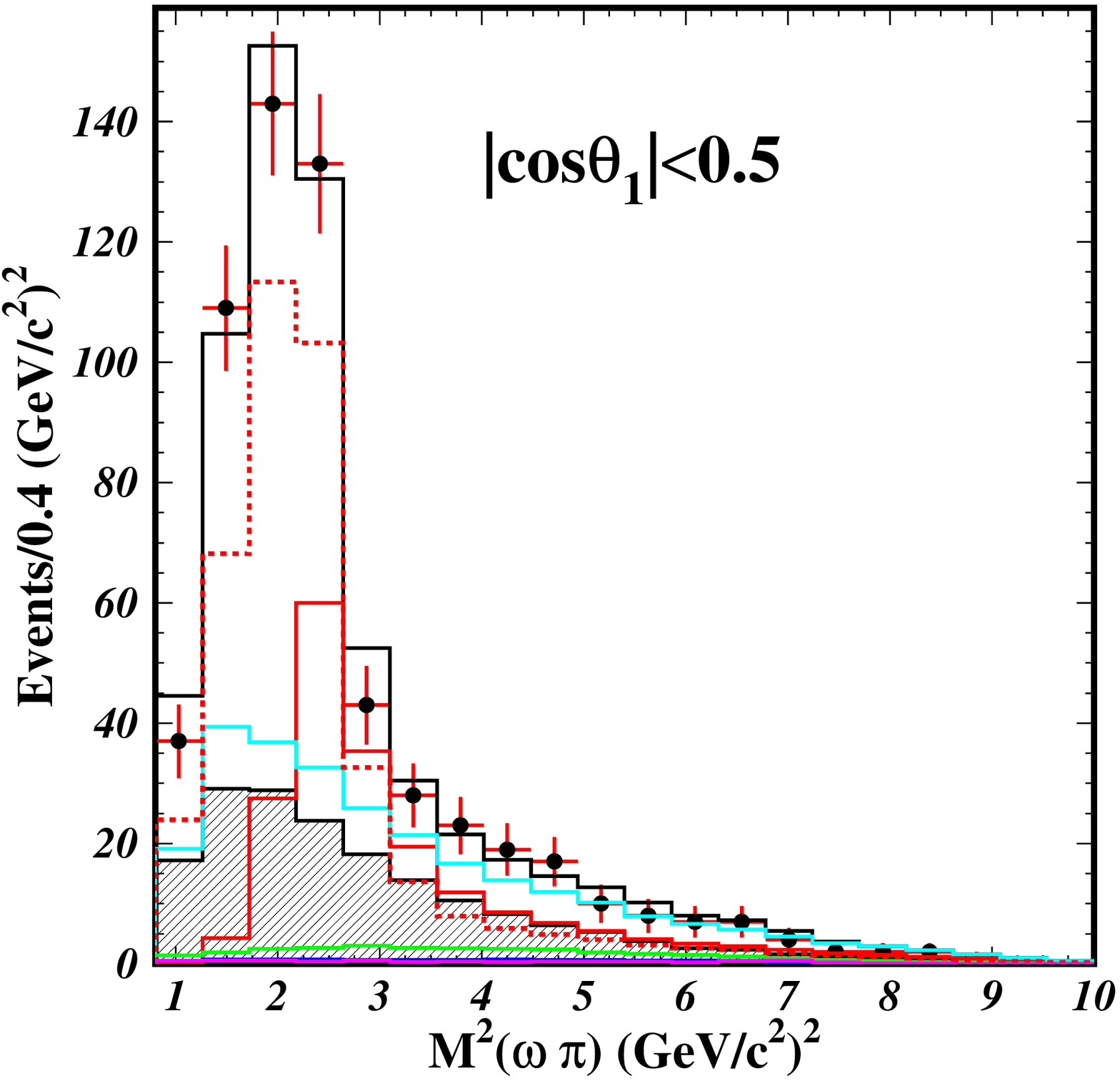} \\
\includegraphics[scale=.33]{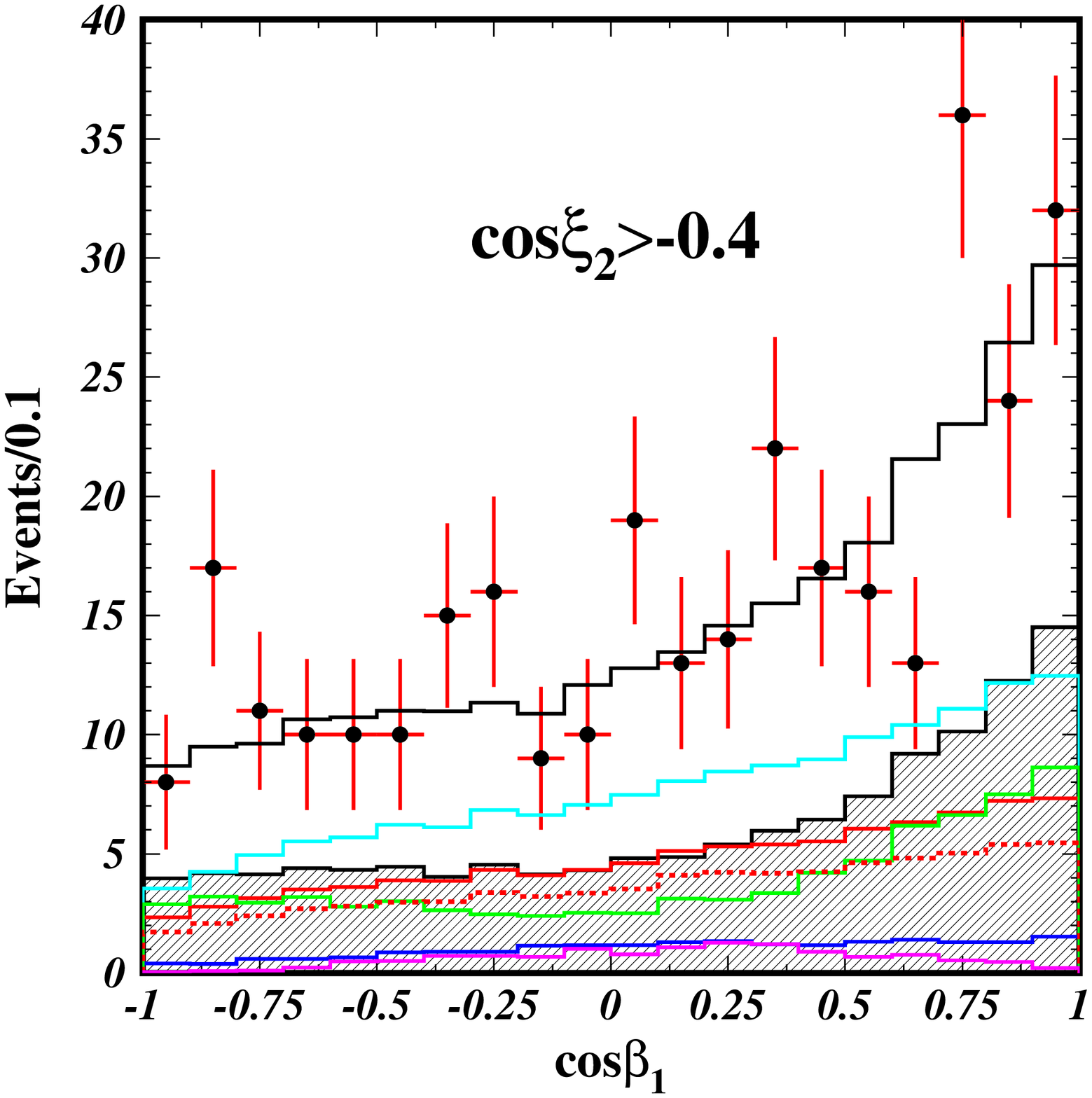} &
\includegraphics[scale=.33]{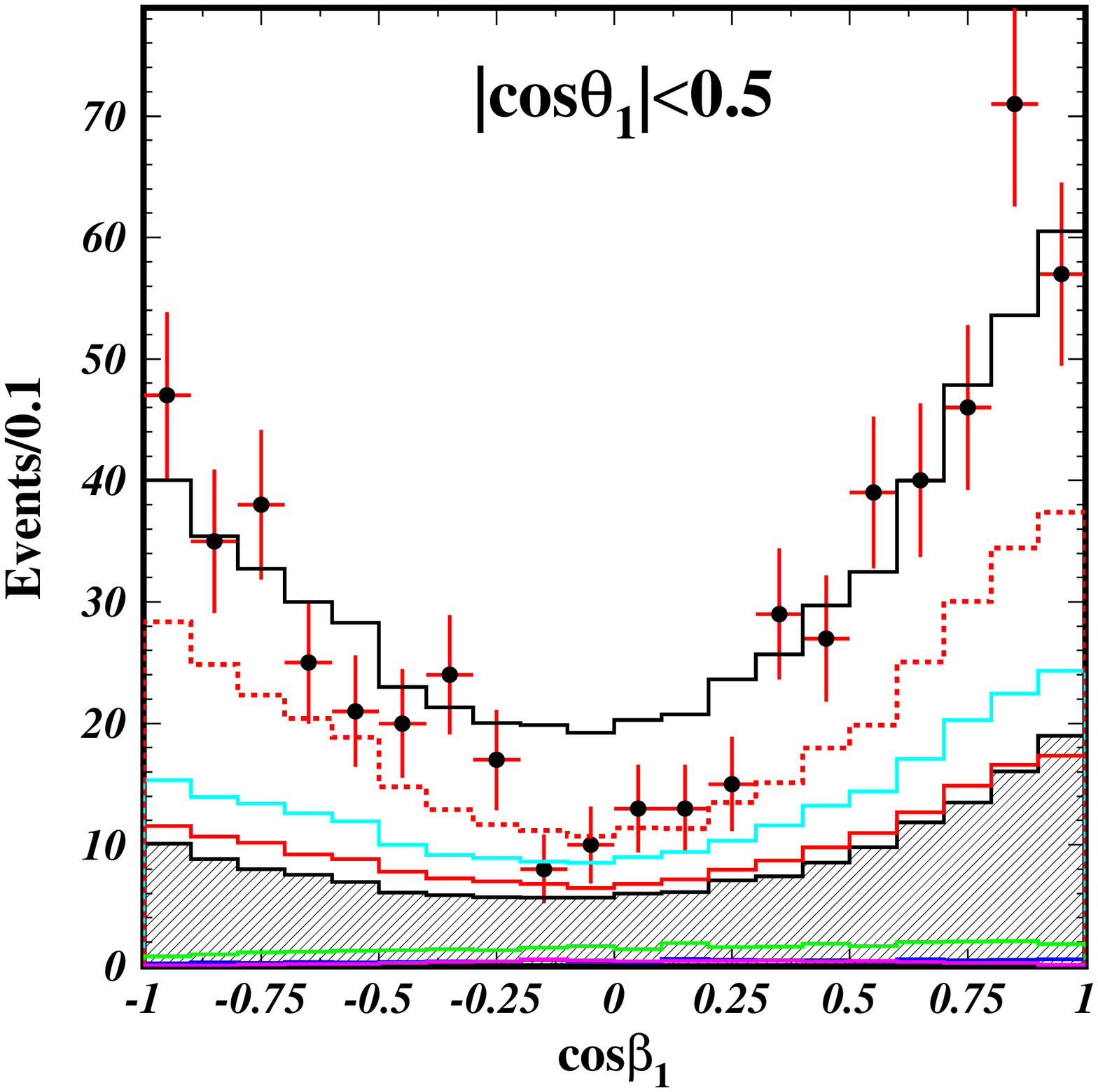} \\
\includegraphics[scale=.33]{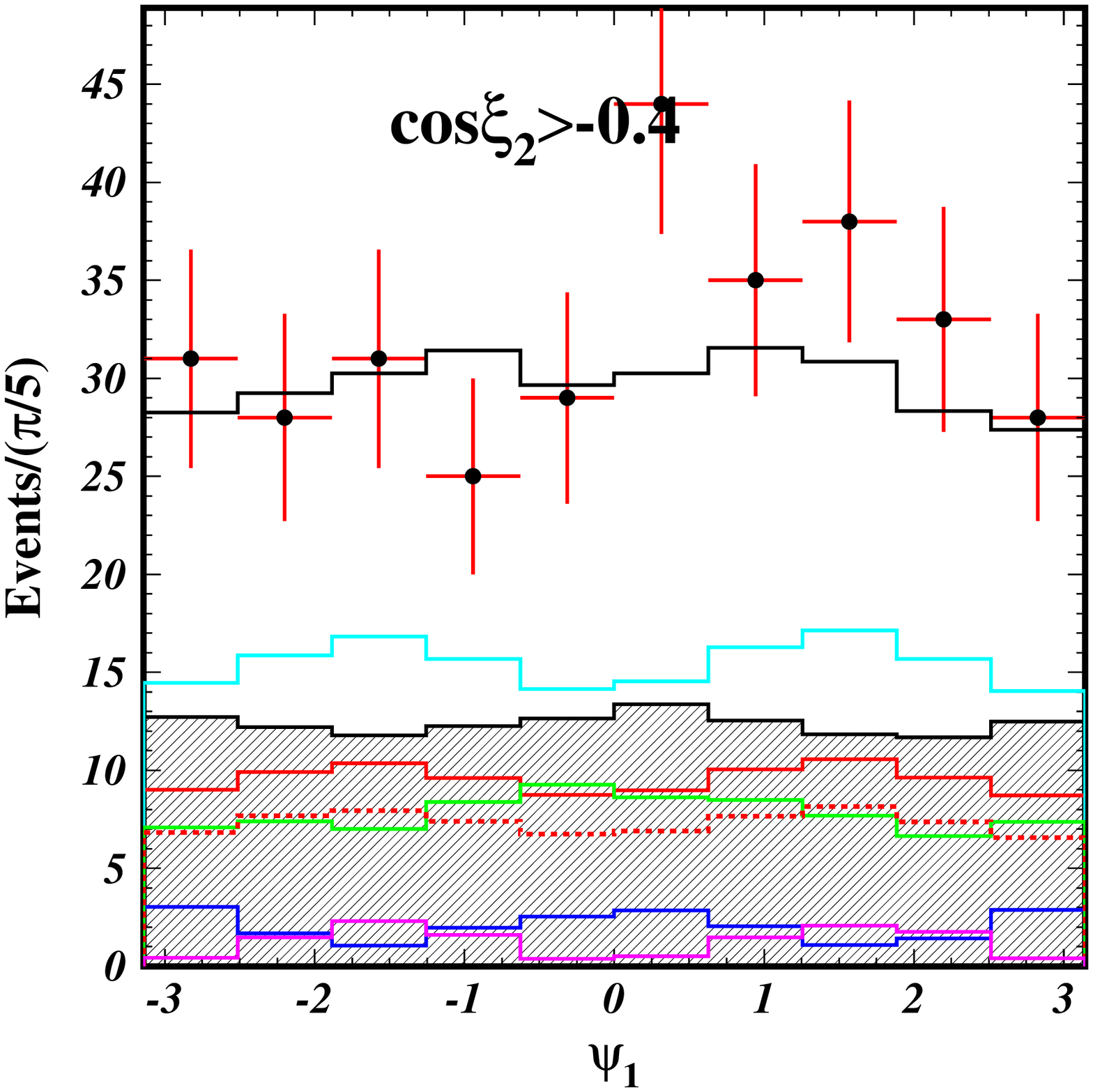} &
\includegraphics[scale=.33]{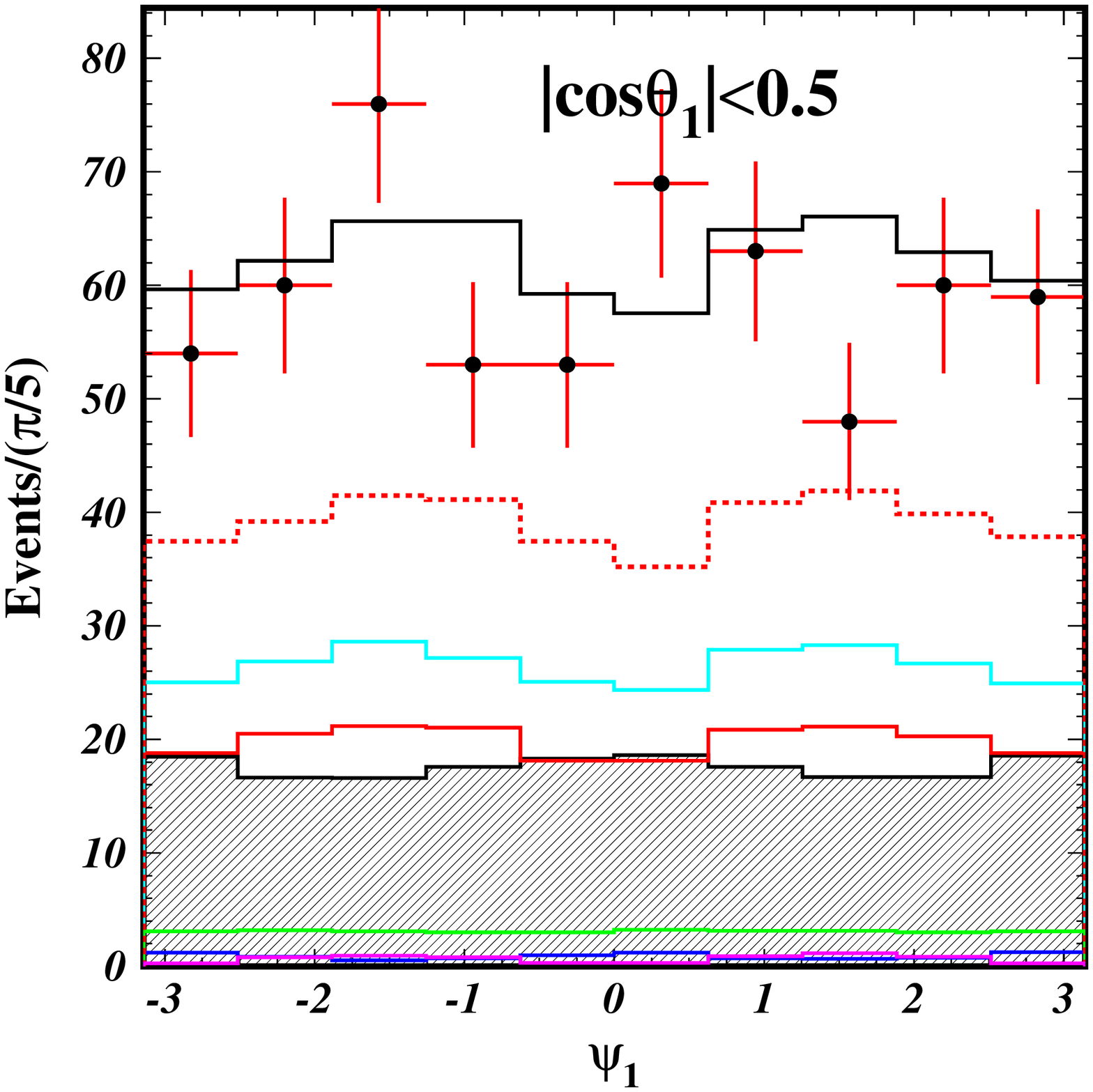} \\
\end{tabular}
\caption{(color online). Distribution of three $\boldsymbol{\omega \pi}$ variables for $D^{*+} \omega \pi^-$ candidates in two different subregions of the signal region (points with error bars), defined by $\cos\xi_2>-0.4$ or $|\cos\theta_1|>0.5$ ($D^{**}$ enriched) and $|\cos\theta_1|<0.5$ ($D^{**}$ depleted). The histograms represent the results of the fit (black), including the following components: $\rho(770)$ (cyan), $\rho(1450)$ (red), $\rho(770)$ and $\rho(1450)$ together (red dashed), $D_1(2430)^0$ (green), $D_1(2420)^0$ (blue), $D^*_2(2460)^0$ (magenta) and background (hatched).}
\label{fig:fitddst}
\end{center}
\end{figure*}

\begin{figure*}[!htbp]
\begin{center}
\begin{tabular}{c c}
\includegraphics[scale=.33]{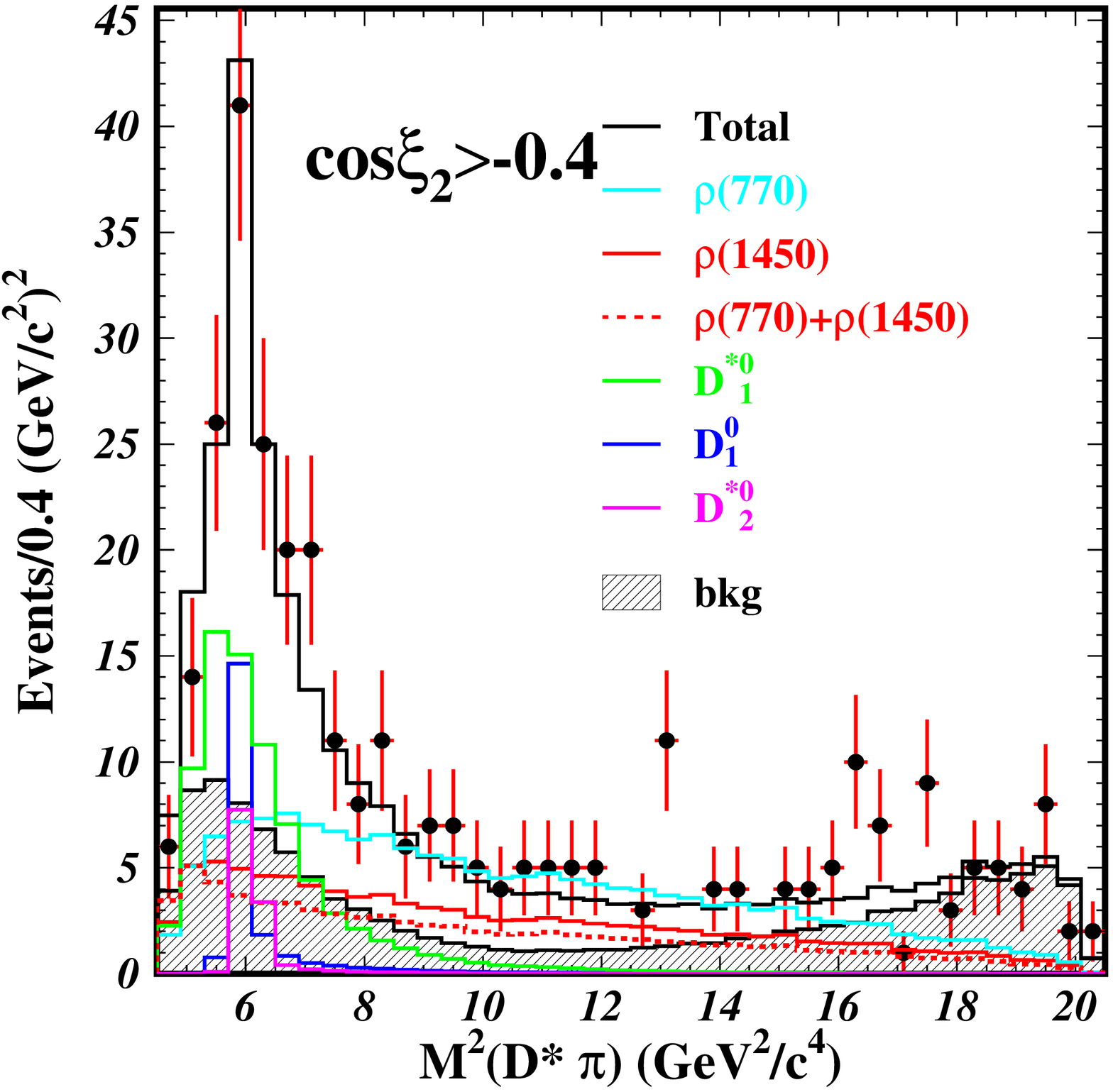} &
\includegraphics[scale=.33]{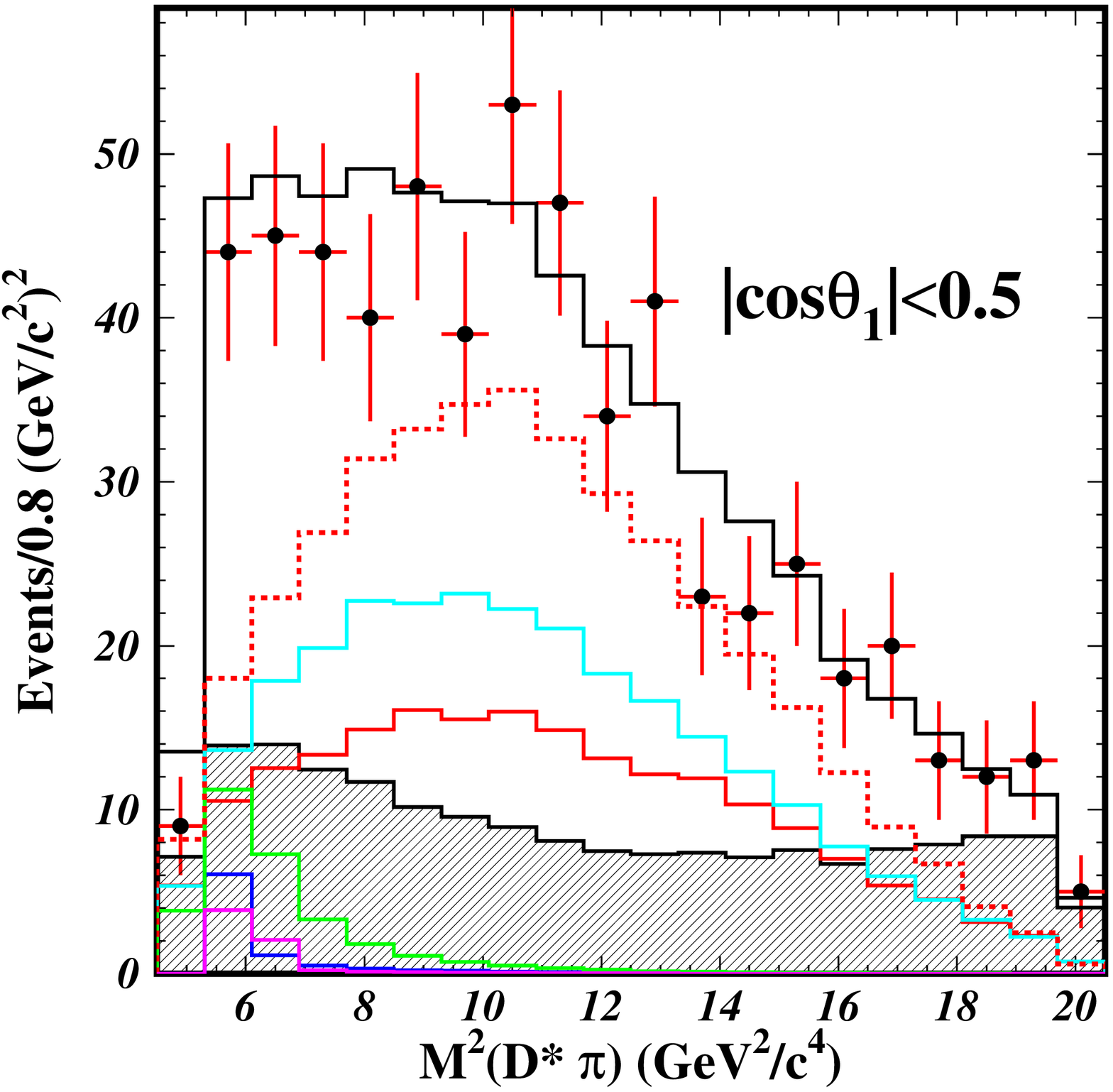} \\
\includegraphics[scale=.33]{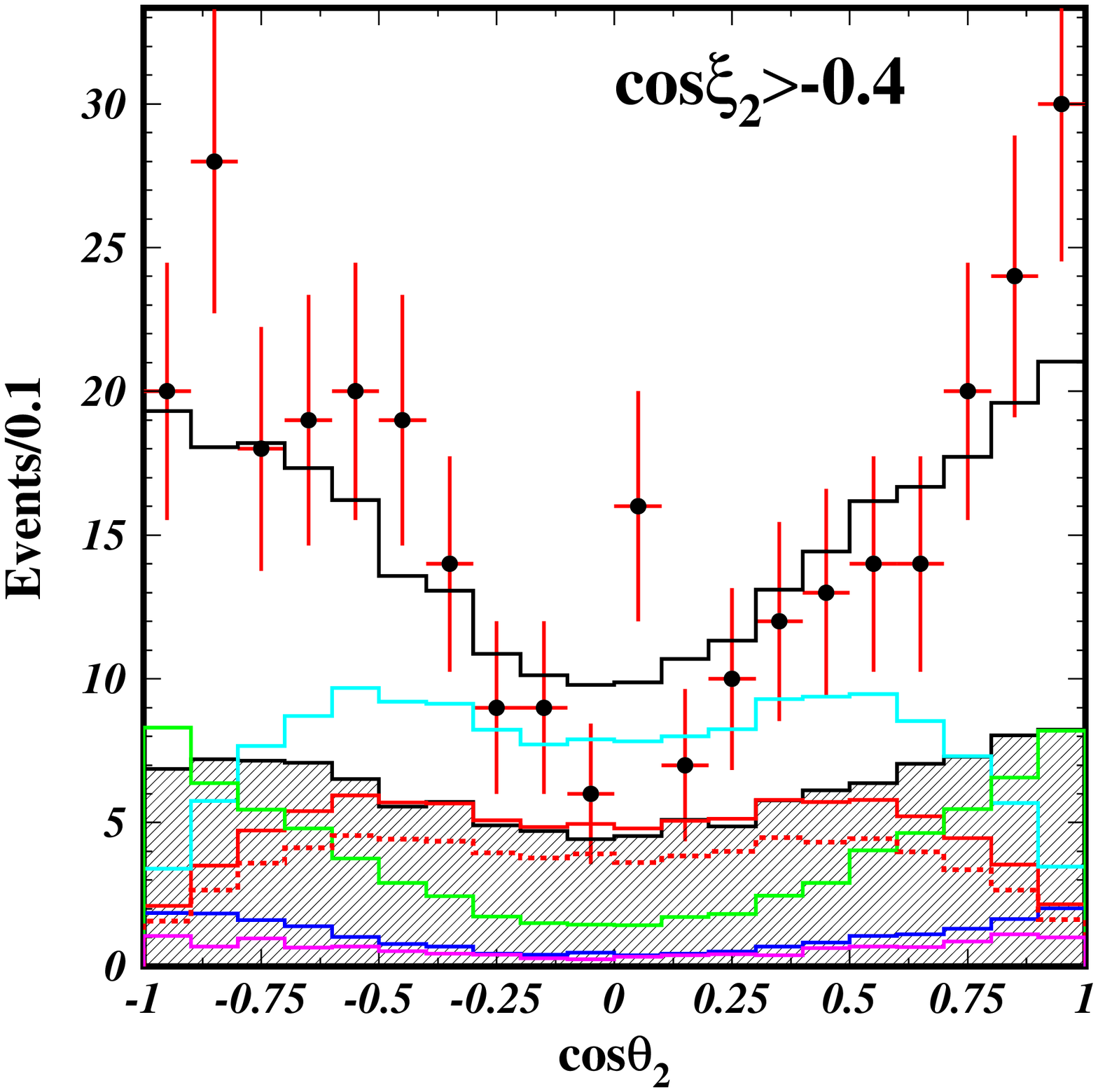} &
\includegraphics[scale=.33]{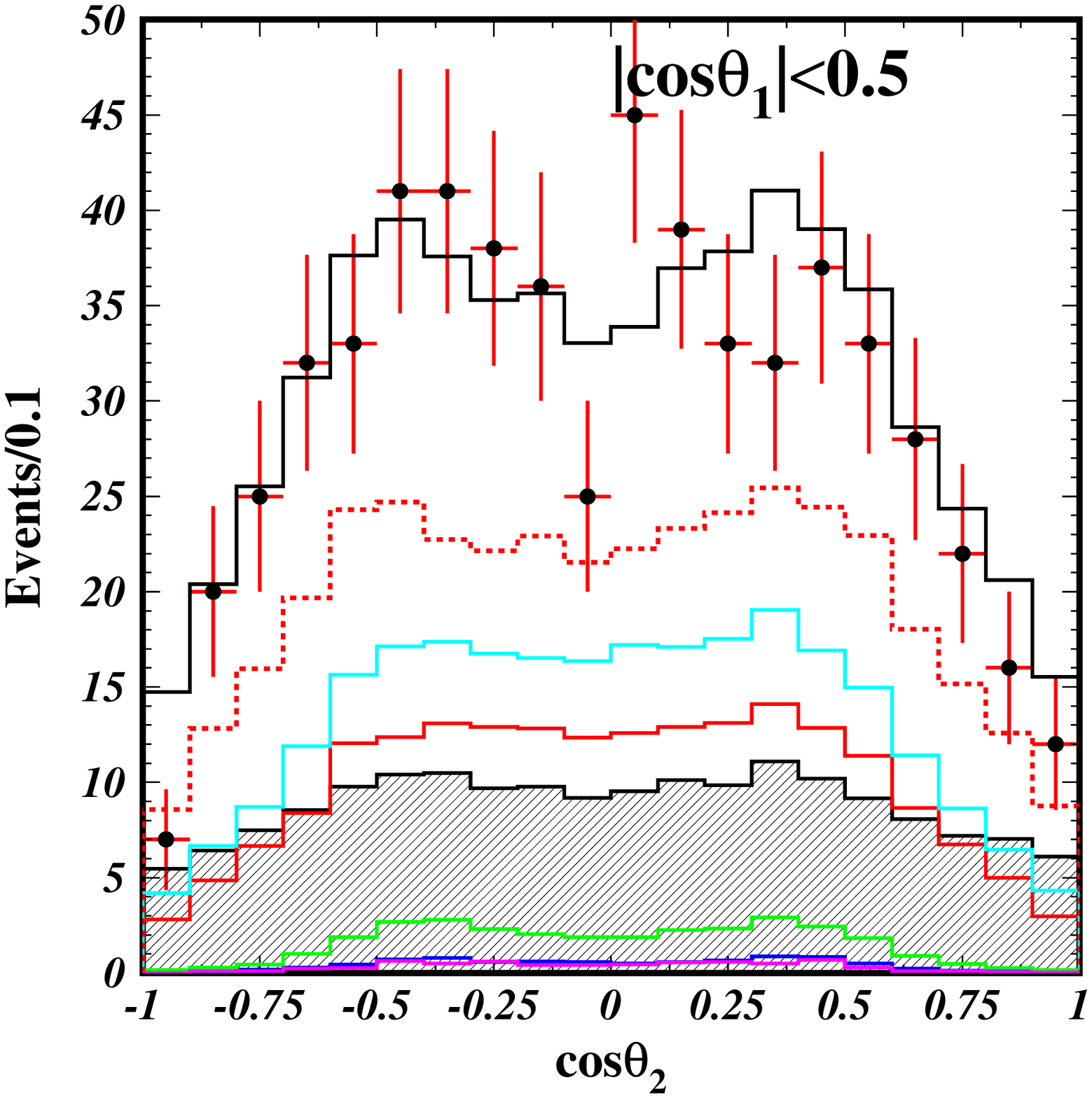} \\
\includegraphics[scale=.33]{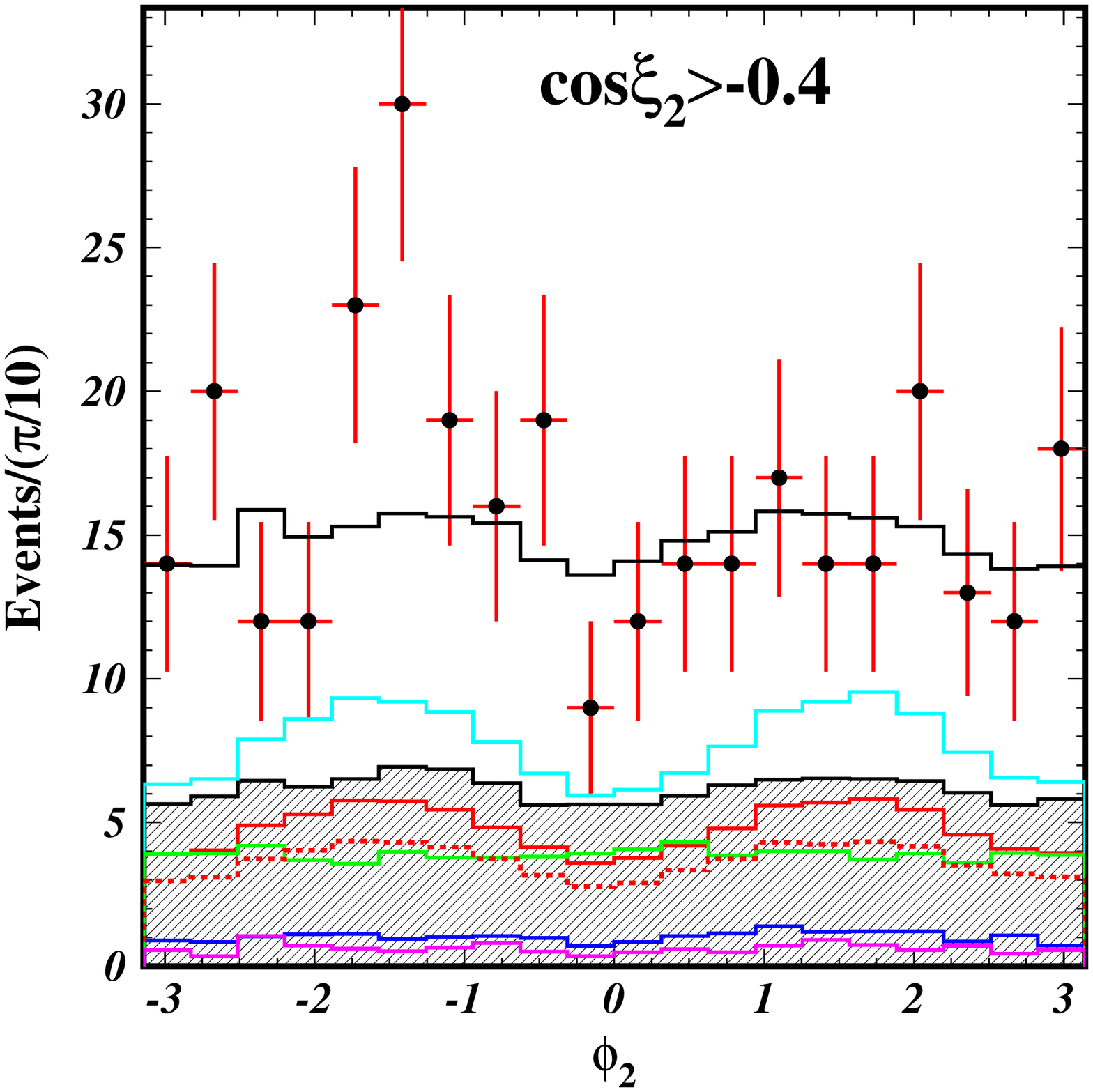} &
\includegraphics[scale=.33]{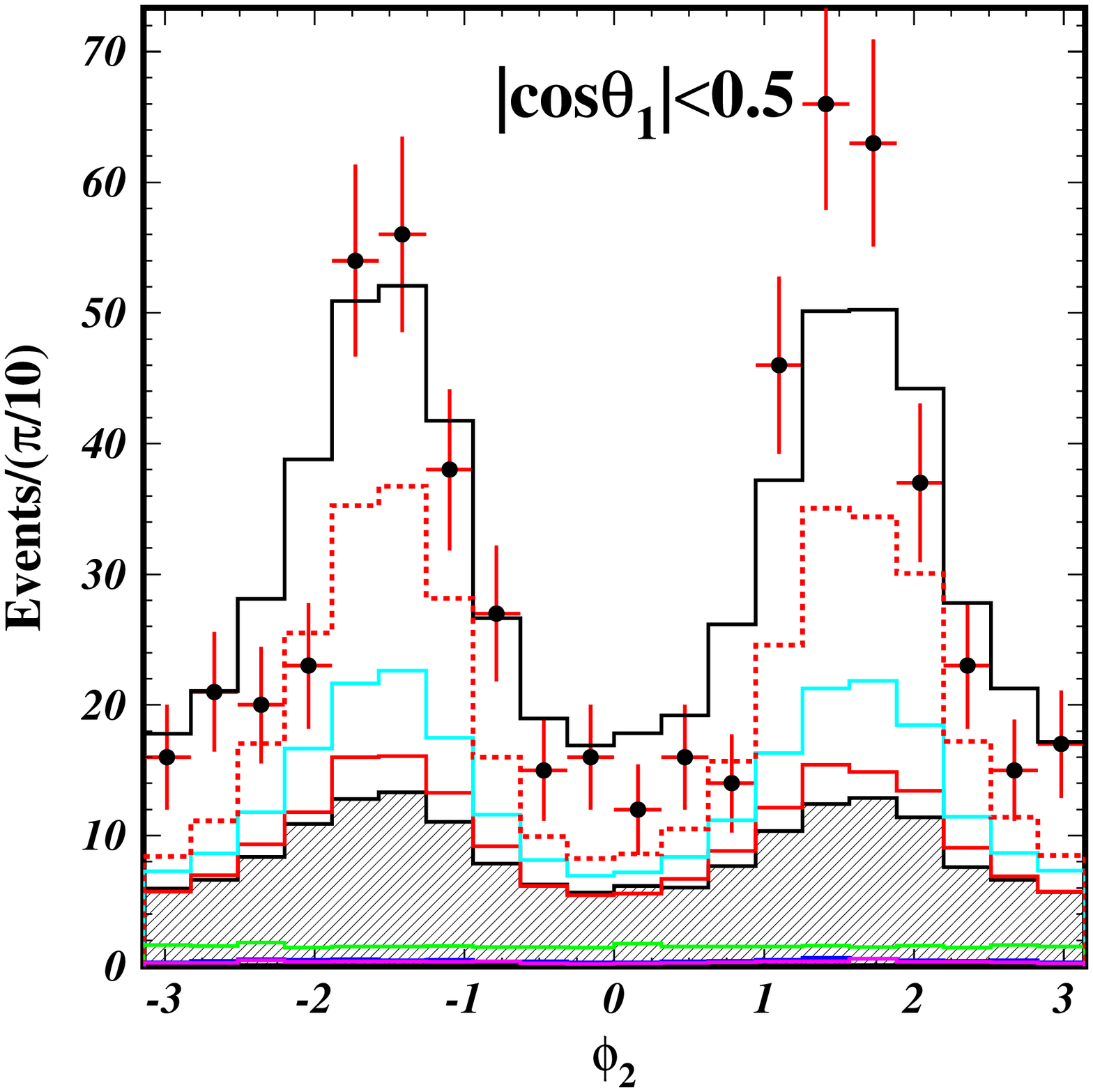} \\
\end{tabular}
\caption{(color online). Distribution of three $\boldsymbol{D^*\pi}$ variables for $D^{*+} \omega \pi^-$ candidates in two different subregions of the signal region (points with error bars), defined by $\cos\xi_2>-0.4$ ($D^{**}$ enriched) and $|\cos\theta_1|<0.5$ ($D^{**}$ depleted). The histograms represent the results of the fit (black), including the following components: $\rho(770)$ (cyan), $\rho(1450)$ (red), $\rho(770)$ and $\rho(1450)$ together (red dashed), $D_1(2430)^0$ (green), $D_1(2420)^0$ (blue), $D^*_2(2460)^0$ (magenta) and background (hatched).}
\label{fig:fitompi}
\end{center}
\end{figure*}

Figures~\ref{fig:fitresompi} and \ref{fig:fitresdstpi} show the distributions of the kinematic variables related to the $\omega \pi$ and $D^* \pi$ systems, respectively, for the $D^{*+} \omega \pi^-$ candidates in the signal region. The results of the fit with the nominal model are superimposed. All plots demonstrate a reasonable description of the data by the fit. A more detailed comparison is shown in Figs.~\ref{fig:fitddst} and \ref{fig:fitompi} for
regions enriched ($\cos\xi_2>-0.4$) and depleted ($|\cos\theta_1|<0.5$) with $D^{**}$ mesons.

To ensure that our fit results correspond to the global minimum, we repeat the signal fit $1000$ times with randomly selected starting values for the fit parameters. None of these fits have better likelihoods than those presented above. For the nominal fit, two local minima are found. One of them, which is $3.3\sigma$ away from the global minimum, corresponds to a very large decay fraction for the $\rho(1450)$, $f_{\rho(1450)}=(157.3 \pm 23.1)\%$, in comparison with the decay fraction for the off-shell $\rho(770)$, $f_{\rho(770)}=(87.5 \pm 13.1)\%$, and a relative phase $\phi_{\rho(1450)}=-2.52 \pm 0.05$. This result is inconsistent with the $e^+e^-$ data \cite{karda}.
For the other one, all the fit parameters coincide with the values presented in Table~\ref{tab:models} within their statistical errors with the exception of the relative phases in the $D_1(2430)^0$ description: the $S$- and $D$-wave phases are shifted by $\pi/2$, whereas the $P$-wave phase remains unchanged. Since this second local minimum is more than $3.5\sigma$ away from the global minimum, it is not considered as a second possible solution for the final results.

\subsection{\bf Statistical uncertainties}
\label{sec:D}

The statistical uncertainties on the fractions of the different intermediate states as well as the upper limit for the $b_1(1235)$ fraction are determined with using a frequency method.
Another objective of this procedure is to estimate how well the nominal signal model describes the data.
Assuming adequate agreement between the data and the nominal signal model, we generate $1000$ statistically independent samples, which are a proper mixture of signal and background events distributed according to the PDF of Eq.~(\ref{pdf}). All the major characteristics such as the reconstruction efficiency and statistics are taken into account. The numbers of signal and background events for each pseudoexperiment are generated according to distributions by statistics.

We fit the obtained MC samples and determine the fractions of quasi-two-body channels for each sample. The distributions of these fractions are then fit with a Gaussian $G(x;\mu,\sigma)$ or bifurcated Gaussian (Gaussian with different standard deviation values $\sigma_1$ and $\sigma_2$ on left and right side of the mean value $\mu$) $G(x; \mu,\sigma_1,\sigma_2)$.
The standard deviations, $\sigma$ or $\sigma_1$ and $\sigma_2$, are considered as the statistical errors for the fractions of the corresponding submode.
The $90\%$ confidence level upper limit for the $b_1(1235)^-$ contribution is obtained directly from the distribution of the $b_1(1235)^-$ decay fractions in the pseudoexperiments.

To measure the goodness of the fit, we utilize two different approaches. The first operates with a mixed sample \cite{gof} combining the experimental data sample and pseudoexperiments with ten times higher statistics than in the experiment. This method allows one to estimate the consistency of the nominal signal model and data in the multidimensional amplitude analysis with the small data sample when the $\chi^2$ method with binning in the multidimensional phase space is not valid.
Following the algorithm described in detail in Ref.~\cite{gof}, we conclude that our nominal model and the data are consistent at $49\%$ confidence level.

For the second technique, we define two $\chi^2$ variables calculated in the $\omega \pi$ and $D^*\pi$ bases, respectively.
For each pair of kinematic variables $j$ and $k$ describing the $\omega \pi$ (or $D^* \pi$) production, we consider $10 \times 10=100$ two-dimensional bins and compute
\begin{linenomath*}
\begin{equation}
\chi^2_{j,k}\,=\,\sum^{100}_{i =1} \frac{(N_{{\rm fit}\,i}-N_{{\rm obs}\,i})^2}{N_{{\rm obs}\,i}}{.}
\label{chid2}
\end{equation}
\end{linenomath*}
In Eq.~(\ref{chid2}), $N_{{\rm fit}\,i}$ is the expected number of events in bin $i$ based on the PDF of Eq.~(\ref{pdf}) and
$N_{{\rm obs}\,i}$ is the number of observed events in that bin.
Then we obtain the total $\chi^2$ as the sum of $\chi^2_{j,k}$ over all possible pairs of variables $j$ and $k$.
In $90\%$ ($78\%$) of the pseudoexperiments this $\chi^2$, calculated with the $\omega \pi$ ($D^*\pi$) variables, has a value smaller than in the data, indicating an acceptable fit quality.

\subsection{\bf Systematic uncertainties}
\label{sec:e}

Two types of uncertainties are considered besides the statistical errors. These are systematic and model uncertainties.

The systematic uncertainty comes from the background description and the efficiency of the selection requirements.
To estimate the uncertainty in the parametrization of the distribution of background events, we use two alternative parametrizations. The first is determined in terms of the $D^{**}$ production variables instead of the $\omega \pi$ basis used in the nominal fit. In the second, we use alternative background functions: a sum of Legendre polynomials instead of a sum of exponential functions used in the  nominal fit and alternative correlation functions. The full parametrization for the nominal background fit is presented in Appendix~\ref{sec:appA}.
The uncertainty related to the efficiency of the definition of the selection requirements is dominated by the variation of the signal region in the $(\Delta E, M(\pi^+\pi^-\pi^0))$ plane. To estimate this uncertainty, we modify the signal region shape from the rectangle to an ellipse, taking into account the correlation between $\Delta E$ and $M(\pi^+\pi^-\pi^0)$. This modification increases the signal-to-background ratio by a factor of about $1.5$.
The contributions to the uncertainty from the background description and the reconstruction efficiency are added in quadrature to obtain the overall systematic uncertainty.

The uncertainties in the parametrization of the signal matrix element determine the model error. There are three sets of such uncertainties. The first is related to the number of contributions to the matrix element. We include an additional nonsignificant $b_1(1235)^-$ amplitude described in Appendix~\ref{sec:appB} and then try several fits: first, we modify the $b_1(1235)^-$ model by removing the $D$-wave contribution to the decay $b_1(1235)^- \to \omega \pi^-$; then we consider the relative helicity phases $\phi_{\pm}$ as free parameters during the fit, independent of the values of the helicity phases defined in the $\rho(770)$ and $\rho(1450)$ amplitudes.
The model error due to the $b_1(1235)$ contribution is assigned as the maximum difference between the values obtained from these fits and the nominal one.
Furthermore, we include in the signal model the contributions from $\rho(1700)^-$, off-shell $D^0$ resonances and $S$-wave nonresonant amplitudes. All of them are nonsignificant.
The second set of errors arises due to the assumption of the signal shape.
We take into account the mixing effect between the $D_1(2430)^0$ and $D_1(2420)^0$ states. Moreover, we modify the transition form factors in the matrix element: we substitute the effective form factor $A(q^2)$, describing the $\rho(770)^- \to \omega \pi^-$ transition for the $P$-wave Blatt-Weisskopf factor $B_P(q^2)$ (see Appendix~\ref{sec:appB}) and we modify the shape of the Isgur-Wise function $h(w)$ describing the production of the $\rho$-meson-like states (see Appendix~\ref{sec:appC}). For the latter, we apply the parametrization that corresponds to the requirements of analyticity and is used in the {\it BABAR} $\bar{B}^0 \to D^{*+} e^- \bar{\nu}_e$ analysis \cite{semil}.
The third set of errors is related to the model parameters that are fixed in the fit. We vary the mass and the width of each resonance [except the $\rho(1450)$] within their known PDG uncertainties \cite{pdg}.
We also vary the parameters $R_1$ , $R_2$ and $\rho^2$ of the invariant form factors describing the $\rho$-meson-like amplitudes (see Appendix~\ref{sec:appC}) within their uncertainties obtained by the {\it BABAR} collaboration \cite{semil}.
Moreover, we vary the parameter $r=1.6$ $({\rm GeV}/c)^{-1}$ used in the Blatt-Weisskopf factors and the form factor $A(q^2)$ (see Appendix~\ref{sec:appB}) in the range from $0.8$ to $2.5$ $({\rm GeV}/c)^{-1}$.

\begin{table*}[t!]
\caption{Summary of systematic and dominant model uncertainties in the parameters of amplitude analysis.}
\label{tab:system}
\begin{ruledtabular}
\begin{tabular}{c c c c c c c}
Contribution & Parameter & \multicolumn{2}{c}{Systematic} & \multicolumn{3}{c}{Dominant model} \\
& & \multicolumn{2}{c}{uncertainties} & \multicolumn{3}{c}{uncertainties} \\
\hline
& & Background & Signal &  $A(q^2)$  &
$r$ & Mixing \\
& & description & region & form factor & parameter & effect \\
\hline
$\rho(770)^-D^{*+}$ & Resonance fraction, \%  & ${}^{+0.8}_{-0.1}$ & $+4.9$  & $+6.1$  & ${}^{+8.7}_{-24.0}$ &  $-0.9$  \\
   \hline
$\rho(1450)^-D^{*+}$ & Resonance phase & ${}^{+0.01}_{-0.02}$ & $+0.07$ & $+0.24$ & ${}^{+0.22}_{-0.17}$ & $0.00$  \\
   & Resonance coupling &  $\pm 0.01$ & $-0.02$ & $+0.08$ & ${}^{+0.10}_{-0.01}$ & $+0.01$ \\
   & Mass, MeV$/c^2$ & $\pm 1$ & $+11$ & $-17$ & ${}^{+1}_{-42}$  & $0$ \\
   & Width, MeV$/c^2$  & ${}^{+2}_{-4}$ & $+3$ &  $+69$  & ${}^{+55}_{-6}$ & $+2$ \\
   & Resonance fraction, \% & $\pm 1.9$ & $-4.4$ & $+9.9$ &  ${}^{+17.4}_{-0.8}$   & $+0.7$  \\
   & $\phi_+$ phase & $\pm 0.05$ & $+0.07$ & $+0.06$ &  $\pm 0.06$  & $0.00$ \\
   & $\phi_-$ phase & $-0.02$ & $+0.02$ & $+0.05$ &  $\pm 0.03$  & $-0.01$ \\
&  FCC fraction, \% &$-0.2$  &$-3.6$ & $+0.3$ & ${}^{+0.3}_{-1.8}$  & $-0.5$ \\
\hline
$D_1(2430)^0 \omega$ & Resonance phase & $-0.07$ & $+0.18$ & $-0.29$ & ${}^{+0.39}_{-0.32}$ & $+0.03$ \\
   & $S$-wave phase & $+0.04$ & $-0.26$ & $+0.04$  & ${}^{+0.04}_{-0.02}$  & $-0.05$   \\
   & $P$-wave phase & ${}^{+0.03}_{-0.04}$ & $-0.26$ & $+0.13$ & ${}^{+0.08}_{-0.05}$ &  $-0.04$  \\
   & Resonance fraction, \%  & $+0.1$ & $+2.7$ & $-0.4$ & ${}^{+1.1}_{-0.2}$   & $+1.3$    \\
   & $S$-wave fraction, \%  & ${}^{+4.2}_{-0.7}$ & $+0.9$ & $-0.3$ &  $-1.0$ &  $+1.2$   \\
   & $P$-wave fraction, \% & ${}^{+1.2}_{-5.5}$ & $+2.1$ & $-0.3$  & ${}^{+2.9}_{-0.1}$  &   $+0.8$  \\
   & $D$-wave fraction, \% & $-0.8$ & $+3.0$ & $+0.5$ &  ${}^{+0.2}_{-2.0}$  &   $-2.1$  \\
   & Long. polarization, \% & ${}^{+4.6}_{-1.2}$ & $-4.4$ & $+0.4$ & ${}^{+0.6}_{-3.5}$  &  $-1.8$ \\
   \hline
$D_1(2420)^0 \omega$ & Resonance phase & ${}^{+0.08}_{-0.03}$ & $+0.08$ & $-0.23$ & ${}^{+0.32}_{-0.27}$ & $+0.05$ \\
   & $S$-wave phase & ${}^{+0.03}_{-0.17}$ & $+0.09$ & $+0.11$ & ${}^{+0.05}_{-0.07}$ & $+0.04$  \\
   & $P$-wave phase & $+0.07$ & $-0.37$ & $+0.02$  & ${}^{+0.02}_{-0.04}$ & $+0.03$  \\
   & Resonance fraction, \%  & $+0.2$ & $+0.4$ &$-0.2$ & ${}^{+0.0}_{-0.1}$ & $+0.5$  \\
   & Long. polarization, \%  & $-3.7$ & $-2.0$ & $-0.9$ & $+2.0$ &  $-2.8$  \\
   \hline
$D^*_2(2460)^0 \omega$ & Resonance phase & $\pm 0.03$ & $-0.12$ & $-0.24$ & $\pm 0.30$  & $+0.03$ \\
   & $P$-wave phase & ${}^{+0.02}_{-0.11}$ & $-0.10$ & $+0.04$ & ${}^{+0.02}_{-0.12}$ & $-0.04$ \\
   & $D$-wave phase & ${}^{+0.01}_{-0.06}$ &$-0.37$ & $+0.08$ & $\pm 0.07$ & $-0.08$  \\
   & Resonance fraction, \% & ${}^{+0.0}_{-0.1}$ & $0.0$ & $0.0$ & ${}^{+0.0}_{-0.1}$ &  $+0.1$   \\
   & Long. polarization, \%  & ${}^{+0.2}_{-2.0}$ & $+2.0$ & $+1.5$ & ${}^{+1.4}_{-0.3}$  & $-1.5$  \\
\end{tabular}
\end{ruledtabular}
\end{table*}
The total model error is obtained by adding all model errors in quadrature.
The sources of systematic and dominant model uncertainties that affect the results of the amplitude analysis are summarized in Table~\ref{tab:system}.

To account for the systematic and model uncertainties in the upper limit of the $b_1(1235)^-$, we determine the $b_1(1235)^-$ contribution with all above described sources of errors [including the $b_1(1235)$ mass and width variation] and use the largest value to evaluate the upper limit. The main effect is due to the removal of the $D$ wave in the $b_1(1235)^- \to \omega \pi^-$ decay.

An additional effect appears due to the interference between $D^{*+} \pi^+ \pi^- \pi^0 \pi^-$ background events and $D^{*+} \omega \pi^-$ signal events. Figure~\ref{fig:bkgstruct} (a) shows that most of the $D^*4\pi$ events lie in the range $M(4\pi) < 2\,{\rm GeV}/c^2$. The investigation of $e^+e^-$ annihilation into a $4 \pi$ system \cite{ee4pi} at these energies as well as the study of the resonant structure in the decay $\tau \to 3\pi \pi^0 \nu_{\tau}$ \cite{tau4pi} demonstrate the dominance of the $a_1(1260) \pi$ and $\omega \pi$ intermediate states.
We assume that our $D^*4\pi$ background is also dominated by $a_1(1260)\pi$ production. In such a case, the interference with the $\omega \pi$ system should be negligible.

\section{\bf DISCUSSION AND CONCLUSION}

This analysis is devoted to the study of the three-body $\bar{B}^0 \to D^{*+} \omega \pi^-$ decay.
We obtain the total branching fraction
\begin{linenomath*}
\begin{equation*}
\mathcal{B}\,=\,(2.31 \pm 0.11\, ({\rm stat.}) \pm 0.14\, ({\rm syst.})) \times 10^{-3}\nonumber{,}
\end{equation*}
\end{linenomath*}
consistent within errors with the CLEO \cite{cleo} and {\it BABAR} \cite{babar} measurements but with a slightly smaller central value.

A full amplitude analysis of the final state has been performed.
A summary of the results with systematic and model uncertainties on parameters and statistical significances of resonant contributions is presented in Table~\ref{tab:final}.
\begin{table*}
\caption{Summary of the final results of the  $\bar{B}^0 \to D^{*+} \omega \pi^-$ amplitude analysis. The first error is statistical, the second is systematic and the third is the model error. The statistical significance, taking into account systematic effects, is given by $\sqrt{2(\Delta \mathcal{L})}$, where $\Delta \mathcal{L}$ is the difference between the negative log-likelihood values with the signal from resonance fixed at zero and the nominal signal yield.}
\label{tab:final}
\begin{ruledtabular}
\begin{tabular}{c c c c}
Contribution & Parameter & Value & Significance\\
\hline
 & Total branching fraction, $10^{-3}$ & $2.31 \pm 0.11 \pm 0.14$ & \\
 & FCC branching fraction, $10^{-3}$ & $1.90 \pm 0.11 {}^{+0.11}_{-0.13} {}^{+0.02}_{-0.06}$ & $29.8\sigma$ \\
 & SCC branching fraction, $10^{-4}$ & $<0.7$ ($90\%\,{\rm C.L.}$) & \\
 \hline
$\rho(770)^-D^{*+}$ & Resonance phase & $0$ (fixed) & \\
   & Resonance coupling  & $1$ (fixed) & \\
   & Resonance branching fraction, $10^{-3}$   & $1.48 \pm 0.27 {}^{+0.15}_{-0.09} {}^{+0.21}_{-0.56}$ & $10.5\sigma$\\
   \hline
$\rho(1450)^-D^{*+}$ & Resonance phase &  $2.56 \pm 0.12 {}^{+0.07}_{-0.02} {}^{+0.24}_{-0.17}$ & \\
   & Resonance coupling & $0.18 {}^{+0.02}_{-0.06} {}^{+0.00}_{-0.02} {}^{+0.10}_{-0.01}$  & \\
   & Mass, MeV$/c^2$     & $1544 \pm 22 {}^{+11}_{-1} {}^{+1}_{-46}$ & \\
   & Width, MeV$/c^2$          & $303 {}^{+31}_{-52}{}^{+3}_{-4} {}^{+69}_{-6}$ & \\
   & Resonance branching fraction, $10^{-3}$  & $1.07 {}^{+0.15}_{-0.31} {}^{+0.06}_{-0.13} {}^{+0.40}_{-0.02}$ & $15.0\sigma$ \\
& $\phi_+$ phase & $0.87 \pm 0.29 {}^{+0.12}_{-0.07} \pm 0.06$ & \\
& $\phi_-$ phase & $-0.02 \pm 0.13 \pm 0.02 \pm 0.05$  & \\
   \hline
$D_1(2430)^0 \omega$ & Resonance phase & $1.24 \pm 0.28 {}^{+0.19}_{-0.07} {}^{+0.39}_{-0.32}$ & \\
   & $S$-wave phase & $-0.05 \pm  0.25 {}^{+0.04}_{-0.26} {}^{+0.04}_{-0.07}$ & \\
   & $P$-wave phase &  $2.24 \pm 0.29 {}^{+0.03}_{-0.26} {}^{+0.13}_{-0.06}$ & \\
   & Resonance branching fraction, $10^{-4}$  & $2.5 \pm 0.4 {}^{+0.7}_{-0.2} {}^{+0.4}_{-0.1}$ & $8.6\sigma$ \\
   & $S$-wave fraction, \%  & $38.9 \pm 10.8 {}^{+4.3}_{-0.7} {}^{+1.2}_{-1.1}$ & \\
   & $P$-wave fraction, \%  & $33.1 \pm 9.5 {}^{+2.4}_{-5.5} {}^{+3.0}_{-4.0}$ & \\
   & $D$-wave fraction, \%  & $28.3 \pm 8.9 {}^{+3.0}_{-0.8} {}^{+3.9}_{-2.9}$ & \\
   & Long. polarization, \%  & $63.0 \pm 9.1 \pm 4.6 {}^{+4.6}_{-3.9}$ & \\
   \hline
$D_1(2420)^0 \omega$ & Resonance phase & $2.12 \pm 0.34 {}^{+0.11}_{-0.03} {}^{+0.33}_{-0.27} $ & \\
   & $S$-wave phase & $-0.07 \pm 0.43 {}^{+0.09}_{-0.17} {}^{+0.12}_{-0.08}$ & \\
   & $P$-wave phase & $-0.25 \pm 0.46 {}^{+0.07}_{-0.37} \pm 0.04 $  & \\
   & Resonance branching fraction, $10^{-4}$  & $0.7 \pm 0.2 {}^{+0.1}_{-0.0} \pm 0.1$ & $5.5\sigma$\\
   & Long. polarization, \%  & $67.1 \pm 11.7 {}^{+0.0}_{-4.2} {}^{+2.3}_{-2.8}$ & \\
   \hline
$D^*_2(2460)^0 \omega$ & Resonance phase & $2.31 \pm 0.50 {}^{+0.03}_{-0.12} \pm 0.11$ & \\
   & $P$-wave phase  & $-0.77 \pm 0.62 {}^{+0.02}_{-0.15} {}^{+0.04}_{-0.15}$ & \\
   & $D$-wave phase & $-1.85 \pm 0.59 {}^{+0.01}_{-0.37} {}^{+0.08}_{-0.11}$ & \\
   & Resonance branching fraction, $10^{-4}$  & $0.4 \pm 0.1 {}^{+0.0}_{-0.1} \pm 0.1$ & $5.0\sigma$\\
   & Long. polarization, \%  & $76.0 {}^{+18.3}_{-8.5} \pm 2.0 {}^{+2.9}_{-2.0}$ & 
\end{tabular}
\end{ruledtabular}
\end{table*}
This is the first consistent study of the $\rho(770)$ and $\rho(1450)$ states in $B$-meson decays. Large signals correspond to the off-shell $\rho(770)^-$ meson and $\rho(1450)^-$ resonance with significances of $10.5\sigma$ and $15.0\sigma$ calculated from the negative log-likelihood values and taking into account systematic effects. 
However, model uncertainties are of about $40\%$.
There is no accurate description yet of the shape of the $\rho(1450)$ resonance.
This leads to an ambiguity in discriminating the $\rho$-meson-like states and to large model errors in the definition of their resonance branching fractions. Nevertheless,
the coherent contribution of these resonances
is determined with smaller model uncertainties.
The statistical significance of this fraction is $29.8\sigma$.
This combined decay fraction gives a dominant contribution to the total branching fraction.

We also measure the relative coupling and the relative phase between the $\rho$-meson-like states.
Neglecting final state interactions for $\rho$-meson-like production, we can compare this production with the $e^+e^-$ SND data \cite{karda}. In the SND analysis \cite{karda} as well as in our analysis, a small change of the resonance shape leads to significant shifts in the fitted resonance parameters.
However, within the isotopic invariance and CVC fitted $\rho$-meson-like resonance parameters are compatible to those observed in processes proceeding through a virtual photon in $e^+e^-$ collisions \cite{karda,tsai}.

The phase difference between the $\rho(770)^-$ and $\rho(1450)^-$ amplitudes is observed to be close to $\pi$ as predicted in Ref.~\cite{cleggdon}.
In the frame of our signal model, we measure the $\rho(1450)^-$ mass and width. Our measurements also show evidence for nontrivial final-state interaction phases in the helicity amplitudes of the $\rho$-meson-like states, off-shell $\rho(770)^-$ and $\rho(1450)^-$, with a significance of $3.3\sigma$. Such effect is observed within the validity of the factorization at relatively low $q^2$. We restrict the description of these resonances by the requirement that the helicity phases $\phi_+$ and $\phi_-$ in the amplitude of the off-shell $\rho(770)^-$ are equal to the corresponding phases in the amplitude of the $\rho(1450)^-$. Similar phases were measured by the CLEO collaboration in $B \to D^* \rho \to D^* \pi\pi$ decays \cite{btodstrho}. Our results as well as those of CLEO show that $\phi_+ > \phi_-$, although statistical uncertainties are large.

In addition to the $\rho$-meson-like states, the $b_1(1235)^-$ resonance could be produced as a possible intermediate state in the color-favored channel.
Such a contribution is generated by SCC and is expected to vanish in the limit of perfect isospin symmetry.
Our measurements do not require any SCC contribution and an upper limit for the product of branching fractions of $\mathcal{B}(\bar{B}^0 \to D^{*+} b_1(1235)^-) \times \mathcal{B}(b_1(1235)^- \to \omega \pi^-)$
has been obtained. This result is the first search for SCC in $B$-meson decays.

Color-suppressed decays $\bar{B}^0 \to D_1(2430)^0 \omega$ and  $\bar{B}^0 \to D_1(2420)^0 \omega$ are observed in our study with significances of $8.6\sigma$ and $5.5\sigma$, taking into account systematic effects. The measurements show the relative dominance of the broad $D_1(2430)^0$ production in comparison with the narrow $D_1(2420)^0$.
Heavy quark symmetry predicts the absence of $D_1(2420)^0$ signal in the limit $\Lambda_{QCD}/m_c \to 0$ \cite{fddst}, where $\Lambda_{QCD}$ is the QCD scale and $m_c$ is the mass of the $c$ quark.
The production of the $D_1(2420)^0$ state can be explained by finite corrections of order $\Lambda_{QCD}/m_c$ in the $D_1(2420)^0$ production.
Moreover, the dominance of broad resonances in the color-suppressed channel can result in comparable production of the broad and narrow states in the $B^- \to D^{**0} \pi^-$ decays \cite{b-todstpi}.

The nonfactorizable $\bar{B}^0 \to D^*_2(2460)^0 \omega$ decay has been observed with a statistical significance of $5.0\sigma$. In SCET theory \cite{scet}, the equality of branching fractions and strong phases in the decays $\bar{B}^0 \to D^*_2(2460)^0 M$ and $\bar{B}^0 \to D_1(2420)^0 M$, where $M=\pi,\rho,K$ or $M=K^*$ with longitudinal polarization, is predicted.
Our result with $M=\omega$ also does not contradict this prediction. However, our errors on the resonance branching fractions and phases are large.

In our analysis, we obtain the partial wave fractions for the intermediate resonances.
For the $\rho$-meson-like states, we fix the relative normalizations $R_1$, $R_2$ and parameter $\rho^2$
in the helicity amplitudes
at values obtained from the semileptonic $B \to D^* l \nu$ analysis \cite{semil}. These normalizations determine the relative partial wave fractions with the dominant $S$-wave production (see Table~\ref{tab:models}).
Another effect takes place for the $D_1(2430)^0$ resonance: all partial waves---$S$, $P$ and $D$ waves---have close probabilities of about $30\%$. A similar tendency is observed in the $D_1(2420)^0$ and $D^*_2(2460)^0$ production. However, the statistical accuracy is not sufficient to obtain significant numerical values (see Table~\ref{tab:models}).

We also measure for the first time the longitudinal polarization of the $\omega$ in case of $D^{**}$ production.
The results have large errors, but they imply nontrivial nonfactorizable QCD effects in the color-suppressed channel \cite{ccs} and can be compared with the measurement of the polarization in the decay $\bar{B}^0 \to D^{*0} \omega$ \cite{b0todstomega}, $\mathcal{P}_{D^*}=(66.5 \pm 5.0)\%$.
All these polarization results, except for the  $D^*_2(2460)^0$, show significant deviations from unity.
The $D^*_2(2460)^0$ result should be considered separately because this tensor state is generated only due to nonfactorizable contributions.

\acknowledgments
We thank the KEKB group for the excellent operation of the
accelerator; the KEK cryogenics group for the efficient
operation of the solenoid; and the KEK computer group,
the National Institute of Informatics, and the
PNNL/EMSL computing group for valuable computing
and SINET4 network support.  We acknowledge support from
the Ministry of Education, Culture, Sports, Science, and
Technology (MEXT) of Japan, the Japan Society for the
Promotion of Science (JSPS), and the Tau-Lepton Physics
Research Center of Nagoya University;
the Australian Research Council and the Australian
Department of Industry, Innovation, Science and Research;
Austrian Science Fund under Grants No.~P 22742-N16 and No. P 26794-N20;
the National Natural Science Foundation of China under Contracts
No.~10575109, No.~10775142, No.~10875115, No.~11175187, and  No.~11475187;
the Ministry of Education, Youth and Sports of the Czech
Republic under Contract No.~LG14034;
the Carl Zeiss Foundation, the Deutsche Forschungsgemeinschaft
and the VolkswagenStiftung;
the Department of Science and Technology of India;
the Istituto Nazionale di Fisica Nucleare of Italy;
National Research Foundation (NRF) of Korea Grants
No.~2011-0029457, No.~2012-0008143, No.~2012R1A1A2008330,
No.~2013R1A1A3007772;
the Basic Research Lab program under NRF Grant No.~KRF-2011-0020333,
No.~KRF-2011-0021196, Center for Korean J-PARC Users, No.~NRF-2013K1A3A7A06056592;
the Brain Korea 21-Plus program and the Global Science Experimental Data
Hub Center of the Korea Institute of Science and Technology Information;
the Polish Ministry of Science and Higher Education and
the National Science Center;
the Ministry of Education and Science of the Russian Federation and
the Russian Foundation for Basic Research;
the Slovenian Research Agency;
the Basque Foundation for Science (IKERBASQUE) and
the Euskal Herriko Unibertsitatea (UPV/EHU) under program UFI 11/55 (Spain);
the Swiss National Science Foundation; the National Science Council
and the Ministry of Education of Taiwan; and the U.S.\
Department of Energy and the National Science Foundation.
This work is supported by a Grant-in-Aid from MEXT for
Science Research in a Priority Area (``New Development of
Flavor Physics'') and from JSPS for Creative Scientific
Research (``Evolution of Tau-lepton Physics'').

\numberwithin{equation}{section}

\begin{appendices}
\renewcommand\appendixname{APPENDIX}

\section{BACKGROUND DESCRIPTION}
\label{sec:appA}

In this section, we present the procedure of the background description in the $(\Delta E, M(\pi^+\pi^-\pi^0))$ plane.

To describe the combinatorial background without real $\omega$ events, we use Region ${\rm IV}$, which includes a SCF component.
In this region, a negative log-likelihood function to be minimized is given by
\begin{linenomath*}
\begin{equation}
\mathcal{L}_{{\rm IV}}(\vec{a})\,=\, -\sum^{}_{({\rm IV})}{\rm ln}\left(\frac{B_{\rm IV}(\vec{a},\vec{x})\epsilon(\vec{x})}{\sum_{{\rm CR}}B_{\rm IV}(\vec{a},\vec{x})}\right){,}
\label{NLL4}
\end{equation}
\end{linenomath*}
where the sum $\sum^{}_{({\rm IV})}$ is over the events in Region ${\rm IV}$, the sum $\sum_{{\rm CR}}$ is calculated over $\bar{B}^0 \to D^{*+} \omega \pi^-$ CR events, which are uniformly generated over the phase space and then reconstructed in Region ${\rm I}$ with the above-described selection procedure, and $\epsilon(\vec{x})$ is the reconstruction efficiency for the $\bar{B}^0 \to D^{*+} \omega \pi^-$ CR events in Region ${\rm I}$.

After the estimation of $B_{\rm IV}$, the minimization procedure is performed for events in Region ${\rm III}$. In addition to the events described by the function $B_{\rm IV}$, this region includes $\bar{B}^0 \to D^{*+} \pi^+ \pi^- \pi^0 \pi^-$ events without $\omega$ in the intermediate state and another SCF component.
The negative log-likelihood function in Region ${\rm III}$ is
\begin{linenomath*}
\begin{align}
\mathcal{L}_{\rm III}(\vec{b})\,=&\, -\sum_{({\rm III})}{\rm ln}\left(\frac{S_{\rm III}}{S_{\rm IV}}\frac{N_{\rm IV}}{N_{\rm III}}\frac{B_{\rm IV}(\vec{a},\vec{x})\epsilon(\vec{x})}{\sum_{\rm CR}B_{\rm IV}(\vec{a},\vec{x})}+{} \right.\nonumber \\ &\left.\left(1-\frac{S_{\rm III}}{S_{\rm IV}}\frac{N_{\rm IV}}{N_{\rm III}}\right)\frac{B_{\rm III}(\vec{b},\vec{x})\epsilon(\vec{x})}{\sum_{\rm CR}B_{\rm III}(\vec{b},\vec{x})}\right){,}
\label{NLL3}
\end{align}
\end{linenomath*}
where $S_{\rm III}$ ($S_{\rm IV}$) is the size of Region ${\rm III}$ (${\rm IV}$), $N_{\rm III}$ ($N_{\rm IV}$) is the number of events in Region ${\rm III}$ (${\rm IV}$), the sum $\sum^{}_{(\rm{III})}$ is over the events in Region ${\rm III}$ and the sum $\sum_{\rm CR}$ is calculated over CR events. The vector $\vec{a}$ in the function $B_{\rm IV}$ is obtained from Region ${\rm IV}$ and fixed in Region ${\rm III}$. The vector $\vec{b}$ is free in Region ${\rm III}$.

A similar procedure is performed in Region ${\rm II}$. In addition to the events described by the function $B_{\rm IV}$, Region ${\rm II}$ includes the combinatorial background with a real $\omega$ and another SCF component. The shape function $B_{\rm II}$ describes these events together with the additional SCF as in to Region ${\rm III}$. The minimization function $\mathcal{L}_{{\rm II}}(\vec{c})$ is similar to $\mathcal{L}_{{\rm III}}(\vec{b})$.

Functions $B_{\rm II}$, $B_{\rm III}$ and $B_{\rm IV}$ describe specific background components defined above and SCF events in Regions ${\rm II}$, ${\rm III}$ and ${\rm IV}$. All these background contributions are present in signal Region ${\rm I}$.
However, the signal region also includes additional SCF in comparison with the SCF level obtained from the sideband regions.
This additional SCF component is determined in MC simulation that shows the same phase space distribution in $\vec{x}$ for all SCF events in each region of the
$(\Delta E, M(\pi^+\pi^-\pi^0))$ plane.
We can repeat a fit in any of the sideband Regions $j={\rm II}$, ${\rm III}$ or ${\rm IV}$, taking into account this contribution, and thus obtain more precisely the function $B_j$, which now describes this SCF and is used further in the signal fit. We choose Region ${\rm II}$ with the function $B_{\rm II}$.

We use the following empirical parametrization to describe the distribution of background events:
\begin{linenomath*}
\begin{align}
B_j(\vec{x})\,=&\,
F_1(M^2(\omega\pi),\cos\xi_1) F_2(\cos\theta_1) \times  \nonumber \\ & F_3(\phi_1) F_4(\cos\beta_1) F_5(\psi_1){,}
\label{bkgpdf}
\end{align}
\end{linenomath*}
where the function $F_1(M^2(\omega\pi),\cos\xi_1)$ describes the correlation between the $M^2(\omega\pi)$ and $\cos\xi_1$ variables:
\begin{widetext}
\begin{linenomath*}
\begin{equation}
\label{b1cor}
F_1(M^2(\omega\pi),\cos\xi_1)\,=\,
(e^{c_1 \sqrt{\Delta} \cos\xi_1}+ c_2 e^{c_3 \sqrt{\Delta} \cos\xi_1})((1-\cos\xi_1)^3 e^{c_4 \Delta}+c_5 \sqrt{\Delta(1+\cos\xi_1)}  e^{c_6 \Delta}){,}
{}
\end{equation}
\end{linenomath*}
\end{widetext}
and the functions $F_2(\cos\theta_1), F_3(\phi_1), F_4(\cos\beta_1), F_5(\psi_1)$ describe one-dimensional projections of the other variables:
\begin{linenomath*}
\begin{align}
\label{b3}
F_2(\cos\theta_1)\,=&\,e^{c_{7}\cos\theta_1}+c_{8}e^{c_{9}\cos\theta_1}{,}\nonumber\\
F_3(\phi_1)\,=&\,1+c_{10}\sin^2(\phi_1){,}\nonumber\\
F_4(\cos\beta_1)\,=&\,e^{c_{11}\cos\beta_1}+c_{12}e^{c_{13}\cos\beta_1}{,}\nonumber\\
F_5(\psi_1)\,=&\,1+c_{14}\sin^2(\psi_1){.}
\end{align}
\end{linenomath*}
Here, $c_i$ are free parameters, $\Delta = M^2(\omega \pi)-M^2_0(\omega\pi)$ and the lower boundary $M^2_0(\omega \pi)=0.7\,({\rm GeV}/c^2)^2$ differs from the kinematic limit $(m_{\omega}+m_{\pi})^2$ because the $\omega$ invariant mass is not constrained to its nominal value.

\section{RESONANT AMPLITUDES}
\label{sec:appB}

In this section, we present all resonant amplitudes used in the fit.
The notations $p^2=M^2(\pi^+\pi^-\pi^0)$, where $\pi^+\pi^-\pi^0$ is the $\omega$ decay product system, and $q^2=M^2(\omega\pi^{\mp})$ ($q^2=M^2(D^{*\pm}\pi^{\mp})$) for the $\rho$-meson-like ($D^{**}$) production in the $\bar{B}^0 (B^0) \to D^{* \pm} \omega \pi^{\mp}$ decay, are used.
The magnitudes of the three-momenta of the $\omega$ decay product system and $D^{* \pm}$ in the $\omega\pi^{\mp}$ and $D^{*\pm}\pi^{\mp}$ rest frames are denoted as $p_{3\pi}$ and $p_{D^*}$, respectively.
The magnitude of the $\omega$ three-momentum in the $\omega\pi^{\mp}$ rest frame when $M(\pi^+\pi^-\pi^0)$ is equal to the $\omega$ nominal mass is denoted as $p_{\omega}$. The magnitude of the $\omega$ ($D^{*\pm}$) three-momentum in the $\omega\pi^{\mp}$ ($D^{*\pm}\pi^{\mp}$) rest frame, when $M(\omega\pi^{\mp})$ ($M(D^{*\pm}\pi^{\mp})$) is equal to the nominal mass of the $\rho$-meson-like ($D^{**}$) resonance and $M(\pi^+\pi^-\pi^0)$ is equal to the $\omega$ nominal mass, is denoted as $p_{0,\omega}$ ($p_{0,D^*}$). The Blatt-Weisskopf penetration factors $B_L(p)$ \cite{blwe} used in the resonant matrix element description are defined for $L=S$, $P$, $D$ and $F$ partial waves as
\begin{linenomath*}
\begin{eqnarray}
B_S(p)\,&=&\,1{,} \nonumber \\
B_P(p)\,&=&\,\sqrt{\frac{1+x^2_0}{1+x^2}}{,} \nonumber \\
B_D(p)\,&=&\,\sqrt{\frac{(x^2_0-3)^2+9x^2_0}{(x^2-3)^2+9x^2}}{,} \nonumber \\
B_F(p)\,&=&\,\sqrt{\frac{x^2_0(x^2_0-15)^2+9(2x^2_0-5)^2}{x^2(x^2-15)^2+9(2x^2-5)^2}}{,}
\end{eqnarray}
\end{linenomath*}
where $x=rp$, $x_0=rp_0$, $r=1.6\,({\rm GeV}/c)^{-1}$ is the hadron radius and $p$ and $p_0$ are the magnitudes of the daughter particle three-momenta in the mother particle rest frame for the case when the resonance invariant mass squared is equal to $q^2$ and the nominal mass squared, respectively.

\begin{center}{\bf $\mathbf{\bar{B}^0 (B^0)\to D^{*\pm} \rho(770)^{\mp} \to D^{* \pm} \omega \pi^{\mp}}$}\end{center}

Since the off-shell $\rho(770)^{\mp}$ has $J^P = 1^-$, the pair $D^{*\pm}$ and $\omega\pi^{\mp}$ can be produced in three partial waves: $S$, $P$ and $D$. $S$ and $D$ waves violate $C$- and $P$- parities and have the additional phase $\pi/2$ in comparison with a $P$ wave.
The $\omega$ and $\pi^{\mp}$ pair is produced in a $P$ wave via the strong decay $\rho(770)^{\mp} \to \omega \pi^{\mp}$.

The resonance matrix element $M_{\rho \pm}$ describing the $\rho(770)^{\mp}$ contribution in the $\bar{B}^0 (B^0) \to D^{* \pm} \omega \pi^{\mp}$ decay is
\begin{linenomath*}
\begin{align}
M_{\rho\pm}\,=&\,\frac{\sqrt{q^2}p_{3\pi}A(p_{3\pi})}{D_{\rho}(q^2)}\left(\vphantom{\frac{1}{f_{J,J+1}}}f_P(q^2)\mathcal{A}_{PP}\pm {} \nonumber \right. \\ & \left. \vphantom{\frac{1}{f_{J,J+1}}} i f_S(q^2) \mathcal{A}_{SP} \pm i f_D(q^2) \mathcal{A}_{DP}\right){,}
\label{mrho}
\end{align}
\end{linenomath*}
where $A(p_{3\pi})$ is the effective form factor describing the $\rho^{\mp} \to \omega \pi^{\mp}$ transition, $f_S(q^2)$, $f_P(q^2)$ and $f_D(q^2)$ are the
partial wave form factors obtained in Appendix \ref{sec:appC},
$\mathcal{A}_{SP}$, $\mathcal{A}_{PP}$ and $\mathcal{A}_{DP}$ are
the angular dependencies shown in Table~\ref{tanrho} that correspond to the definite partial waves in the $\bar{B}^0 (B^0) \to D^{*\pm} \rho(770)^{\mp}$ and $\rho(770)^{\mp} \to \omega \pi^{\mp}$ decays, and
$D_{\rho}(q^2)$ is the Breit-Wigner (BW) denominator, describing the $\rho(770)^{\mp}$ shape:
\begin{linenomath*}
\begin{equation}
D_{\rho}(q^2)\,=\,q^2-m^2_{\rho}+i m_{\rho}\Gamma_{\rho}(q^2){.}
\label{drho}
\end{equation}
\end{linenomath*}
Here, $m_{\rho}$ is the $\rho(770)^{\mp}$ mass and $\Gamma_{\rho}(q^2)$ is the $q^2$-dependent width.

The form factor $A(p_{3\pi})$ restricts an overly rapid growth of the matrix element of the decay $\rho^{\mp} \to \omega \pi^{\mp}$ with $p_{3\pi}$ and is chosen as \cite{achasov}
\begin{linenomath*}
\begin{equation}
A(p_{3\pi})\,=\,\frac{1}{1+(r p_{3\pi})^2}{.}
\label{aff}
\end{equation}
\end{linenomath*}
The width $\Gamma_{\rho}(q^2)$ for events with $q^2 > (m_{\omega} + m_{\pi})^2$, where $m_{\pi}$ is the mass of the charged pion, is parametrized as \cite{achasov}
\begin{linenomath*}
\begin{equation}
\Gamma_{\rho}(q^2)\,=\,\frac{m_{\rho}}{\sqrt{q^2}}\frac{k_{\pi}^3}{k_{0,\pi}^3}B^2_P(k_{\pi})\Gamma_{\rho} +
\frac{g^2_{\omega\rho\pi}A^2(p_{\omega})}{12 \pi} \frac{\sqrt{q^2}}{m_{\rho}}p_{\omega}^3{.}
\label{rhomatrix}
\end{equation}
\end{linenomath*}
Here, $g_{\omega\rho \pi}$ is a coupling constant, which is equal to $16\,({\rm GeV}/c^2)^{-1}$ \cite{lublinsky},  $k_{\pi}$ is the magnitude of the momentum of the $\pi^{\mp}$ in the $\rho^{\mp} \to \pi^{\mp} \pi^0$ decay computed in the $\rho(770)^{\mp}$ rest frame, $k_{0,\pi}$ is the same magnitude, when $\sqrt{q^2}=m_{\rho}=(775\pm  1)\,{\rm MeV}/c^2$, and $\Gamma_{\rho}=(149\pm 1)\,{\rm MeV}/c^2$ is the $\rho(770)^{\mp}$ width \cite{pdg}.
The first term in Eq.~(\ref{rhomatrix}) corresponds to the dominant $\rho(770)^{\mp}$ decay mode to the $\pi^{\mp}\pi^0$ system and the second term describes the $\omega \rho \pi$ interaction.
For events with $q^2 \le (m_{\omega} +m_{\pi})^2$, we use $\Gamma_{\rho}(q^2)=\Gamma_{\rho}$.

\begin{table}[!t]
\caption{Angular dependencies corresponding to the $\omega\pi^{\mp}$ quantum numbers $J^P=1^-$.
$L_1$ ($L_2$) is the relative orbital angular momentum between the $D^{*\pm}$ and $\omega \pi^{\mp}$ ($\omega$ and $\pi^{\mp}$). The notations $c_{\alpha}=\cos\alpha$ and $s_{\alpha}=\sin\alpha$ are used. The angles $\theta$, $\phi$, $\beta$, $\psi$, $\xi$ correspond to the $\omega\pi^{\mp}$ angular basis.}
\begin{ruledtabular}
\begin{tabular}{l l l}
& & \\
$L_1$ & $L_2$ & $\mathcal{A}_{L_1\,L_2}$ \\
\\
\hline
& & \\
$S$ & $P$ & $-s_{\theta}s_{\phi}c_{\beta}s_{\xi}+s_{\theta}c_{\phi}s_{\beta}s_{\psi}-s_{\theta}s_{\phi}s_{\beta}c_{\psi}c_{\xi}$ \\
& & \\
$P$ & $P$ & $s_{\theta}s_{\phi}s_{\beta}s_{\psi}c_{\xi}+s_{\theta}c_{\phi}s_{\beta}c_{\psi}$ \\
& & \\
$D$ & $P$ & $2s_{\theta}s_{\phi}c_{\beta}s_{\xi}+s_{\theta}c_{\phi}s_{\beta}s_{\psi}- s_{\theta}s_{\phi}s_{\beta}c_{\psi}c_{\xi}$ \\
& & \\
\end{tabular}
\label{tanrho}
\end{ruledtabular}
\end{table}

The magnitude and phase, corresponding to this resonant amplitude, are fixed at values $1$ and $0$, respectively. The free parameters are the relative helicity phases $\phi_{\pm}$ in the form factors $f_S(q^2)$, $f_P(q^2)$ and $f_D(q^2)$.

\begin{center}{\bf $\mathbf{\bar{B}^0 (B^0) \to D^{*\pm} \rho(1450)^{\mp} \to D^{*\pm} \omega \pi^{\mp}}$}\end{center}

The resonant matrix element corresponding to the $\rho(1450)^{\mp}$ intermediate state has a form similar to Eq.~(\ref{mrho}) except for the form factor $A(p_{3\pi})$ and the width
$\Gamma_{\rho}(q^2)$. Since the $\rho(1450)^{\mp}$ is on-shell, we use the Blatt-Weisskopf form factor $B_P(p_{3\pi})$ instead of $A(p_{3\pi})$ \cite{achasov}.
The width $\Gamma_{\rho(1450)}(q^2)$ for events with $q^2 > (m_{\omega} + m_{\pi})^2$ is parametrized as \cite{achasov}
\begin{linenomath*}
\begin{align}
\Gamma_{\rho(1450)}(q^2)\,=&\,\frac{m_{\rho(1450)}}{\sqrt{q^2}}\frac{k_{\pi}^3}{k_{0,\pi}^3}B^2_P(k_{\pi})\frac{\Gamma_{\rho(1450)}}{2} +{} \nonumber \\
&\frac{\sqrt{q^2}}{m_{\rho(1450)}}\frac{p_{\omega}^3}{p_{0,\omega}^3}B^2_P(p_{\omega})\frac{\Gamma_{\rho(1450)}}{2}{,}
\label{rho1450matrix}
\end{align}
\end{linenomath*}
where $k_{\pi}$ is the same as in Eq.~(\ref{rhomatrix}) but computed in the $\rho(1450)^{\mp}$ rest frame and $k_{0,\pi}$ is calculated as $k_{\pi}$ but with $\sqrt{q^2}=m_{\rho(1450)}$.
The first term in Eq.~(\ref{rho1450matrix}) corresponds to the $\rho(1450)^{\mp} \to \pi^{\mp} \pi^0$ decay while the second describes the $\rho(1450)^{\mp} \to \omega \pi^{\mp}$ decay. We assume that the $\rho(1450)^{\mp}$ resonance decays to these final states with equal probabilities.
For events with $q^2 \le (m_{\omega} +m_{\pi})^2$, we use $\Gamma_{\rho(1450)}(q^2)=\Gamma_{\rho(1450)}$.

We assume that the relative helicity phases $\phi_{\pm}$ for the $\rho(1450)^{\mp}$ production are the same as for the off-shell $\rho(770)^{\mp}$.
This assumption does not contradict the common description of the matrix element
because of the validity of the factorization hypothesis. Since the typical values of $q^2$ are close to each other for the $\rho(770)^{\mp}$ and $\rho(1450)^{\mp}$, we can neglect the difference between the appropriate FSI helicity phases for these resonances.

The free parameters for the $\rho(1450)^{\mp}$ amplitude obtained from the fit are the relative magnitude and phase, the mass and width of the $\rho(1450)^{\mp}$, and the helicity phases $\phi_{\pm}$, which are the same as in the $\rho(770)^{\mp}$ amplitude.

\begin{center}{\bf $\mathbf{\bar{B}^0 (B^0) \to D^{*\pm} b_1(1235)^{\mp} \to D^{*\pm} \omega \pi^{\mp}}$}\end{center}

The $b_1(1235)^{\mp}$ resonance has quantum numbers $J^P=1^+$. As such, its wave function has an additional phase $\pi/2$.
The resonant matrix element is written as
\begin{linenomath*}
\begin{align}
M_{b_1\pm}\,=&\,\frac{i}{D_{b_1}(q^2)}\left[\vphantom{\frac{1}{f_{J,J+1}}}m^2_{b_1} B_S(p_{3\pi}) \left(\vphantom{\frac{1}{f_{J,J+1}}}\pm f_P(q^2)\mathcal{A}_{PS}+ {} \nonumber \right. \right. \nonumber \\& \left. \left. i f_S(q^2) \mathcal{A}_{SS}+i f_D(q^2) \mathcal{A}_{DS}\vphantom{\frac{1}{f_{J,J+1}}}\right)- {} \nonumber \right. \\
        &\left. {} \,a_{DS} e^{i \phi_{DS}} P_1(p_{3\pi}) B_D(p_{3\pi}) \times {} \nonumber \right. \\
         &\left. {} \left(\vphantom{\frac{1}{f_{J,J+1}}}\pm f_P(q^2)\mathcal{A}_{PD}+i f_S(q^2) \mathcal{A}_{SD}+ {} \nonumber \right. \right. \\
         & \left. \left. i f_D(q^2) \mathcal{A}_{DD}\vphantom{\frac{1}{f_{J,J+1}}}\right)\vphantom{\frac{1}{f_{J,J+1}}}\right]{,}
\label{mb1}
\end{align}
\end{linenomath*}
where $D_{b_1}(q^2)$ is the BW denominator defined in Eq.~(\ref{drho}) and describing the $b_1(1235)^{\mp}$ shape, $a_{DS}$ and $\phi_{DS}$ are the parameters describing the admixture of $S$ and $D$ waves in the amplitude of the $b_1(1235)^{\mp}$ decay and $P_1(p_{3\pi})$ is the momentum factor corresponding to the $D$ wave in the $b_1(1235)^{\mp}$ decay. This factor can be defined for the intermediate resonance with arbitrary integer spin $J$ as
\begin{linenomath*}
\begin{equation}
P_J(p_{3\pi})\,=\,\frac{\sqrt{q^2} p_{3\pi}^2}{\sqrt{p^2_{3\pi}+p^2}+\frac{J+1}{J}\sqrt{p^2}}{.}
\label{tilde_f}
\end{equation}
\end{linenomath*}
The form factors $f_S(q^2)$, $f_P(q^2)$ and $f_D(q^2)$ are determined in Appendix \ref{sec:appC}, and $\mathcal{A}_{SS}$, $\mathcal{A}_{PS}$, $\mathcal{A}_{DS}$, $\mathcal{A}_{SD}$, $\mathcal{A}_{PD}$ and $\mathcal{A}_{DD}$ are the angular dependencies defined in Table~\ref{tanb1} that correspond to the $\omega\pi^{\mp}$ quantum numbers $J^P=1^+$.

The parameters $a_{DS}$ and $\phi_{DS}$ are fixed at the values measured by the Brookhaven
E852 collaboration \cite{brookhaven}. There, the amplitude ratio was found to be $|D/S|=0.269 \pm 0.013$ and the phase difference $\phi_{DS}=0.18 \pm 0.08\,{\rm rad}$ \cite{brookhaven}.
To relate the parameter $a_{DS}$ to the ratio $|D/S|$, the helicity amplitude $M_{++}$, corresponding to the positive helicities of the $b_1(1235)^{\mp}$ and the $\omega$ in the decay $b_1(1235)^{\mp} \to \omega \pi^{\mp}$, is written in terms of partial waves:
\begin{linenomath*}
\begin{equation}
M_{++}\,=\,M^S_{++} + M^D_{++}\,=\,S/\sqrt{3} + D/\sqrt{6}{,}
\end{equation}
\end{linenomath*}
where $M^S_{++}$ and $M^D_{++}$ are the terms corresponding to the $S$ and $D$ waves, respectively.
To calculate these terms, we denote the polarization four-vectors of the $b_1(1235)^{\mp}$ and the $\omega$ as $\varepsilon_{\mu}$ and $v_{\mu}$, respectively. In such a case the terms are written as
\begin{linenomath*}
\begin{eqnarray}
M^S_{++}\,&=&\,m^2_{b_1} \varepsilon^{+}_{\mu}v^{+*\mu}{,}\nonumber \\
M^D_{++}\,&=&\,-a_{DS}e^{i\phi_{DS}}P_1(p_{0,\omega})\varepsilon^{+}_{\mu}v^{+*\mu}{,}
\end{eqnarray}
where $p^2=m^2_{\omega}$ in $P_1(p_{0,\omega})$.
\end{linenomath*}
Taking into account that $\varepsilon^{+}_{\mu}v^{+*\mu}=-1$, we have
\begin{linenomath*}
\begin{align}
a_{DS}\,=&\,\frac{1}{\sqrt{2}}\frac{m^2_{b_1}}{P_1(p_{0,\omega})}\frac{|D|}{|S|}{,}
\end{align}
\end{linenomath*}
and obtain $a_{DS}=5.2 \pm 0.3$.
\begin{table}[!t]
\caption{Angular dependencies corresponding to the  $\omega\pi^{\mp}$ quantum numbers $J^P=1^+$.
$L_1$ ($L_2$) is the relative orbital angular momentum between the $D^{*\pm}$ and $\omega \pi^{\mp}$ ($\omega$ and $\pi^{\mp}$).
The notations $c_{\alpha}=\cos\alpha$ and $s_{\alpha}=\sin\alpha$ are used. The angles $\theta$, $\phi$, $\beta$, $\psi$, $\xi$ correspond to the $\omega\pi$ angular basis.}
\begin{ruledtabular}
\begin{tabular}{l l l}
& & \\
$L_1$ & $L_2$ & $\mathcal{A}_{L_1\,L_2}$ \\
\\
\hline
& & \\
$S$ & $S$ & $-c_{\theta}c_{\beta}c_{\xi}+s_{\theta}c_{\phi}c_{\beta}s_{\xi}-s_{\theta}s_{\phi}s_{\beta}s_{\psi}+$ \\
& & $+s_{\theta}c_{\phi}s_{\beta}c_{\psi}c_{\xi}+c_{\theta}s_{\beta}c_{\psi}s_{\xi}$ \\
& & \\
$P$ & $S$ & $-c_{\theta}s_{\beta}s_{\psi}s_{\xi}+s_{\theta}s_{\phi}s_{\beta}c_{\psi}-s_{\theta}c_{\phi}s_{\beta}s_{\psi}c_{\xi}$ \\
& & \\
$D$ & $S$ & $2 c_{\theta}c_{\beta}c_{\xi}+s_{\theta}c_{\phi}c_{\beta}s_{\xi}-s_{\theta}s_{\phi}s_{\beta}s_{\psi}+$ \\
& & $+s_{\theta}c_{\phi}s_{\beta}c_{\psi}c_{\xi}-2 c_{\theta}s_{\beta}c_{\psi}s_{\xi}$ \\
& & \\
$S$ & $D$ & $2 c_{\theta}c_{\beta}c_{\xi}-2 s_{\theta}c_{\phi}c_{\beta}s_{\xi}-s_{\theta}s_{\phi}s_{\beta}s_{\psi}+$ \\
& & $+s_{\theta}c_{\phi}s_{\beta}c_{\psi}c_{\xi}+c_{\theta}s_{\beta}c_{\psi}s_{\xi}$ \\
& & \\
$P$ & $D$ & $2 c_{\theta}s_{\beta}s_{\psi}s_{\xi}+s_{\theta}s_{\phi}s_{\beta}c_{\psi}-s_{\theta}c_{\phi}s_{\beta}s_{\psi}c_{\xi}$ \\
& & \\
$D$ & $D$ & $-4 c_{\theta}c_{\beta}c_{\xi}-2 s_{\theta}c_{\phi}c_{\beta}s_{\xi}-s_{\theta}s_{\phi}s_{\beta}s_{\psi}+$ \\
& & $+s_{\theta}c_{\phi}s_{\beta}c_{\psi}c_{\xi}-2 c_{\theta}s_{\beta}c_{\psi}s_{\xi}$ \\
& & \\
\end{tabular}
\label{tanb1}
\end{ruledtabular}
\end{table}

The width $\Gamma_{b_1}(q^2)$ for events with $q^2 > (m_{\omega}+m_{\pi})^2$ is parametrized via the $b_1(1235)^{\mp}\to\omega\pi^{\mp}$ decay:
\begin{linenomath*}
\begin{align}
\Gamma_{b_1}(q^2)\,=&\,\frac{m_{b_1}}{\sqrt{q^2}}\frac{p_{\omega}}{p_{0,\omega}}\Gamma_{b_1}\times {} \nonumber\\&\frac{m^4_{b_1} B^2_S(p_{\omega})+2 a^2_{DS} P^2_1(p_{\omega})B^2_D(p_{\omega})}{m^4_{b_1} + 2 a^2_{DS} P^2_1(p_{0,\omega})}{,}
\end{align}
\end{linenomath*}
where $m_{b_1(1235)}=(1230\pm 3)\,{\rm MeV}/c^2$, $\Gamma_{b_1}=(142\pm 9)\,{\rm MeV}/c^2$ \cite{pdg} and the factor $2$ accounts for the normalization of the $D$ wave relative to the $S$ wave.
For events with $q^2\le(m_{\omega}+m_{\pi})^2$, we use $\Gamma_{b_1}(q^2)=\Gamma_{b_1}$.

Since the typical values of $q^2$ for this resonant decay is close to the values corresponding to the $\rho(770)^{\mp}$ and
$\rho(1450)^{\mp}$ amplitudes, we assume that the FSI helicity phases $\phi_{\pm}$ in this decay are the same as for the $\rho(770)^{\mp}$ and $\rho(1450)^{\mp}$ contributions.
The free parameters for this contribution are the  relative magnitude and phase.

\begin{center}{\bf $\mathbf{\bar{B}^0 (B^0) \to D_1(2430)^0 (\bar{D}_1(2430)^0)  \omega \to D^{*\pm} \pi^{\mp} \omega}$, \\ $\mathbf{\bar{B}^0 (B^0) \to D_1(2420)^0 (\bar{D}_1(2420)^0) \omega \to D^{*\pm} \pi^{\mp} \omega}$}\end{center}

The notations $D'_1$ and $D_1$ for the $D_1(2430)^0$ $(\bar{D}_1(2430)^0)$ and $D_1(2420)^0$ $(\bar{D}_1(2420)^0)$, respectively, are used in this subsection.

The observable $D'_1$ and $D_1$ states are not charge-conjugation eigenstates but rather the admixtures between the pure states with $J^P_j=1^+_{1/2}$ and $J^P_j=1^+_{3/2}$, where the quantum number $j$ is the total angular momentum of the $u$ quark \cite{mixing}.
Mixing in the $jj$ coupling scheme is written as
\begin{linenomath*}
\begin{eqnarray}
M_{D_1 \pm}\,&=&\,\frac{1}{D_{D_1}(q^2)}\left(\vphantom{\frac{1}{f_{J,J+1}}}a_{1/2} e^{i \phi_{1/2}} \sin\omega M_{1/2 \pm}+ {} \nonumber \right.\\
&& \left. \,a_{3/2} e^{i \phi_{3/2}} \cos\omega e^{-i \varphi} M_{3/2 \pm}\vphantom{\frac{1}{f_{J,J+1}}}\right){,} \nonumber \\
M_{D'_1 \pm}\,&=&\,\frac{1}{D_{D'_1}(q^2)}\left(a_{1/2} e^{i \phi_{1/2}} \cos\omega M_{1/2 \pm}- {} \nonumber \right. \\
&&\left. \,a_{3/2} e^{i \phi_{3/2}} \sin\omega e^{i \varphi} M_{3/2 \pm}\vphantom{\frac{1}{f_{J,J+1}}}\right){,}
\end{eqnarray}
\end{linenomath*}
where $\omega$ and $\varphi$ are the mixing angles, $a_{1/2}$, $a_{3/2}$, $\phi_{1/2}$ and $\phi_{3/2}$ are the relative magnitudes and phases between the pure matrix elements $M_{1/2 \pm}$ and $M_{3/2\pm}$, which correspond to the $J^P_j=1^+_{1/2}$ and $J^P_j=1^+_{3/2}$ quantum numbers, respectively.

The pure matrix elements $M_{1/2 \pm}$ and $M_{3/2 \pm}$ are
\begin{linenomath*}
\begin{eqnarray}
M_{1/2 \pm}\,&=&\, i m^2_{D'_1} B_S(p_{D^*}) [\pm f_P(q^2)\mathcal{A}_{PS}+ {} \nonumber \\
&&\,i f_S(q^2) \mathcal{A}_{SS}+i f_D(q^2) \mathcal{A}_{DS}]{,}\nonumber  \\
M_{3/2 \pm}\,&=&\,- i P_1(p_{D^*}) B_D(p_{D^*}) [\pm f_P(q^2)\mathcal{A}_{PD}+ {} \nonumber \\
&&\,i f_S(q^2) \mathcal{A}_{SD}+i f_D(q^2) \mathcal{A}_{DD}]{,}
\label{md1}
\end{eqnarray}
\end{linenomath*}
where $P_1(p_{D^*})$ is defined in Eq.~(\ref{tilde_f}) with $p^2=m^2_{D^*}$ and the angular dependencies $\mathcal{A}_{SS}$, $\mathcal{A}_{PS}$, $\mathcal{A}_{DS}$, $\mathcal{A}_{SD}$, $\mathcal{A}_{PD}$ and $\mathcal{A}_{DD}$ have the form shown in Table~\ref{tanb1} except for one feature: the angular basis ($\theta$, $\phi$, $\beta$, $\psi$, $\xi$) describes here the $D^{**}$ production. The transition form factors
$f_S(q^2)$, $f_P(q^2)$ and $f_D(q^2)$ are given in Appendix \ref{sec:appC}. Since the mixing effect is predicted and confirmed to be small, we use in Eq.~(\ref{md1}) the physical mass $m_{D'_1}$ instead of the mass of the pure $j=1/2$ state.

The $q^2$-dependent widths $\Gamma_{D_1}(q^2)$ and $\Gamma_{D'_1}(q^2)$ of the $D_1$ and $D'_1$ states are parametrized via their decays to $D^{*\pm}\pi^{\mp}$:
\begin{linenomath*}
\begin{align}
\Gamma_{D'_1}(q^2)\,=&\,\frac{m_{D'_1}}{\sqrt{q^2}}B^2_{S}(p_{D^*})\frac{p_{D^*}}{p_{0,D^*}}\Gamma_{D'_1}{,}\nonumber\\
\Gamma_{D_1}(q^2)\,=&\,\frac{m_{D_1}}{\sqrt{q^2}}B^2_{D}(p_{D^*})\frac{P^2_1(p_{D^*})}{P^2_1(p_{0,D^*})}\frac{p_{D^*}}{p_{0,D^*}}\Gamma_{D_1}{,}
\end{align}
\end{linenomath*}
where $P_1(p_{D^*})$ and $P_1(p_{0,D^*})$ are defined as in Eq.~(\ref{md1}), $m_{D_1}=(2421\pm 1)\,{\rm MeV}/c^2$, $\Gamma_{D_1}=(27\pm 3)\,{\rm MeV}/c^2$, $m_{D'_1}=(2427\pm 36)\,{\rm MeV}/c^2$ and $\Gamma_{D'_1}=(384\pm 117)\,{\rm MeV}/c^2$ are fixed \cite{pdg}.

The free parameters describing these resonant amplitudes are the mixing angles $\omega$ and $\varphi$, the relative magnitudes and phases $a_{1/2}$, $a_{3/2}$, $\phi_{1/2}$ and $\phi_{3/2}$, and the relative normalizations and phases defined in the $f_S(q^2)$, $f_P(q^2)$ and $f_D(q^2)$ form factors.

\begin{center}{\bf $\mathbf{\bar{B}^0 (B^0) \to D^*_2(2460)^0  (\bar{D}^*_2(2460)^0)\omega \to D^{*\pm} \pi^{\mp} \omega}$}\end{center}

The notation $D^*_2$ for the $D^*_2(2460)^0$ ($\bar{D}^*_2(2460)^0$) is used in this subsection.

Since the $D^*_2$ state has the quantum numbers $J^P=2^+$, the $P$ and $F$ waves describing its production violate $P$- and $C$-parities, and thus have
the additional phase $\pi/2$ in comparison with the $D$ wave. The resonant matrix element is
\begin{linenomath*}
\begin{align}
M_{D^*_2 \pm}\,=&\,\frac{p_{D^*}^2B_D(p_{D^*})}{D_{D^*_2}(q^2)}\left(\vphantom{\frac{1}{f_{J,J+1}}}f_D(q^2)\mathcal{A}_{DD}\pm {} \nonumber\right.\\
&\,\left. i f_P(q^2) \mathcal{A}_{PD} \pm i f_F(q^2) \mathcal{A}_{FD}\vphantom{\frac{1}{f_{J,J+1}}}\right){,}
\label{md2}
\end{align}
\end{linenomath*}
where $\mathcal{A}_{PD}$, $\mathcal{A}_{DD}$ and $\mathcal{A}_{FD}$ are the angular dependencies describing each partial wave and shown in Table~\ref{tand2}, and the
transition form factors $f_P(q^2)$, $f_D(q^2)$ and $f_F(q^2)$ are parametrized in Appendix~\ref{sec:appC}.

\begin{table}[!t]
\begin{ruledtabular}
\caption{Angular dependencies corresponding to the $D^{*\pm}\pi^{\mp}$ quantum number $J^P=2^+$.
$L_1$ ($L_2$) is the relative orbital angular momentum between the $D^{*\pm}\pi^{\mp}$ and $\omega$ ($D^{*\pm}$ and $\pi^{\mp}$). The notations $c_{\alpha}=\cos\alpha$ and $s_{\alpha}=\sin\alpha$ are used. The angles $\theta$, $\phi$, $\beta$, $\psi$, $\xi$ correspond to the $D^{**}$ angular basis.}
\begin{tabular}{l l l}
& & \\
$L_1$ & $L_2$ & $\mathcal{A}_{L_1\,L_2}$ \\
\\
\hline
& & \\
$P$ & $D$ & $c_{\theta}s_{\beta}s_{\psi}s_{2\xi}+s_{\theta}c_{\phi}s_{\beta}s_{\psi}c_{2\xi}-s_{\theta}s_{\phi}s_{\beta}c_{\psi}c_{\xi}$ \\
& & \\
$D$ & $D$ & $s_{\theta}s_{\phi}s_{\beta}s_{\psi}+s_{\theta}c_{\phi}s_{\beta}c_{\psi}c_{\xi}$ \\
& & \\
$F$ & $D$ & $-3/2 c_{\theta}s_{\beta}s_{\psi}s_{2\xi}+s_{\theta}c_{\phi}s_{\beta}s_{\psi}c_{2\xi}-s_{\theta}s_{\phi}s_{\beta}c_{\psi}c_{\xi}$ \\
& & \\
\end{tabular}
\label{tand2}
\end{ruledtabular}
\end{table}

The $q^2$-dependent width $\Gamma_{D^*_2}(q^2)$ is determined via decays of the $D^*_2$ to $D^{*\pm}\pi^{\mp}$ and $D^{\pm}\pi^{\mp}$ with the probabilities of $40\%$ and $60\%$:
\begin{linenomath*}
\begin{align}
\Gamma_{D^*_2}(q^2)\,&=\,\frac{2}{5}\frac{\sqrt{q^2}}{m_{D^*_2}}\frac{p_{D^*}^5}{p_{0,D^*}^5}B^2_D(p_{D^*})\Gamma_{D^*_2}\, + {} \nonumber \\
&
\frac{3}{5}\frac{m_{D^*_2}}{\sqrt{q^2}}\frac{k_D^5}{k_{0,D}^5}B^2_D(k_D)\Gamma_{D^*_2} {,}
\end{align}
\end{linenomath*}
where $k_D$ is the $D^{\pm}$-meson momentum magnitude in the $D^*_2 \to D^{\pm} \pi^{\mp}$ decay computed in the $D^*_2$ rest frame, $k_{0,D}$ is the same momentum when $\sqrt{q^2}=m_{D^*_2}=(2463\pm 1)\,{\rm MeV}/c^2$,
and $\Gamma_{D^*_2}=(49\pm 1)\,{\rm MeV}/c^2$ \cite{pdg}.

The free parameters, describing this tensor contribution, are an overall magnitude and phase as well as normalizations and relative phases of the partial wave form factors in the matrix element.

\section{PARTIAL WAVE FORM FACTORS}
\label{sec:appC}

In this section, we obtain full expressions of the partial wave form factors used in the resonant matrix elements.
The symbols $p_{3\pi,B}$ and $p_{D^*,B}$ are used in this section for the magnitudes of the three-momenta of the $\omega$ decay product system and $D^{*\pm}$ in the $B$ meson rest frame, respectively.

The partial wave form factors $f_S(q^2)$, $f_P(q^2)$ and $f_D(q^2)$ describing the $\omega\pi^{\mp}$ resonance  production in Eqs.~(\ref{mrho}) and (\ref{mb1}) can be expressed in terms of three helicity amplitudes [$H_0(q^2)$ and $H_{\pm}(q^2)$], which correspond to three polarization states of the $D^{*\pm}$ (one longitudinal and two transverse), and two transverse helicity phases $\phi_{\pm}$ defined relative to the longitudinal amplitude $H_0(q^2)$:
\begin{linenomath*}
\begin{align}
f_S(q^2)\,=&\,\sqrt{\frac{q^2}{3}}\frac{H_+(q^2)e^{i\phi_+}+H_-(q^2)e^{i\phi_-}+H_0(q^2)}{\sqrt{3}}{,} \nonumber \\
f_P(q^2)\,=&\,\sqrt{\frac{q^2}{2}}\frac{H_+(q^2)e^{i\phi_+}-H_-(q^2)e^{i\phi_-}}{\sqrt{2}}{,}\nonumber \\
f_D(q^2)\,=&\,\sqrt{\frac{q^2}{6}}\frac{H_+(q^2)e^{i\phi_+}+H_-(q^2)e^{i\phi_-}-2 H_0(q^2)}{\sqrt{6}}{.}
\label{helampls}
\end{align}
\end{linenomath*}
Here, the additional factors $\sqrt{q^2/2}$, $\sqrt{q^2/3}$ and $\sqrt{q^2/6}$ are introduced to take into account the $q^2$-dependent vertex of the $\omega\pi$ production in the factorization assumption and the relative normalization fractions of the angular dependencies shown in Tables~\ref{tanrho} and \ref{tanb1}.
The helicity amplitudes $H_0(q^2)$ and $H_{\pm}(q^2)$ can be written in terms of three invariant form factors $A_1(q^2)$, $A_2(q^2)$ and $V(q^2)$ \cite{neubert,semil}:
\begin{linenomath*}
\begin{align}
H_0(q^2)\,=&\,-\frac{(m^2_B-m^2_{D^*}-q^2)(m_B+m_{D^*})}{2 m_{D^*} \sqrt{q^2}}A_1(q^2)+ {} \nonumber \\
&\,\frac{2 p^2_{D^*,B} m^2_B}{m_{D^*} \sqrt{q^2}(m_B+m_{D^*})} A_2(q^2){,} \nonumber \\
H_{\pm}(q^2)\,=&\,-(m_B+m_{D*}) A_1(q^2) \pm {} \nonumber \\
&\,\frac{2 p_{D^*,B} m_B}{m_B+m_{D^*}} V(q^2){.}
\label{helicity}
\end{align}
\end{linenomath*}
The invariant form factors $A_1(q^2)$, $A_2(q^2)$ and $V(q^2)$ describe the $\bar{B}^0 (B^0) \to D^{*\pm}$ transition and can be related to the Isgur-Wise function $h(w)$ ($w$ is the invariant four-velocity transfer) under the assumption of heavy quark symmetry \cite{neubert,semil}:
\begin{linenomath*}
\begin{eqnarray}
A_1(q^2)\,&=&\,\left(1-\frac{q^2}{(m_B+m_{D^*})^2}\right)\frac{m_B+m_{D^*}}{2\sqrt{m_B m_{D^*}}} h(w){,} \nonumber \\
A_2(q^2)\,&=&\,R_2\frac{m_B+m_{D^*}}{\sqrt{m_B m_{D^*}}} h(w){,} \nonumber \\
V(q^2)\,&=&\,R_1\frac{m_B+m_{D^*}}{\sqrt{m_B m_{D^*}}} h(w){,}
\label{invhqs}
\end{eqnarray}
\end{linenomath*}
where $R_1$ and $R_2$ are the relative factors,
and the Isgur-Wise function $h(w)$ can be parametrized as \cite{neubert,semil}
\begin{linenomath*}
\begin{equation}
h(w) = 1-{\mathbf \rho^2} (w-1)
\end{equation}
\end{linenomath*}
with
\begin{linenomath*}
\begin{equation}
w = \frac{m^2_B+m^2_{D^*}-q^2}{2 m_B m_{D^*}}{.}
\end{equation}
\end{linenomath*}
The values $R_1$, $R_2$ and $\rho^2$ used in our analysis were measured by the {\it BABAR} collaboration in the $\bar{B}^0 \to D^{*+} e^- \bar{\nu}_e$ decay \cite{semil}:
\begin{linenomath*}
\begin{eqnarray}
R_1\,&=&\,1.40 \pm 0.06{,} \nonumber \\
R_2\,&=&\,0.87 \pm 0.04{,} \nonumber \\
\rho^2\,&=&\,0.79 \pm 0.06 {.}
\label{hqs}
\end{eqnarray}
\end{linenomath*}

The partial wave form factors $f_S(q^2)$, $f_P(q^2)$, $f_D(q^2)$ and $f_F(q^2)$ describing the $D^{**}$ resonance production and introduced in Eqs.~(\ref{md1}) and (\ref{md2}) contain the momentum dependencies, corresponding to the definite angular orbital momenta of the decay products in the $B$ meson rest frame. These dependencies can be explicitly extracted.
The form factors for the $D_1(2430)^0$  ($\bar{D}_1(2430)^0$) and $D_1(2420)^0$ ($\bar{D}_1(2420)^0$) production can be written as \cite{jhep}
\begin{linenomath*}
\begin{eqnarray}
f_S(q^2)\,&=&\,-\frac{R_S}{\sqrt{3}} m^2_B B_S(q^2) e^{i \phi_S}{,} \nonumber \\
f_P(q^2)\,&=&\,\frac{R_P}{\sqrt{2}} m_B p_{3\pi,B} B_P(q^2) e^{i \phi_P}{,}\nonumber \\
f_D(q^2)\,&=&\,\frac{1}{\sqrt{6}} P_1(p_{3\pi,D^{**}}) B_D(q^2){,}
\label{fp2d}
\end{eqnarray}
\end{linenomath*}
where $R_S$ and $R_P$ ($\phi_S$ and $\phi_P$) are magnitudes (phases) of $S$- and $P$-wave amplitudes defined relative to $D$-wave amplitude, $P_1(p_{3\pi,D^{**}})$ is defined in Eq.~(\ref{tilde_f}) and $p_{3\pi,D^{**}}$ is the three-momentum magnitude of the $\omega$ decay product defined in the $D^{*\pm}\pi^{\mp}$ rest frame.
Similar expressions can be written for the $D^*_2 (2460)^0$ ($\bar{D}^*_2(2460)^0$) production \cite{jhep}:
\begin{linenomath*}
\begin{eqnarray}
f_P(q^2)\,&=&\,- \frac{R_P}{\sqrt{3}}\frac{m^2_B p_{3\pi,B}}{m_{D^*_2}} B_P(q^2) e^{i\phi_P}{,} \nonumber \\
f_D(q^2)\,&=&\,\frac{R_D}{\sqrt{2}}\frac{m_B p^2_{3\pi,B}}{m_{D^*_2}} B_D(q^2)e^{i\phi_D}{,} \nonumber \\
f_F(q^2)\,&=&\,\frac{P_2(p_{3\pi,D^{**}}) p_{3\pi,B}}{2m_{D^*_2}} B_F(q^2){,}
\end{eqnarray}
\end{linenomath*}
where
$R_P$ and $R_D$ ($\phi_P$ and $\phi_D$) are magnitudes (phases) of $P$- and $D$-wave amplitudes defined relative to $F$-wave amplitude,
$P_2(p_{3\pi,D^{**}})$ is defined in Eq.~(\ref{tilde_f}) with $J=2$ and $m_{D^*_2}$ is the mass of the $D^*_2 (2460)^0$ ($\bar{D}^*_2(2460)^0$) state.

\end{appendices}


\begin{thebibliography}{99}

\bibitem[1]{neubert}
M.~Neubert,
Phys.\ Rep.\ {\bf 245}, 259 (1994).

\bibitem[2]{uraltsev}
N.~Uraltsev,
Phys.\ Lett.\ B\ {\bf 501}, 86 (2001).

\bibitem[3]{pdg}
K.A.~Olive {\em et al.}
(Particle Data Group),
Chin. Phys. C \textbf{38}, 090001 (2014).

\bibitem[4]{babar_spectr}
P.~del Amo Sanchez {\em et al.}
({\it BABAR} Collaboration),
Phys.\ Rev.\ D\ {\bf 82}, 111101 (2010).

\bibitem[5]{lhcb_spectr}
R.~Aaij {\em et al.}
(LHCb Collaboration),
J. High Energy Phys. \ { 09} (2013) 145.

\bibitem[6]{mixing}
F.E.~Close and E.S.~Swanson,
Phys.\ Rev.\ D\ {\bf 72}, 094004 (2005).

\bibitem[7]{dsmasses}
P.~Colangelo, F.~De~Fazio, and R.~Ferrandes,
Mod.\ Phys.\ Lett.\ A\ {\bf 19}, 2083 (2004);
E.S.~Swanson,
Phys.\ Rep.\ {\bf 429}, 243 (2006),
and references therein.

\bibitem[8]{balagura}
V.~Balagura {\em et al.}
(Belle Collaboration),
Phys.\ Rev.\ D\ {\bf 77}, 032001 (2008).

\bibitem[9]{ckm}
N.~Cabibbo,
Phys.\ Rev.\ Lett.\ {\bf 10}, 531 (1963);
M.~Kobayashi and T.~Maskawa,
Prog.\ Theor.\ Phys.\ {\bf 49}, 652 (1973).

\bibitem[10]{bsemil}
D.~Liventsev {\em et al.}
(Belle Collaboration),
Phys.\ Rev.\ D\ {\bf 77}, 091503 (2008);
B.~Aubert {\em et al.}
({\it BABAR} Collaboration),
Phys.\ Rev.\ Lett.\ {\bf 100}, 151802 (2008).

\bibitem[11]{b0todstpi}
A.~Kuzmin {\em et al.}
(Belle Collaboration),
  Phys.\ Rev.\  D\ {\bf 76}, 012006 (2007);
K.~Abe {\em et al.}
(Belle Collaboration), arXiv:hep-ex/0412072.

\bibitem[12]{b-todstpi}
K.~Abe {\em et al.}
(Belle Collaboration),
Phys.\ Rev.\ D\ {\bf 69}, 112002 (2004);
B.~Aubert {\em et al.}
({\it BABAR} Collaboration),
Phys.\ Rev.\ D\ {\bf 79}, 112004 (2009).

\bibitem[13]{lhcbtod3pi}
R.~Aaji {\em et al.}
(LHCb Collaboration),
Phys.\ Rev.\ D\ {\bf 87}, 092001 (2013).

\bibitem[14]{lhcb0todpipi}
R.~Aaij {\em et al.}
(LHCb Collaboration),
arXiv:1505.01710,
submitted to Phys.\ Rev.\ D.

\bibitem[15]{wilson}
K.G.~Wilson,
Phys.\ Rev.\ {\bf 179}, 1499 (1969).

\bibitem[16]{fddst}
S.~Veseli and I.~Dunietz,
Phys.\ Rev.\ D\ {\bf 54}, 6803 (1996);
H.Y.~Cheng, C.K.~Chua, and C.W.~Hwang,
Phys.\ Rev.\ D\ {\bf 69}, 074025 (2004).

\bibitem[17]{scet}
S.~Mantry,
Phys.\ Rev.\ D\ {\bf 70}, 114006 (2004).

\bibitem[18]{revddst}
H.Y.~Cheng and C.K.~Chua,
Phys.\ Rev.\ D\ {\bf 74}, 034020 (2006).

\bibitem[19]{fscc}
S.~Weinberg,
Phys.\ Rev.\ {\bf 112}, 1375 (1958).

\bibitem[20]{betadecay}
T.~Sumikama {\em et al.},
Phys.\ Lett.\ B\ {\bf 664}, 235 (2008).

\bibitem[21]{taudecay}
B.~Aubert {\em et al.}
({\it BABAR} Collaboration),
Phys.\ Rev.\ Lett.\ {\bf 103}, 041802 (2009);
P.~del Amo Sanchez {\em et al.}
({\it BABAR} Collaboration),
Phys.\ Rev.\ D\ {\bf83}, 032002 (2011);
B.~Aubert {\em et al.}
({\it BABAR} Collaboration),
Phys.\ Rev.\ D\ {\bf 77}, 112002 (2008).

\bibitem[22]{rhoradial}
T.~Barnes,\ F.E.~Close,\ P.R.~Page, and E.S.~Swanson,
Phys.\ Rev.\ D\ {\bf 55}, 4157 (1997).

\bibitem[23]{rhohybrid}
F.E.~Close and P.R.~Page,
Phys.\ Rev.\ D\ {\bf 56}, 1584 (1997).

\bibitem[24]{rhomixture}
A.~Donnachie and Yu.S.~Kalashnikova,
Phys.\ Rev.\ D\ {\bf 60}, 114011 (1999).

\bibitem[25]{factortest}
J.~Korner and G.~Goldstein,
Phys.\ Lett.\ B\ {\bf 89}, 105 (1979).

\bibitem[26]{b0todstomega}
J.P.~Lees {\em et al.}
({\it BABAR} Collaboration),
Phys.\ Rev.\ D\ {\bf 84}, 112007 (2011);
{\bf 87}, 039901(E) (2013).

\bibitem[27]{btophiks}
M.~Prim {\em et al.}
(Belle Collaboration),
Phys.\ Rev.\ D\ {\bf 88}, 072004 (2013);
R.~Aaji {\em et al.}
(LHCb Collaboration),
J. High Energy Phys. {05} (2014) 069.

\bibitem[28]{scetddst}
A.E.~Brechman, S.~Mantry, and I.W.~Stewart,
Phys.\ Lett.\ B\ {\bf 608}, 77 (2005).

\bibitem[29]{cleo}
J.P.~Alexander {\em et al.}
(CLEO Collaboration),
Phys.\ Rev.\ D\ {\bf 64}, 092001 (2001).

\bibitem[30]{babar}
B.~Aubert {\em et al.}
({\it BABAR} Collaboration),
Phys.\ Rev.\ D\ {\bf 74}, 012001 (2006).

\bibitem[31]{kekb}
S.~Kurokawa and E.~Kikutani,
Nucl.\ Instrum.\ Methods\ Phys.\ Res.,\ Sect.\ A\ {\bf 499}, 1 (2003), and other papers included in this volume;
T.~Abe {\em et al.},
Prog.\ Theor.\ Exp.\ Phys.\ {\bf 2013}, 03A001 (2013) and following articles up to 03A011.

\bibitem[32]{belledet}
A.~Abashian {\em et al.}
(Belle Collaboration),
Nucl.\ Instrum \ Methods\ Phys.\ Res.,\ Sect.\ A\ {\bf 479},
117 (2002); also see detector section in J.~Brodzicka {\em et al.},
Prog.\ Theor.\ Exp.\ Phys. {\bf 2012}, 04D001 (2012).

\bibitem[33]{evtgen}
D.J.~Lange,
Nucl.\ Instrum.\ Methods\ Phys.\ Res.,\ Sect.\ A\ A\ {\bf 462}, 152 (2001).

\bibitem[34]{photos}
E.~Barberio and Z.~Was, Comput.\ Phys.\  Commun.\ {\bf 79},
291 (1994).

\bibitem[35]{geant}
R.~Brun {\em et al.},
GEANT 3.21, CERN DD/EE/84-1, 1984.

\bibitem[36]{jhep}
D.V.~Matvienko, A.S.~Kuzmin, and S.I~Eidelman,
J. High Energy Phys. {09} (2011) 129.

\bibitem[37]{brandt}
S.~Brandt, Ch.~Peyrou, R.~Sosnowski, and A.~Wroblewski,
Phys.\ Lett.\ {\bf 12}, 57 (1964).

\bibitem[38]{pi+}
We neglect the symmetry relative to variables describing the kinematics of the $\pi^+$ mesons in the final state because of the narrowness of the $D^{*+}$.

\bibitem[39]{NBB}
We assume equal rates for
$\Upsilon(4S) \to B^+B^-$ and $\Upsilon(4S) \to B^0\bar{B}^0$, and $N_B$ is equal to the number of $B$ pairs produced.

\bibitem[40]{tracking}
W.~Dungel {\em et al.},
(Belle Collaboration),
Phys.\ Rev.\  D\ {\bf 82}, 112007 (2010).

\bibitem[41]{pi0}
S.~Ryu {\em et al.},
(Belle Collaboration),
Phys.\ Rev.\  D\ {\bf 89}, 072009 (2014).

\bibitem[42]{pid}
E.~Nakano,
Nucl.\ Instrum.\ Methods\ Phys.\ Res.,\ Sect.\ A\ {\bf 494},
402 (2002);

\bibitem[43]{NBBsys}
K.~Abe {\em et al.}
(Belle Collaboration),
Phys.\ Rev.\  Lett.\ {\bf 87}, 101801 (2001).

\bibitem[44]{isobar}
G.N.~Fleming,
Phys.\ Rev.\ {\bf 135}, B551 (1964).

\bibitem[45]{karda}
M.N.~Achasov {\em et al.}
(SND Collaboration),
Phys.\ Rev.\ D\ {\bf 88}, 054013 (2013).

\bibitem[46]{gof}
M.~Williams,
J. Instrum. {\bf 5}, 09004 (2010) and references therein.

\bibitem[47]{semil}
B.~Aubert {\em et al.}
({\it BABAR} Collaboration),
Phys.\ Rev.\ D\ {\bf 74}, 092004 (2006).

\bibitem[48]{ee4pi}
R.~Akhmetshin {\em et al.}
(CMD-2 Collaboration),
Phys.\ Lett.\ B\ {\bf 466}, 392 (1999).

\bibitem[49]{tau4pi}
K.~Edwards {\em et al.}
(CLEO Collaboration),
Phys.\ Rev.\ D\ {\bf 61}, 072003 (2000).

\bibitem[50]{tsai}
Y.-S.~Tsai,
Phys.\ Rev.\ D\ {\bf 4}, 2821 (1971);
{\bf 13}, 771(e) (1976).

\bibitem[51]{cleggdon}
A.B.~Clegg and A.~Donnachie,
Z.\ Phys.\ C\ {\bf 62}, 455 (1994).

\bibitem[52]{btodstrho}
S.E.~Csorna {\em et al.}
(CLEO Collaboration),
Phys.\ Rev.\ D\ {\bf 67}, 112002 (2003).

\bibitem[53]{ccs}
H.Y.~Cheng, C.K.~Chua, and A.~Soni,
Phys.\ Rev.\ D\ {\bf 71}, 014030 (2005).

\bibitem[54]{blwe}
J.~Blatt and V.~Weisskopf, Theoretical Nuclear Physics,(John Wiley and Sons, New York, 1952), p. 361 (1952);
F.~Von Hippel, C.~Quigg,
Phys.\ Rev.\ D\ {\bf 5}, 624 (1972).

\bibitem[55]{achasov}
M.N.~Achasov {\em et al.}
(SND Collaboration),
Phys.\ Lett.\ B\ {\bf 486}, 29 (2000).

\bibitem[56]{lublinsky}
M.~Lublinsky,
Phys.\ Rev.\ D\ {\bf 55}, 249 (1997).

\bibitem[57]{brookhaven}
M.~Nozar {\em et al.}
(E852 Collaboration),
Phys.\ Lett.\ B\ {\bf 541}, 35 (2002).

\end{thebibliography}
\end{document}